\documentclass[journal]{IEEEtran}

\newcommand{\ITWTwoOTwoFivSty}{
\documentclass[conference,letter]{IEEEtran}
\addtolength{\topmargin}{1.8mm} 
\addtolength{\textheight}{-3mm}}

\usepackage{mdwmath}
\usepackage{mdwtab}
\usepackage{tikz}
\usepackage{mathrsfs}
\usepackage{graphicx}
\usepackage{amsmath}
\usepackage{latexsym}
\usepackage{amssymb}
\usepackage{amsmath,bm}
\usepackage{xcolor}
\usepackage{comment}
\usepackage{cite}
\usepackage{url}
\usepackage[T3,T1]{fontenc}
\DeclareSymbolFont{tipa}{T3}{cmr}{m}{n}
\DeclareMathAccent{\invbreve}{\mathalpha}{tipa}{16}

\global\long\def\L{\mathsf{L}}
\global\long\def\D{\mathsf{D}}
\global\long\def\Sgen{\mathcal{S}_\mathsf{gen}}
\global\long\def\Kgen{\mathcal{K}_\mathsf{gen}}

\newcommand{\theTa}{\omega}

\begin{document}
%
\title{
A Framework for Distributed Source Encryption using Mutual Information
  Security Criterion and 
the Strong Converse Theorem   
}
\author{%
	\IEEEauthorblockN{
       Yasutada Oohama and Bagus Santoso\\
    }
	\IEEEauthorblockA{
        University of Electro-Communications, Tokyo, Japan
    \\ 
        Email: \url{{oohama,santoso.bagus}@uec.ac.jp}
    }
%
}
\maketitle

\begin{abstract}
We reinvestigate the general distributed secure source 
coding based on the common key cryptosystem proposed 
by Oohama and Santoso (ITW 2021). They proposed 
a framework of distributed soure encryption and 
derived the necessary and sufficient conditions 
to have reliable and secure transmission. 
However, the bounds of the rate region, which 
specifies both necessary and sufficient conditions 
to have reliable and secure transmission under 
the proposed cryptosystem, were derived based 
on a self-tailored \emph{non-standard} security 
criterion. In this paper we adopt the \emph{standard 
security criterion}, i.e., \emph{standard mutual 
information}. We further improve 
the framework of Oohama and Santoso by relaxing  
some constraints on encoder and decoder functions. 
For our new framework and security criterion, 
we establish the necessary and sufficient conditions 
for reliable and secure transmission on
the fixed value of constrains for reliability 
and security. 
We further establish the the strong converse 
theorem. Information spectrum method and 
a variant of Birkhoff-von Neumann theorem play 
an important role in deriving the result. 
\end{abstract}

\newcommand{\Draft}{

the mutual information as a 
security criterion. This criterion directly measures 
a information leakage on plain texts from cipher text.

In this paper, we successfully 
determine the rate region under 
the fixed value of constrains 
for reliability and security. 
We further establish the  
For the security criterion we 
establish the rate region based on  
the \emph{standard security criterion},
i.e., \emph{standard mutual information}. 
A variant of Birkhoff-von Neumann 
theorem plays an important role in deriving our result.     

}

\newcommand{\AbstOld}{
We consider the distributed secure source coding based 
on the common key cryptosystem. This cryptosystem was 
posed and investigated by Oohama and Santoso (2022). 
In this paper we adopt the mutual information as a 
security criterion. This criterion directly measures 
a information leakage on plain texts from cipher text. 
Under this secure criterion, we establish the strong 
converse theorem. A variant of Birkhoff-von Neumann 
theorem plays an important role in deriving the 
result.}

\newcommand{\AbstOldTwo}{

This criterion directly measures a information 
leakage on plain texts from cipher text.      
Under this secure criterion, we establish 
the strong converse theorem. 

Our result yields that    

}

\IEEEpeerreviewmaketitle

\newcommand{\qed}{\hfill$\square$}
\newcommand{\suchthat}{\mbox{~s.t.~}}
\newcommand{\markov}{\leftrightarrow}

\newcommand{\argmax}{\mathop{\rm argmax}\limits}
\newcommand{\argmin}{\mathop{\rm argmin}\limits}

\newcommand{\ExP}{\rm e}

\newcommand{\Cls}{class NL}
\newcommand{\vSpa}{\vspace{0.3mm}}
\newcommand{\Prmt}{\zeta}
\newcommand{\pj}{\omega_n}

\newfont{\bg}{cmr10 scaled \magstep4}
\newcommand{\bigzerol}{\smash{\hbox{\bg 0}}}
\newcommand{\bigzerou}{\smash{\lower1.7ex\hbox{\bg 0}}}
\newcommand{\nbn}{\frac{1}{n}}
\newcommand{\ra}{\rightarrow}
\newcommand{\la}{\leftarrow}
\newcommand{\ldo}{\ldots}
\newcommand{\typi}{A_{\epsilon }^{n}}
\newcommand{\bx}{\hspace*{\fill}$\Box$}
\newcommand{\pa}{\vert}
\newcommand{\ignore}[1]{}


\newtheorem{proposition}{Proposition}
\newtheorem{definition}{Definition}
\newtheorem{theorem}{Theorem}
\newtheorem{lemma}{Lemma}
\newtheorem{corollary}{Corollary}
\newtheorem{remark}{Remark}
\newtheorem{property}{Property}
\newtheorem{condition}{Condition}

\newcommand{\defeq}{:=}

\newcommand{\Qed}{\hbox{\rule[-2pt]{3pt}{6pt}}}
\newcommand{\beq}{\begin{equation}}
\newcommand{\eeq}{\end{equation}}
\newcommand{\beqa}{\begin{eqnarray}}
\newcommand{\eeqa}{\end{eqnarray}}
\newcommand{\beqno}{\begin{eqnarray*}}
\newcommand{\eeqno}{\end{eqnarray*}}
\newcommand{\ba}{\begin{array}}
\newcommand{\ea}{\end{array}}

\newcommand{\vc}[1]{\mbox{\boldmath $#1$}}
\newcommand{\lvc}[1]{\mbox{\scriptsize \boldmath $#1$}}
\newcommand{\svc}[1]{\mbox{\tiny \boldmath $#1$}}

\newcommand{\wh}{\widehat}
\newcommand{\wt}{\widetilde}
\newcommand{\ts}{\textstyle}
\newcommand{\ds}{\displaystyle}
\newcommand{\scs}{\scriptstyle}
\newcommand{\vep}{\varepsilon}
\newcommand{\rhp}{\rightharpoonup}
\newcommand{\cl}{\circ\!\!\!\!\!-}
\newcommand{\bcs}{\dot{\,}.\dot{\,}}
\newcommand{\eqv}{\Leftrightarrow}
\newcommand{\leqv}{\Longleftrightarrow}

\newcommand{\irr}[1]{{\color[named]{Red}#1\normalcolor}}

\newcommand{\hugel}{{\arraycolsep 0mm
                    \left\{\ba{l}{\,}\\{\,}\ea\right.\!\!}}
\newcommand{\Hugel}{{\arraycolsep 0mm
                    \left\{\ba{l}{\,}\\{\,}\\{\,}\ea\right.\!\!}}
\newcommand{\HUgel}{{\arraycolsep 0mm
                    \left\{\ba{l}{\,}\\{\,}\\{\,}\vspace{-1mm}
                    \\{\,}\ea\right.\!\!}}
\newcommand{\huger}{{\arraycolsep 0mm
                    \left.\ba{l}{\,}\\{\,}\ea\!\!\right\}}}
\newcommand{\Huger}{{\arraycolsep 0mm
                    \left.\ba{l}{\,}\\{\,}\\{\,}\ea\!\!\right\}}}
\newcommand{\HUger}{{\arraycolsep 0mm
                    \left.\ba{l}{\,}\\{\,}\\{\,}\vspace{-1mm}
                    \\{\,}\ea\!\!\right\}}}

\newcommand{\hugebl}{{\arraycolsep 0mm
                    \left[\ba{l}{\,}\\{\,}\ea\right.\!\!}}
\newcommand{\Hugebl}{{\arraycolsep 0mm
                    \left[\ba{l}{\,}\\{\,}\\{\,}\ea\right.\!\!}}
\newcommand{\HUgebl}{{\arraycolsep 0mm
                    \left[\ba{l}{\,}\\{\,}\\{\,}\vspace{-1mm}
                    \\{\,}\ea\right.\!\!}}
\newcommand{\hugebr}{{\arraycolsep 0mm
                    \left.\ba{l}{\,}\\{\,}\ea\!\!\right]}}
\newcommand{\Hugebr}{{\arraycolsep 0mm
                    \left.\ba{l}{\,}\\{\,}\\{\,}\ea\!\!\right]}}
\newcommand{\HUgebr}{{\arraycolsep 0mm
                    \left.\ba{l}{\,}\\{\,}\\{\,}\vspace{-1mm}
                    \\{\,}\ea\!\!\right]}}

\newcommand{\hugecl}{{\arraycolsep 0mm
                    \left(\ba{l}{\,}\\{\,}\ea\right.\!\!}}
\newcommand{\Hugecl}{{\arraycolsep 0mm
                    \left(\ba{l}{\,}\\{\,}\\{\,}\ea\right.\!\!}}
\newcommand{\hugecr}{{\arraycolsep 0mm
                    \left.\ba{l}{\,}\\{\,}\ea\!\!\right)}}
\newcommand{\Hugecr}{{\arraycolsep 0mm
                    \left.\ba{l}{\,}\\{\,}\\{\,}\ea\!\!\right)}}

\newcommand{\hugepl}{{\arraycolsep 0mm
                    \left|\ba{l}{\,}\\{\,}\ea\right.\!\!}}
\newcommand{\Hugepl}{{\arraycolsep 0mm
                    \left|\ba{l}{\,}\\{\,}\\{\,}\ea\right.\!\!}}
\newcommand{\hugepr}{{\arraycolsep 0mm
                    \left.\ba{l}{\,}\\{\,}\ea\!\!\right|}}
\newcommand{\Hugepr}{{\arraycolsep 0mm
                    \left.\ba{l}{\,}\\{\,}\\{\,}\ea\!\!\right|}}

\newcommand{\MEq}[1]{\stackrel{
{\rm (#1)}}{=}}

\newcommand{\MLeq}[1]{\stackrel{
{\rm (#1)}}{\leq}}

\newcommand{\ML}[1]{\stackrel{
{\rm (#1)}}{<}}

\newcommand{\MGeq}[1]{\stackrel{
{\rm (#1)}}{\geq}}

\newcommand{\MG}[1]{\stackrel{
{\rm (#1)}}{>}}

\newcommand{\MPreq}[1]{\stackrel{
{\rm (#1)}}{\preceq}}

\newcommand{\MSueq}[1]{\stackrel{
{\rm (#1)}}{\succeq}}

\newcommand{\MSubeq}[1]{\stackrel{
{\rm (#1)}}{\subseteq}}

\newcommand{\MSupeq}[1]{\stackrel{
{\rm (#1)}}{\supseteq}}

\newcommand{\MRarrow}[1]{\stackrel{
{\rm (#1)}}{\Rightarrow}}

\newcommand{\MLarrow}[1]{\stackrel{
{\rm (#1)}}{\Leftarrow}}

\newcommand{\MLRarrow}[1]{\stackrel{
{\rm (#1)}}{\Leftrightarrow}}

%
%
\newcommand{\SZZpp}{

\newcommand{\vcc}{{c}^n}
\newcommand{\vck}{{k}^n}
\newcommand{\vcx}{{x}^n}
\newcommand{\vcy}{{y}^n}
\newcommand{\vcz}{{z}^n}
\newcommand{\vckone}{{k}_1^n}
\newcommand{\vcktwo}{{k}_2^n}
\newcommand{\vcxone}{{x}^n}

\newcommand{\vcxtwo}{{x}_2^n}
\newcommand{\vcyone}{{y}_1^n}
\newcommand{\vcytwo}{{y}_2^n}

\newcommand{\cvcx}{\check{x}^n}
\newcommand{\cvcy}{\check{y}^n}
\newcommand{\cvcz}{\check{z}^n}
\newcommand{\cvcxone}{\check{x}^n}

\newcommand{\cvcxtwo}{\check{x}_2^n}

\newcommand{\hvcx}{\widehat{x}^n}
\newcommand{\hvcy}{\widehat{y}^n}
\newcommand{\hvcz}{\widehat{z}^n}
\newcommand{\hvckone}{\widehat{k}_1^n}
\newcommand{\hvcktwo}{\widehat{k}_2^n}

\newcommand{\hvcxone}{\widehat{x}^n}

\newcommand{\hvcxtwo}{\widehat{x}_2^n}

\newcommand{\lvcc}{{c}^n}
\newcommand{\lvck}{{k}^n}
\newcommand{\lvcx}{{x}^n}
\newcommand{\lvcy}{{y}^n}
\newcommand{\lvcz}{{z}^n}

\newcommand{\lvckone}{{k}_1^n}
\newcommand{\lvcktwo}{{k}_2^n}
\newcommand{\lvcxone}{{x}^n}

\newcommand{\lvcxtwo}{{x}_2^n}
\newcommand{\lvcyone}{{y}_1^n}
\newcommand{\lvcytwo}{{y}_2^n}

\newcommand{\clvcxone}{\check{x}^n}

\newcommand{\clvcxtwo}{\check{x}_2^n}

\newcommand{\hlvckone}{\widehat{k}_1^n}
\newcommand{\hlvcktwo}{\widehat{k}_2^n}

\newcommand{\hlvcxone}{\widehat{x}^n}

\newcommand{\hlvcxtwo}{\widehat{x}_2^n}

\newcommand{\rvcc}{{C}^n}
\newcommand{\rvck}{{K}^n}
\newcommand{\rvcx}{{X}^n}
\newcommand{\rvcy}{{Y}^n}
\newcommand{\rvcz}{{Z}^n}
\newcommand{\rvccone}{{C}_1^n}
\newcommand{\rvcctwo}{{C}_2^n}
\newcommand{\rvckone}{{K}_1^n}
\newcommand{\rvcktwo}{{K}_2^n}
\newcommand{\rvcxone}{{X}^n}

\newcommand{\rvcxtwo}{{X}_2^n}
\newcommand{\rvcyone}{{Y}_1^n}
\newcommand{\rvcytwo}{{Y}_2^n}
\newcommand{\hrvcx}{\widehat{X}^n}
\newcommand{\hrvcxone}{\widehat{X}_1^n}
\newcommand{\hrvcxtwo}{\widehat{X}_2^n}

\newcommand{\lrvcc}{{C}^n}
\newcommand{\lrvck}{{K}^n}
\newcommand{\lrvcx}{{X}^n}
\newcommand{\lrvcy}{{Y}^n}
\newcommand{\lrvcz}{{Z}^n}
\newcommand{\lrvckone}{{K}_1^n}
\newcommand{\lrvcktwo}{{K}_2^n}

\newcommand{\lrvcxone}{{X}^n}
\newcommand{\lrvcxtwo}{{X}_2^n}
\newcommand{\lrvcyone}{{Y}_1^n}
\newcommand{\lrvcytwo}{{Y}_2^n}
\newcommand{\rvcci}{{C}_i^n}
\newcommand{\rvcki}{{K}_i^n}
\newcommand{\rvcxi}{{X}_i^n}
\newcommand{\rvcyi}{{Y}_i^n}
\newcommand{\hrvcxi}{\widehat{X}_i^n}
\newcommand{\vcki}{{k}_i^n}
\newcommand{\vcsi}{{s}_i^n}
\newcommand{\vcti}{{t}_i^n}
\newcommand{\vcvi}{{v}_i^n}
\newcommand{\vcwi}{{w}_i^n}
\newcommand{\vcxi}{{x}_i^n}
\newcommand{\vcyi}{{y}_i^n}

\newcommand{\vcs}{{s}^n}
\newcommand{\vct}{{t}^n}
\newcommand{\vcv}{{v}^n}
\newcommand{\vcw}{{w}^n}
}

\newcommand{\crvcx}{\check{\vc X}}
\newcommand{\crvcxi}{\check{\vc X}_i}
\newcommand{\crvcxone}{\check{\vc X}_1}
\newcommand{\crvcxtwo}{\check{\vc X}_2}

   \newcommand{\lcrvcx}{\check{\vc X}}
\newcommand{\lcrvcxone}{\check{\lvc X}_1}
\newcommand{\lcrvcxtwo}{\check{\lvc X}_2}

  \newcommand{\cvcc}{\check{c}^m}
 \newcommand{\lcvcc}{\check{c}^m}
 \newcommand{\crvcc}{\check{C}^m}
\newcommand{\lcrvcc}{\check{C}^m}

  \newcommand{\cvccone}{\check{c}_1^{m_1}}
 \newcommand{\crvccone}{\check{C}_1^{m_1}}
  \newcommand{\cvcctwo}{\check{c}_2^{m_2}}
 \newcommand{\crvcctwo}{\check{C}_2^{m_2}}

\newcommand{\vca}{{\vc a}}
\newcommand{\vcb}{{\vc b}}
\newcommand{\vcc}{{\vc c}}
\newcommand{\vck}{{\vc k}}
\newcommand{\vcx}{{\vc x}}
\newcommand{\vcy}{{\vc y}}
\newcommand{\vcz}{{\vc z}}
\newcommand{\vckone}{{\vc k}_1}
\newcommand{\vcktwo}{{\vc k}_2}
\newcommand{\vcxone}{{\vc x}_1}
\newcommand{\vcxtwo}{{\vc x}_2}
\newcommand{\vcyone}{{\vc y}_1}
\newcommand{\vcytwo}{{\vc y}_2}

\newcommand{\cvcx}{\check{\vc x}}
\newcommand{\cvcy}{\check{\vc y}}
\newcommand{\cvcz}{\check{\vc z}}
\newcommand{\cvcxone}{\check{\vc x}_1}
\newcommand{\cvcxtwo}{\check{\vc x}_2}
\newcommand{\cvcxi}  {\check{\vc x}_i}

\newcommand{\hvcx}{\widehat{\vc x}}
\newcommand{\hvcy}{\widehat{\vc y}}
\newcommand{\hvcz}{\widehat{\vc z}}
\newcommand{\hvckone}{\widehat{\vc k}_1}
\newcommand{\hvcktwo}{\widehat{\vc k}_2}
\newcommand{\hvcki}  {\widehat{\vc k}_i}
\newcommand{\hvcxone}{\widehat{\vc x}_1}
\newcommand{\hvcxtwo}{\widehat{\vc x}_2}
\newcommand{\hvcxi}  {\widehat{\vc x}_i}

\newcommand{\lvca}{{\lvc a}}
\newcommand{\lvcb}{{\lvc b}}
\newcommand{\lvcc}{{\lvc c}}
\newcommand{\lvck}{{\lvc k}}
\newcommand{\lvcx}{{\lvc x}}
\newcommand{\lvcy}{{\lvc y}}
\newcommand{\lvcz}{{\lvc z}}

\newcommand{\lvckone}{{\lvc k}_1}
\newcommand{\lvcktwo}{{\lvc k}_2}
\newcommand{\svckone}{{\svc k}_1}
\newcommand{\svcktwo}{{\svc k}_2}

\newcommand{\lvcki}  {{\lvc k}_i}
\newcommand{\lvcxi}  {{\lvc x}_i}
\newcommand{\lvcxone}{{\lvc x}_1}
\newcommand{\lvcxtwo}{{\lvc x}_2}

\newcommand{\lvcyone}{{y}_1}
\newcommand{\lvcytwo}{{y}_2}

\newcommand{\clvcxone}{\check{\lvc x}_1}
\newcommand{\clvcxtwo}{\check{\lvc x}_2}
\newcommand{\clvcxi}{\check{\lvc x}_i}

\newcommand{\hlvckone}{\widehat{k}_1}
\newcommand{\hlvcktwo}{\widehat{k}_2}
\newcommand{\hlvcxone}{\widehat{x}_1}
\newcommand{\hlvcxtwo}{\widehat{x}_2}

\newcommand{\rvcc}{{\vc C}}
\newcommand{\rvck}{{\vc K}}
\newcommand{\rvcx}{{\vc X}}

\newcommand{\rvcy}{{\vc Y}}
\newcommand{\rvcz}{{\vc Z}}
\newcommand{\rvccone}{{\vc C}_1}
\newcommand{\rvcctwo}{{\vc C}_2}
\newcommand{\rvckone}{{\vc K}_1}
\newcommand{\rvcktwo}{{\vc K}_2}

\newcommand{\rvcxone}{{\vc X}_1}
\newcommand{\rvcxtwo}{{\vc X}_2}
\newcommand{\rvcyone}{{\vc Y}_1}
\newcommand{\rvcytwo}{{\vc Y}_2}
\newcommand{\rvctxone}{\wt{\vc X}_1}
\newcommand{\rvctxtwo}{\wt{\vc X}_2}

\newcommand{\hrvcxone}{\widehat{\vc X}_1}
\newcommand{\hrvcxtwo}{\widehat{\vc X}_2}

\newcommand{\lrvcc}{{\lvc C}}
\newcommand{\lrvck}{{\lvc K}}
\newcommand{\lrvcx}{{\lvc X}}
\newcommand{\lrvcy}{{\lvc Y}}
\newcommand{\lrvcz}{{\lvc Z}}

\newcommand{\lrvcxi  }{{\lvc X}_i}
\newcommand{\lrvcki  }{{\lvc K}_i}
\newcommand{\lrvckone}{{\lvc K}_1}
\newcommand{\lrvcktwo}{{\lvc K}_2}
\newcommand{\lrvcxone}{{\lvc X}_1}
\newcommand{\lrvcxtwo}{{\lvc X}_2}
\newcommand{\lrvcyone}{{\lvc Y}_1}
\newcommand{\lrvcytwo}{{\lvc Y}_2}

\newcommand{\lrvctxone}{\wt{\lvc X}_1}
\newcommand{\lrvctxtwo}{\wt{\lvc X}_2}

\newcommand{\srvcc}{{\svc C}}
\newcommand{\srvck}{{\svc K}}
\newcommand{\srvcx}{{\svc X}}
\newcommand{\srvcy}{{\svc Y}}
\newcommand{\srvcz}{{\svc Z}}

\newcommand{\srvcxi  }{{\svc X}_i}
\newcommand{\srvckone}{{\svc K}_1}
\newcommand{\srvcktwo}{{\svc K}_2}
\newcommand{\srvcxone}{{\svc X}_1}
\newcommand{\srvcxtwo}{{\svc X}_2}
\newcommand{\srvcyone}{{\svc Y}_1}
\newcommand{\srvcytwo}{{\svc Y}_2}

\newcommand{\rvcci}{{\vc C}_i}
\newcommand{\rvcki}{{\vc K}_i}
\newcommand{\rvcxi}{{\vc X}_i}
\newcommand{\rvcyi}{{\vc Y}_i}
\newcommand{\hrvcxi}{\widehat{\vc X}_i}
\newcommand{\vcki}{{\vc k}_i}
\newcommand{\vcsi}{{\vc s}_i}
\newcommand{\vcti}{{\vc t}_i}
\newcommand{\vcvi}{{\vc v}_i}
\newcommand{\vcwi}{{\vc w}_i}
\newcommand{\vcxi}{{\vc x}_i}
\newcommand{\vcyi}{{\vc y}_i}

\newcommand{\vcs}{{\vc s}}
\newcommand{\vct}{{\vc t}}
\newcommand{\vcv}{{\vc v}}
\newcommand{\vcw}{{\vc w}}

\newcommand{\errP}{\varepsilon}  
\newcommand{\secP}{\delta}
\newcommand{\DcSet}{\cal D}

\newcommand{\CommenT}{

First, allow us to restate the main results of our paper.

We propose a new security metric for encryption schemes which is more
strict compared with the widely-used security metric based on the
mutual information. Let $\Delta$ denote the newly proposed metric and
$I$ denote the mutual information between the plaintexts and the 
ciphertexts. The new metric denoted by $\Delta$ is valid in the sense
that if $I=0$, then $\Delta=0$. This result stated in Property part a)
is the most essential part of our results. We prove the necessary and
sufficient condition to achieve the security for distributed
encryption with correlated keys under the newly proposed metric.

The extended version of this paper with a more detailed explanation is
presented in the following arXiv paper.
Yasutada Oohama, Bagus Santoso,
``Distributed Source Coding with Encryption Using Correlated Keys",
arXiv:2102.06363 [cs.IT].

Below, we will explain the significance of our results in brief.

In the general case, proving the necessary condition under own
proposed metric is more strict than the widely-used metric is
meaningless. However, we claim that our newly proposed metric is a
rare example of a special case that does not fall into the trap of the
general case above.

The first reason is that we already have a concrete scheme that
satisfies the more strict requirement of the proposed metric, i.e.,
the distributed encryption scheme proposed by Santoso and Oohama [1],
[2]. In [1],[2], on the surface, Santoso and Oohama stated that their
distributed encryption scheme is secure under the security metric
based on the mutual information. However, if one looks underneath a
little bit, one can easily discover that actually Santoso and Oohama
proved the security of their scheme based on a metric that is more
strict than the mutual information. And one can easily see that the
more strict security metric they used is equal to our new proposed
metric. Thus, our proposed metric is not a mere theoretical concept,
but an achievable security requirement in the practical world.

The second reason is that we argue that the proposed metric $\Delta$
has a deeper relationship with the already widely-used security metric
in information theory, i.e., the mutual information $I$.  In this
paper, based on our new proposed extension of the Birkhoff-von Neumann
theorem (Lemma 1), we have proven that for any decodable encryption
scheme, the following holds: $I=0$ only if $\Delta=0$ (Property 3(a)).
In other words, an encryption scheme can not be proven to be perfectly
secure based on the mutual information ($I=0$) unless it is perfectly
secure under our new proposed metric ($\Delta=0$). Although we have
not shown the general case where $\Delta$ and $I$ is nearly zero in
this paper yet, since we have shown that the newly proposed metric is
not only acting as the theoretical "upper-bound" of the mutual
information but is also acting as the "lower-bound" of it in the
perfect case, we argue that one can treat our proposed metric as a
valid alternative of security metric in information theory.

Here we will explain another meaning of the statement "for any
decodable encryption: $I=0$ only if $\Delta=0$" (Property 3(a)). As we
state in the paper, one can represent the new proposed metric $\Delta$
as a direct sum of (1) the mutual information $I$ between plaintexts
and ciphertexts, and (2)the divergence between the distribution of
ciphertexts and uniform distribution. Informally speaking, one can see
$\Delta$ as the summation of the degree of dependency between
plaintexts and ciphertexts and the distance between the ciphertexts'
distribution and uniform distribution. Hence, for any encryption
scheme, unless the distribution of the ciphertexts is completely
uniform, the plaintexts and ciphertexts can not be completely
independent. In other words, since the complete independence between
plaintexts and ciphertexts means perfect security, one can also say
that unless one flattens the distribution of ciphertexts into a
completely uniform distribution, one can not achieve perfect security.

Now we will explain the relationship between the above statement of
Property 3(a) and the result of Santoso and Oohama in [1],[2]. In [1],
[2], Santoso and Oohama used the mutual information as the security
metric. Informally, Santoso and Oohama use a certain compression
function to flat the distribution of ciphertexts in each encryption
node before they are released to the public channel such that even in
the case that the encryption keys between distributed sources are
correlated, the security is guaranteed. However, it has not been clear
whether the flattening of the ciphertexts is the only possible method
to guarantee the security of the scheme. In short, the above statement
of Property 3(a) said that in the perfect case, the flattening of the
ciphertexts into a perfect uniform distribution is the only way to
make the scheme achieve perfect security.
}

\newcommand{\IsitTwoSixIntro}{
We consider the distributed secure source coding based 
on the common key cryptosystem. This cryptosystem was 
posed and investigated by Oohama and Santoso 
~\cite{DBLP:conf/itw/OohamaS21}.

The framework covers the secrecy amplification 
problem for distributed encrypted 
sources with correlated keys using 
post-encryption-compression (PEC)~\cite{DBLP:conf/isit/SantosoO17, santosoOhPEC:19}.
The framework is also  closely related to 
several previous works on the PEC, e.g., Johnson et al. \cite{1337277},
Klinc et al. \cite{DBLP:journals/tit/KlincHJKR12}.
Our study also has a close connection with several previous works 
on the Shannon cipher system, e.g. \cite{Sh49}, \cite{Ya91} 
\cite{DBLP:journals/tit/IwamotoOS18}.

In ~\cite{DBLP:conf/itw/OohamaS21},
the reliable and secure rate region which provides 
the necessary and sufficient conditions 
for the security of the proposed framework
were derived under a self-tailored 
\emph{non-standard security criterion}.
A subsequent work~\cite{OohamaIsita2022} 
attempted to derive the condition under 
another non-standard security criterion, 
which is claimed as a natural variant of the standard mutual
information. However, establishing 
the necessary and sufficient conditions 
of the security based on the standard 
security criterion, i.e., \emph{standard 
mutual information}, remained as an open problem.

In their paper, Oohama and Santoso \cite{DBLP:conf/itw/OohamaS25} 
try to solve the above open problem. 
The two correlated information 
sources are separately encrypted 
using distributed correlated key sources. 
The two information sources are independent 
from the two key source. 
Under some condition for which we have some relationship 
between correlated information sources 
and correlated key sources,    
 Oohama and Santoso \cite{DBLP:conf/itw/OohamaS25} 
 determined the reliable and secure rate 
region.  

In this paper we try to solve 
the problem for the general case where we do not 
have the relationship between the two 
information sources and the two correlated 
keys assumed by Oohama and Santoso 
\cite{DBLP:conf/itw/OohamaS25}.   
In this case we prove the weak 
converse coding theorem under 
some assumption on the encryption 
and decryption scheme. When we 
have some further assumption that 
encoded source outputs satisfy   
some information spectrum property, we  
establish the strong converse theorem.
Information spectrum method developed 
by Han \cite{Han98InfSpec} 
is quite useful for deriving the above results.
}

\section{Introduction \label{sec:introduction}}

We consider the distributed secure source coding based 
on the common key cryptosystem. This secure communication 
system was 
posed and investigated by 
Oohama and Santoso~\cite{DBLP:conf/itw/OohamaS21}.
The framework of this secure communication 
system covers the secrecy amplification 
problem for distributed encrypted 
sources with correlated keys using 
post-encryption-compression (PEC)~\cite{DBLP:conf/isit/SantosoO17, santosoOhPEC:19}.
The framework is also  closely related to 
several previous works on the PEC, e.g., Johnson et al. \cite{1337277},
Klinc et al. \cite{DBLP:journals/tit/KlincHJKR12}.
The distributed secure communication system treated in this paper 
also has a close connection with several previous works 
on the Shannon cipher system, e.g. \cite{Sh49}, \cite{Ya91} 
\cite{DBLP:journals/tit/IwamotoOS18}.

In ~\cite{DBLP:conf/itw/OohamaS21},
the reliable and secure rate region which provides 
the necessary and sufficient conditions 
for the security of the proposed framework
were derived under a self-tailored 
\emph{non-standard security criterion}.
A subsequent work~\cite{OohamaIsita2022} 
attempted to derive the condition under 
another non-standard security criterion, 
which is claimed as a natural variant of the standard mutual
information. However, establishing 
the necessary and sufficient conditions 
of the security based on the standard 
security criterion, i.e., \emph{standard 
mutual information}, remained as an open problem.

In this paper, we try to solve the open problem.
Our observation reveals that the failure of previous 
attempts is mainly because the security criterion
is used as the starting point for proving the
\emph{strong converse}, the necessary 
conditions for reliable and secure 
transmission. 
We discover that this method
greatly reduces the flexibility to construct 
the proof. We develop a new technique to prove 
the strong converse
without sacrificing the \emph{standard mutual information}
as the security criterion.
Our technique proceeds
not by using the \emph{mutual information},
which is the security criterion, as the starting point,
but instead 
the \emph{conditional mutual information}.
Based on the new technique we derive 
new results on  the bounds of the rate region,
which specifies both
necessary and sufficient  
conditions to have reliable and secure transmission.
Our main results can be summarized as follows:
\begin{itemize}
    \item[(1)] 
        The outer bound
        matches with the inner bound 
        of the rate region in the following 
        cases:
        \begin{itemize}
            \item[a)] sources 
        are independent, 
            \item[b)] 
                the entropy of each source is less 
                than the entropy 
                of each corresponding key 
                and  the entropy of 
                the combined sources is less than
                the entropy of  the combined keys.
        \end{itemize} 
    \item[(2)] 
        For the general case where the condition b) stated 
        in the first item is not necessary satisfied, we prove that 
        the outer bound matches the inner bound 
        in the sum rate part of the rate region in
        general case.    
    \item[(3)]  In the above general case we prove the weak converse theorem under some assumption on the encryption 
and decryption scheme and two encoded source outputs. When we have some further assumption that the two  
encoded source outputs satisfy some information spectrum property, 
we establish the strong converse theorem.    
\end{itemize}

Information spectrum method \cite{Han98InfSpec} and 
a variant of Birkhoff-von Neumann theorem  play 
an important role in deriving those results.

\vspace{-1mm}

\section{Secure Source Coding Problem}

\vspace{-1mm}
\subsection{Preliminaries}

In this subsection, we show the basic notations and related consensus used in this paper. 

\noindent{}
\textit{Random Sources of Information and Keys: \ }
Let $(X_1,X_2)$ be a pair of random variables from a finite set 
$\mathcal{X}_1\times \mathcal{X}_2$. 
Let $\{(X_{1,t},X_{2,t})\}_{t=1}^\infty$ be a stationary \emph{discrete
memoryless source} (DMS) such that for each $t=1,2,\ldots$, 
the pair $(X_{1,t},X_{2,t})$ takes values in finite set 
$\mathcal{X}_1\times \mathcal{X}_2$ and obeys the same distribution 
as that of $(X_1,X_2)$ denoted by 
$p_{X_1 X_2}=\{p_{X_1 X_2} (x_1,x_2)\}_{(x_1,x_2)
\in\mathcal{X}_1\times 
\mathcal{X}_2}$. The stationary DMS 
$\{(X_{1,t},X_{2,t})\}_{t=1}^\infty$ 
is specified with $p_{X_1X_2}$. 
\newcommand{\Zap}{}{
Also, let $(K_1, K_2)$ be a pair of random variables taken from 
the same  finite set $\mathcal{X}_1\times \mathcal{X}_2$ 
representing the pair of keys used for encryption at 
two separate terminals, of which the detailed description will 
be presented later. Similarly, 
let $\{(K_{1,t},K_{2,t})\}_{t=1}^\infty$ 
be a stationary discrete 
memoryless source such that for each $t=1,2,\ldots$, 
the pair $(K_{1,t},K_{2,t})$ takes values in finite set
$\mathcal{X}_1\times \mathcal{X}_2$ and obeys the same 
distribution as that of $(K_1,K_2)$ denoted by 
$p_{K_1 K_2}=
\{p_{K_1 K_2} (k_1,k_2)\}_{(k_1,k_2)
\in\mathcal{X}_1\times \mathcal{X}_2}$.
The stationary DMS $\{(K_{1,t},K_{2,t})\}_{t=1}^\infty$ 
is specified with $p_{K_1K_2}$.
}

\vspace{1mm}

\noindent{}\textit{Random Variables and Sequences: \ }
We write the sequence of random variables with length $n$ 
from the information source as follows:
${\rvcxone}\defeq X_{1,1}X_{1,2}\cdots X_{1,n}$, 
${\rvcxtwo}\defeq X_{2,1}X_{2,2}\cdots X_{2,n}$.
Similarly, the strings with length $n$ of $\mathcal{X}_1^n$ 
and $\mathcal{X}_2^n$ are written as 
${\vcxone}\defeq x_{1,1}x_{1,2}\cdots x_{1,n}\in\mathcal{X}_1^n$ 
and ${\vcxtwo}
\defeq
x_{2,1}x_{2,2}\cdots x_{2,n}\in\mathcal{X}_2^n$ respectively.
For $({\vcxone}, {\vcxtwo})\in\mathcal{X}_1^n\times \mathcal{X}_2^n$, 
$p_{{\lrvcxone}{\lrvcxtwo}}({\vcxone},{\vcxtwo})$ stands for the 
probability of the occurrence of $({\vcxone}, {\vcxtwo})$. 
When the information source is memoryless 
specified with $p_{X_1X_2}$, we have the following equation holds:
$p_{{\lrvcxone} {\lrvcxtwo}}({\vcxone},{\vcxtwo})=
\prod_{t=1}^n p_{X_1X_2}(x_{1,t},x_{2,t})$.
In this case we write $p_{{\lrvcxone} {\lrvcxtwo}}({\vcxone},{\vcxtwo})$
as $p_{X_1 X_2}^n({\vcxone},{\vcxtwo})$.
Similar notations are used for other random variables and sequences.

\ \par\noindent{}\emph{Consensus and Notations: }
Without loss of generality, throughout this paper,
we assume that $\mathcal{X}_1$ and $\mathcal{X}_2$ are finite fields.
\newcommand{\OmiTzzz}{
The notation $\oplus$ is used to denote the field addition operation,
while the notation $\ominus$ is used to denote the field subtraction 
operation, i.e., $a\ominus b = a \oplus (-b)$ for any
elements $a,b$ of a same finite field. 
For the sake of simplicity, we use the same notation for field 
addition and subtraction for both $\mathcal{X}_1$ and $\mathcal{X}_2$.
}
Throughout this paper all logarithms are taken to the base 2. 

\subsection{Basic System Description}

Let the information sources and keys be generated 
independently by different parties 
$\Sgen$ and $\Kgen$ respectively.
In our setting, we assume the followings.
\begin{itemize}
  \item The sources ${\rvcxone}$ and ${\rvcxtwo}$ are 
              generated by $\Sgen$ and are correlated
	       to each other.
	\item The random keys ${\rvckone}$  and ${\rvcktwo}$ 
              are generated by $\Kgen$ and are 
            correlated to each other.
	\item The sources are independent to the keys.
\end{itemize}

\noindent
\underline{\it Source coding without encryption:} \ The 
two correlated random sources ${\rvcxone}$ and ${\rvcxtwo}$ 
from $\Sgen$ be sent to two separated nodes $\mathsf{E}_1$ 
and $\mathsf{E}_2$ respectively.
Further settings of the system are described as follows. 
Those are also shown in Fig. \ref{fig:mainA}.

\begin{enumerate}
	\item \emph{Encoding Process:} \ 
        For each $i=1,2$, at the node $\mathsf{E}_i$, 
        the encoder function 
        $\phi_{i}^{(n)}: {\cal X}_i^n $ $ \to {\cal M}_i^{(n)}$ 
        observes ${\rvcxi}$ to generate 
        $M_i^{(n)} 
         =\phi_{i}^{(n)}({\rvcxi})$. 
	\item \emph{Transmission:} \ 
        Next, the encoded sources ${M}_i^{(n)}$, $i=1,2$ 
        are sent to the 
        information processing center $\D$ through two \emph{noiseless} 
        channels. 
	\item \emph{Decoding Process:} \ 
        In $\D$, the decoder function observes ${M}_i^{(n)},i=1,2$ 
        to output $(\widehat{\rvcx}_1,\widehat{\rvcx}_2)$,
        using the mapping $\psi^{(n)}$ defined by 
        $\psi^{(n)}:
             {\cal M}_1^{(n)}\times {\cal M}_2^{(n)} 
         \to {\cal X}_1^n    \times {\cal X}_2^n$. 
         Here we set 
        \begin{align*}
         (\widehat{\rvcx}_1,\widehat{\rvcx}_2) \defeq & 
        \psi^{(n)}(M^{(n)}_1,M^{(n)}_2)
         \\ 
         =& \psi^{(n)}\left(
         \phi^{(n)}_{1}(\rvcxone),
         \phi^{(n)}_{2}(\rvcxtwo)\right).
        \end{align*}
\newcommand{\Zapp}{
More concretely, the decoder outputs the unique pair 
$(\widehat{\rvcx}_1,\widehat{\rvcx}_2)$ from 
$(\phi_{1}^{(n)})^{-1}(\tilde{X}_1^{m_1}) \times 
 (\phi_{2}^{(n)})^{-1}(\tilde{X}_2^{m_2})$ 
in a proper manner.}
\end{enumerate}
\newcommand{\OmittZXX}{
}{
\begin{figure}[t]
\centering
\includegraphics[width=0.47\textwidth]{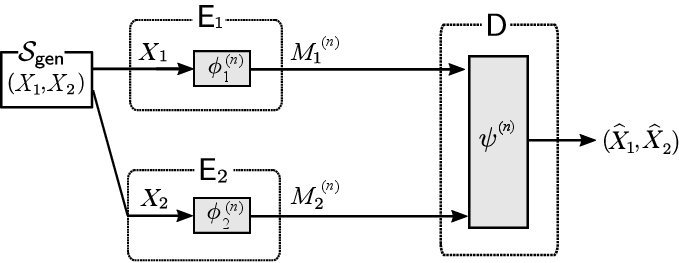}
\caption{Distributed source coding without encryption.
\label{fig:mainA}}
\end{figure}
}

\newcommand{\Omittbbb}{
}{
\begin{figure}[t]
\centering
\includegraphics[width=0.47 \textwidth]{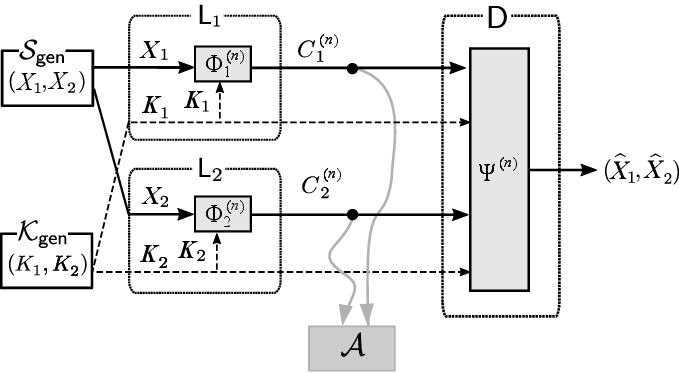}
\caption{Distributed source coding with encryption.
\label{fig:main}}
\vspace*{-5mm}
\end{figure}

}

For the above $(\phi_1^{(n)},
                \phi_2^{(n)},\psi^{(n)})$, we define 
the set $\mathcal{D}^{(n)}$ of correct decoding by 
\begin{align*}
\mathcal{D}^{(n)} & :=
\{(\vcxone,\vcxtwo) 
\in \mathcal{X}_1^{n}\times \mathcal{X}_2^{n}:
\\ 
&\psi^{(n)}( 
\phi_{1}^{(n)}(\vcxone),
\phi_{2}^{(n)}(\vcxtwo))
=(\vcxone, \vcxtwo)\}.
\end{align*}

\newcommand{\OmitZZZ}{
On $|{\cal D}^{(n)}|$, we have the following property. 
\begin{property}\label{pr:prOnDecSet} 
We have the following.  
\begin{align}
& |{\cal D}^{(n)}|=|{\cal X}_1^{m_1}||{\cal X}_2^{m_2}|. 
\label{eqn:CardOne}
\end{align}
\end{property}
Proof of Property \ref{pr:prOnDecSet} is given 
in Appendix \ref{apd:ProofPrOnDecSet}. 
}

\noindent
\underline{\it Distributed source coding with encryption:} \ 
The two correlated random sources 
${\rvcxone}$ and ${\rvcxtwo}$ 
from $\Sgen$ are sent to two separated nodes 
$\mathsf{L}_1$ and $\mathsf{L}_2$, respectively. 
The two random keys ${\rvckone}$ and ${\rvcktwo}$ from $\Kgen$, 
are also sent to $\mathsf{L}_1$ and and 
$\mathsf{L}_2$, respectively. 
Further settings of our system are described as follows. 
Those are also shown in Fig. \ref{fig:main}.
\begin{enumerate}
\item \emph{Source Processing:} \ For each $i=1,2$, 
at the node $\L_i$, ${\rvcx}_i$ is encrypted with 
the key ${\rvck}_i$ using the encryption function 
        $\Phi_i^{(n)}:{\cal X}_i^n \times {\cal X}_i^n$ 
       	$\to {\cal C}_i^{(n)}$. 
        For each $i=1,2$, the ciphertext $C_i^{(n)}$ of ${\rvcx}_i$ 
        is given by $C_i^{(n)}=\Phi_{i}^{(n)}
         ({\rvck}_i,{\rvcx}_i)$. 
        On the encryption function $\Phi_i^{(n)}, i=1,2$, we 
        use the following notation:
        $
         \Phi_i^{(n)}({\rvck}_i, {\rvcx}_i)
        =\Phi^{(n)}_{i, {\lrvck}_i}({\rvcx}_i)
        =\Phi^{(n)}_{i, {\lrvcx}_i}({\rvck}_i).
        $  
        \item \emph{Transmission:} \ Next, the ciphertext 
        $C_i^{(n)},i=1,2$ 
        are sent to the information processing center $\D$ through 
        two \emph{public} communication channels. 
        Meanwhile, the key ${\rvck}_i, i=1,2$, are sent 
        to $\D$ through two \emph{private} communication channels.
        \item \emph{Sink Node Processing:} \ In $\D$, we decrypt 
        the ciphertext $(\widehat{\rvcx}_1,\widehat{\rvcx}_2)$ 
         from $C_i^{(n)},i=1,2,$ using 
        the key ${\rvck}_i,i=1,2$, through the corresponding 
         decryption procedure $\Psi^{(n)}$ defined by 
        $ 
         \Psi^{(n)}:
         {\cal X}_1^n \times {\cal X}_2^n \times 
         {\cal C}_{1}^{(n)} \times {\cal C}_{2}^{(n)} 
         \to {\cal X}_1^n \times {\cal X}_2^n.
        $
        Here we set 
        $
         (\widehat{\rvcx}_1,$ $\widehat{\rvcx}_2)\defeq 
          \Psi^{(n)}({\rvckone},{\rvcktwo},C_1^{(n)},C_2^{(n)}).
        $ 
        More concretely, the decoder outputs the unique pair 
        $(\widehat{\rvcx}_1,\widehat{\rvcx}_2)$ from 
        $(\Phi_{1,\lrvckone}^{(n)})^{-1}({C}_1^{(n)}) \times 
         (\Phi_{2,\lrvckone}^{(n)})^{-1}({C}_2^{(n)})$ 
        in a proper manner. On the decryption function 
        $\Psi^{(n)}$, we use the following notation:
        \begin{align*}
        & \Psi^{(n)}({\rvck}_1,{\rvck}_2,C_1^{(n)},C_2^{(n)})
        =\Psi^{(n)}_{\lrvckone,\lrvcktwo}(C_1^{(n)},C_2^{(n)})
        \\
        &=\Psi^{(n)}_{C_1^{(n)},C_2^{(n)}}
          ({\rvckone},{\rvcktwo}).
        \end{align*}  
\end{enumerate}

Fix any $(\rvckone, \rvcktwo)=(\vckone,\vcktwo) 
\in \mathcal{X}_1^{n}\times \mathcal{X}_2^{n}$.
For this $(\rvckone,\rvcktwo)$ and 
for $(\Phi_1^{(n)},\Phi_2^{(n)},\Psi^{(n)})$,  
we define the set $ \mathcal{D}^{(n)}_{\lvckone,\lvcktwo}$ 
of correct decoding by 
\begin{align*}
\mathcal{D}^{(n)}_{\lvckone,\lvcktwo}
& := \{(\vcxone,\vcxtwo) \in \mathcal{X}_1^{n}
                      \times \mathcal{X}_2^{n}:
\\ 
&\Psi^{(n)}
(\Phi_{1}^{(n)}(\vckone, \vcxone),
(\Phi_{2}^{(n)}(\vcktwo, \vcxtwo))
=(\vcxone, \vcxtwo)\}.
\end{align*}
We require that the cryptosystem 
$
(\Phi_{1}^{(n)}, 
 \Phi_{2}^{(n)},
 \Psi^{(n)})
$
must satisfy the following condition.

\begin{condition}
\label{cond:CondOne}
For each distributed source encryption system 
$(\Phi_{1}^{(n)},\Phi_{2}^{(n)},\Psi^{(n)})$,  
there exists a distributed source coding system 
$(\phi_1^{(n)}, \phi_2^{(n)}, \psi^{(n)})$ such that  
for any $(\vckone, \vcktwo) \in \mathcal{X}_1^{n} 
\times \mathcal{X}_2^{n}$ and for any $(\vcxone, \vcxtwo) 
\in \mathcal{X}_1^{n} \times \mathcal{X}_2^{n}$, 
\begin{align*}
& \Psi_{\lvckone,\lvcktwo}^{(n)}
 (\Phi_{1,\lvckone}^{(n)}(\vcxone),
  \Phi_{2,\lvcktwo}^{(n)}(\vcxtwo)) 
\\
&=\psi^{(n)}
(\phi_{1}^{(n)}(\vcxone), 
                   \phi_{2}^{(n)}(\vcxtwo)). 
\end{align*}
\end{condition}

The above condition implies that 
$$
{\cal D}^{(n)}={\cal D}^{(n)}_{\lvckone,\lvcktwo}, 
\forall (\vckone,\vcktwo) \in {\cal X}_1^n\times {\cal X}_2^n.
$$ 
For each $i=1,2$, we set 
\begin{align}
({\cal D}^{(n)})_i=\{\vcxi:(\vcxone,\vcxtwo) \in  {\cal D}^{(n)}
\mbox{ for some }{\vc x}_{3-i}\}. 
\end{align}
For each $i=1,2$ and each $\vcxi\in ({\cal D}^{(n)})_i$, 
we set 
$$
{\cal D}_{i|3-i}^{(n)}(\vcxi|{\vc x}_{3-i})
\defeq \{\vcxi: (\vcxone,\vcxtwo)\in {\cal D}^{(n)}\}.
$$
We have the following properties on ${\cal D}^{(n)}$. 
\begin{property}
\label{pr:prOne}
If $(\vcxone,\vcxtwo), 
    (\vcx^{\prime}_1,
     \vcx^{\prime}_2) 
\in {\cal D}^{(n)}$ and $(\vcxone,\vcxtwo) 
\neq (\vcx^{\prime}_1,\vcx^{\prime}_2)$, then 
$$
(\Phi^{(n)}_{1,\lvckone}(\vcxone), 
 \Phi^{(n)}_{2,\lvcktwo}(\vcxtwo))
\neq 
(\Phi^{(n)}_{1,\lvckone}(\vcx^{\prime}_1), 
 \Phi^{(n)}_{2,\lvcktwo}(\vcx^{\prime}_2)). 
$$
Specifically, for each $i=1,2$ and 
each $\vcxi \in {\cal D}^{(n)}({\vc x}_{3-i})$,
$$
\Phi^{(n)}_{i,\lvcki}(\vcxi) 
\neq 
\Phi^{(n)}_{i,\lvcki}(\vcx^{\prime}_i). 
$$
\end{property}

Proof of Property \ref{pr:prOne} is given 
in Appendix \ref{apd:ProofPrOne}.
\newcommand{\ProofPrOne}{
\subsection{
Proof of Property \ref{pr:prOne}
}\label{apd:ProofPrOne}

In this appendix we prove Property \ref{pr:prOne}.

\begin{IEEEproof}[Proof of Property \ref{pr:prOne}] 
Under $(\vcxone,\vcxtwo), (\vcx_1^{\prime},\vcx_2^{\prime}) 
\in {\cal D}^{(n)}$ and 
$(\vcxone,\vcxtwo)\neq$ 
$(\vcx_1^{\prime},\vcx_2^{\prime})$, 
we assume that 
\beq
 (\Phi^{(n)}_{1,\lvckone}({\vcxone}), 
  \Phi^{(n)}_{2,\lvcktwo}({\vcxtwo})) 
=(\Phi^{(n)}_{1,\lvckone}({\vcx}_1^{\prime}), 
  \Phi^{(n)}_{2,\lvcktwo}({\vcx}_2^{\prime})).
\label{eqn:Assum} 
\eeq
Then we have the following: 
\begin{align}
&(\vcxone,\vcxtwo)\MEq{a}
\psi^{(n)}
(\phi_1^{(n)}(\vckone),
           \phi_2^{(n)}(\vcktwo) ,
\notag\\
&\MEq{b}
\Psi^{(n)}_{\lvckone,\lvcktwo}
(\Phi^{(n)}_{1,\lvckone}({\vcxone}),  
 \Phi^{(n)}_{2,\lvcktwo}({\vcxtwo}))
\notag\\
& \MEq{c}
\Psi^{(n)}_{\lvckone,\lvcktwo}
(\Phi^{(n)}_{1,\lvckone}({\vcx}_1^{\prime}),
 \Phi^{(n)}_{2,\lvcktwo}({\vcx}_2^{\prime}))
\notag\\
& \MEq{d}
 \psi^{(n)}
(\phi^{(n)}_1({\vcx}_1^{\prime}),
 \phi^{(n)}_2({\vcx}_2^{\prime}))\MEq{e}
             ({\vcx}_1^{\prime},
              {\vcx}_2^{\prime}).
\label{eqn:SdCCv}
\end{align}
Steps (a) and (e) follow from the definition of 
${\cal D}^{(n)}$. Step (c) follows from (\ref{eqn:Assum}).
Steps (b) and (d) follow from the relationship between 
$
(\phi^{(n)}_1,
 \phi^{(n)}_2,
 \psi^{(n)})
$
and 
$
(\Phi^{(n)}_{1,\lvckone},
 \Phi^{(n)}_{2,\lvcktwo},
 \Psi^{(n)}_{  \lvckone,
               \lvcktwo}).$
The equality (\ref{eqn:SdCCv}) contradicts the first assumption.
Hence we must have Property \ref{pr:prOne}.
\end{IEEEproof}
}
From Property \ref{pr:prOne}, we have the following result, which 
is a key result of this paper.    
\begin{lemma}\label{lem:LemOne}
  $\forall (c_1,c_2) 
 \in {\cal C}_1^{(n)} \times 
     {\cal C}_2^{(n)}$, we have the following:  
\begin{align}
&\sum_{\lvcxi \in {\cal D}_{i|3-i}^{(n)}(\lvcx_{3-i})}
 p_{C_i^{(n)}| \lrvcxone \lrvcxtwo}
 (c_i|\vcxone,\vcxtwo)\leq 1\mbox{ for }i=1,2,
\label{eqn:LemOneLEqOne} \\
& \sum_{(\lvcxone, \lvcxtwo) \in {\cal D}^{(n)}}
p_{C_1^{(n)} C_2^{(n)}| \lrvcxone \lrvcxtwo}
(c_1,c_2|\vcxone,\vcxtwo)\leq 1. 
\label{eqn:LemOneLEqTwo}
\end{align}
\end{lemma}

Proof of Lemma \ref{lem:LemOne} is given 
in Appendix \ref{apd:ProofLemOne}. This lemma 
can be regarded as an extension of the Birkhoff-von Neumann
theorem \cite{iwamoto:11}.

\newcommand{\ProofLemOne}{
\subsection{Proof of Lemma \ref{lem:LemOne}}
\label{apd:ProofLemOne}

In this appendix we prove Lemma \ref{lem:LemOne}.
Before proving this lemma we 
give an observation on 
$p_{C_1^{(n)}|\lrvcxone\lrvcxtwo}$ and 
$p_{C_1^{(n)} C_2^{(n)}|\lrvcxone\lrvcxtwo}$.
For $\vcxi\in {\cal X}_i^n$, $i=1,2$, we set 
\begin{align*}
& {\cal A}_{\lvcxi}(c_i) 
 \defeq \left\{
\vcki: \Phi^{(n)}_{i,\lvcxi}({\vck}_i)= 
c_i
\right\}.
\end{align*}
Furthermore, for  
$(\vcxone,\vcxtwo) \in {\cal X}_1^n \times {\cal X}_2^n$, we set 
\begin{align*}
& {\cal A}_{\lvcxone,\lvcxtwo}(c_1,c_2)
\defeq \left\{
(\vckone,\vcktwo): \Phi^{(n)}_{i,\lvcxi}({\vck}_i)= 
c_i,i=1,2 
\right\}.
\end{align*}
We have that for each $(c_i,\vcxone,\vcxtwo) 
\in {\cal C}_i^{(n)} 
\times {\cal X}_1^n \times {\cal X}_2^n$, $i=1,2$,
\begin{align}
&p_{C_i^{(n)} |\lrvcxone\lrvcxtwo}
(c_i|\vcxone,\vcxtwo) 
\nonumber\\  
&={\rm Pr}\left \{\rvcki
\in {\cal A}_{\lvcxi}(c_i)
\Bigl|\rvcxone=\vcxone,\rvcxtwo=\vcxtwo \right\}
\nonumber \\ 
&\MEq{a} {\rm Pr}\left\{\rvcki 
 \in {\cal A}_{\lvcxi} (c_i) \right\}.   
\label{eqn:ddrtrqa}
\end{align}
Step (a) follows from 
$\rvcki \perp (\rvcxone,\rvcxtwo)$.
We have that for each $(c_1,c_2,\vcxone,\vcxtwo) 
\in {\cal C}_1^{(n)} \times {\cal C}_2^{(n)} 
\times {\cal X}_1^n \times {\cal X}_2^n$, 
\begin{align}
&p_{C_1^{(n)} C_2^{(n)}|\lrvcxone\lrvcxtwo}
(c_1,c_2|\vcxone,\vcxone) 
\nonumber\\  
&={\rm Pr}\left \{(\rvckone,\rvcktwo) 
\in {\cal A}_{\lvcxone,\lvcxtwo}(c_1,c_2)
\Bigl|\rvcxone=\vcxone,\rvcxtwo=\vcxtwo \right\}
\nonumber\\  
&\MEq{a} {\rm Pr}\left\{(\rvckone,\rvcktwo)
 \in {\cal A}_{\lvcxone,\lvcxtwo}(c_1,c_2)
\right\}.
\label{eqn:ddrtrqz}
\end{align}
Step (a) follows 
from $(\rvckone,\rvcktwo) \perp (\rvcxone,\rvcxtwo)$.

\begin{IEEEproof}[Proof of Lemma \ref{lem:LemOne}]
Property \ref{pr:prOne}  
implies that 
\begin{align}
&     {\cal A}_{\lvcxi}(c_i)
 \cap {\cal A}_{{\lvcx}^{\prime}_i}
 (c_i)=\emptyset
\notag \\
& \mbox{ for }\vcxi, 
             {\vcx}^{\prime}_i
             \in {\cal D}_{i|3-i}^{(n)}({\vcx}_{3-i}),
             \vcxi \neq {\vcx}^{\prime}_i,
\label{eqn:Serra}\\             
&     {\cal A}_{\lvcxone, 
                \lvcxtwo}(c_1,c_2)
 \cap {\cal A}_{{\lvcx}^{\prime}_1,
                {\lvcx}^{\prime}_2}
 (c_1,c_2)=\emptyset
\notag\\
& \mbox{ for }(\vcxone, 
               \vcxtwo) \neq 
            ({\vcx}^{\prime}_1,
             {\vcx}^{\prime}_2) \in {\cal D}^{(n)}. 
\label{eqn:Serrq}
\end{align}
We first prove (\ref{eqn:LemOneLEqOne}) of
Lemma \ref{lem:LemOne}. For each $i=1,2$, 
we have the following chain of 
equalities:
\begin{align*}
& \sum_{\lvcxi \in {\cal D}_{i|3-i}^{(n)}(\lvcx_{3-i})}
 p_{ C_i^{(n)}| \lrvcxone \lrvcxtwo }
     (c_i| \vcxone, \vcxtwo)
\\
&\MEq{a}
\sum_{\lvcxi \in {\cal D}_{i|3-i}^{(n)}(\lvcx_{3-i})}
 {\rm Pr}\left\{\rvcki
 \in {\cal A}_{\lvcxi}(c_i)
\right\}
\\
&\MEq{b}\Pr \left\{
  \rvcki
  \in \bigcup_{
  \lvcxi \in {\cal D}_{i|3-i}^{(n)}(\lvcx_{3-i}) } 
  {\cal A}_{\lvcxi}(c_i)
  \right\}
\leq 1. 
\end{align*}
Step (a) follows from (\ref{eqn:ddrtrqa}).
Step (b) follows from (\ref{eqn:Serra}). 
We next prove (\ref{eqn:LemOneLEqTwo}) 
of Lemma \ref{lem:LemOne}. 
We have the following chain of 
equalities:
\begin{align*}
& \sum_{(\lvcxone, \lvcxtwo) \in {\cal D}^{(n)}}
 p_{ C_1^{(n)} C_2^{(n)} | \lrvcxone \lrvcxtwo }
 (c_1,c_2| \vcxone, \vcxtwo)
\\
&\MEq{a}\sum_{
(\lvcxone, \lvcxtwo) \in {\cal D}^{(n)}}
 {\rm Pr}\left\{(\rvckone,\rvcktwo)
 \in {\cal A}_{\lvcxone,\lvcxtwo}(c_1,c_2)
\right\}
\\
&\MEq{b}\Pr \left\{
  (\rvckone, \rvcktwo)
  \in \bigcup_{
  (\lvcxone, \lvcxtwo) \in {\cal D}^{(n)}} 
  {\cal A}_{\lvcxone,\lvcxtwo}(c_1,c_2)
  \right\}
\leq 1. 
\end{align*}
Step (a) follows from (\ref{eqn:ddrtrqz}).
Step (b) follows from (\ref{eqn:Serrq}). 
\end{IEEEproof} 
}

\subsection{
Security Criterion and Problem Formulation 
}

In the following arguments all logarithms are taken 
to the base two. 
The adversary ${\cal A}$ tries to estimate 
$({\rvcxone},{\rvcxtwo}) \in \mathcal{X}_1^n \times \mathcal{X}_2^n$ 
from 
$(C_1^{(n)},$ $C_2^{(n)})$.
The mutual information (MI) between $(\rvcxone,\rvcxtwo)$ and 
$(C_1^{(n)},C_2^{(n)})$ 
denoted by 
$$
\Delta_{\rm MI}^{(n)}
\defeq I(C_1^{(n)}C_2^{(n)};\rvcxone \rvcxtwo)
$$
indicates a leakage of information on $(\rvcxone,\rvcxtwo)$ 
from $(C_1^{(n)}$ $,C_2^{(n)})$. 
In this sense it seems to be quite natural 
to adopt the mutual information $\Delta_{\rm MI}^{(n)}$ 
as a security criterion. 

\noindent
\underline{\it Defining Reliability and Security:} 
The decoding process is successful if 
$(\widehat{\rvcx}_1,
  \widehat{\rvcx}_2)
  =(\rvcxone,\rvcxtwo)$ holds.
Hence the decoding error probability 
is given by  
\begin{align*}
&\Pr[\Psi^{(n)}(\rvckone, \rvcktwo,
  \phi_1^{(n)}({\rvckone, \rvcxone}),
  \phi_2^{(n)}({\rvcktwo, \rvcxtwo}))
\\
&\qquad \neq ({\rvcxone},{\rvcxtwo}) ]
\\
&=\Pr[\Psi^{(n)}_{\lrvckone,\lrvcktwo}
(\Phi_{1,\lrvckone}^{(n)}(\rvcxone),
 \Phi_{2,\lrvcktwo}^{(n)}(\rvcxtwo))
\neq ({\rvcxone},{\rvcxtwo}) ]
\\
&=\Pr[\psi^{(n)}
(\phi_{1}^{(n)}(\rvcxone),
 \phi_{2}^{(n)}(\rvcxtwo))
\neq           (\rvcxone,
                \rvcxtwo)]
\\
&=\Pr[(\rvcxone,\rvcxtwo)\notin {\cal D}^{(n)}].
\end{align*}
Since the above quantity depends only on 
$(\phi_1^{(n)}, \phi_2^{(n)},\psi^{(n)})$,
we write the error probability $p_{{\rm e}}$ of decoding as 
\begin{align*}
p_{{\rm e}}^{(n)}=&p_{{\rm e}}^{(n)} 
(\phi_1^{(n)},\phi_2^{(n)},\psi^{(n)}
|{p}_{X_1X_2}^n) 
\\ 
\defeq & \Pr[(\rvcxone,\rvcxtwo) \notin {\cal D}^{(n)})].
\end{align*}
\begin{definition}[Reliable and Secure Rate Pair] 
We fix some positive constant ${\secP}_0$. 
    For a fixed pair $({\errP},{\secP}) 
        \in  (0,1)\times [0,{\secP}_0]$, 
        $(R_1,R_2)$ is said to be an 
        $({\errP},{\secP})$-{\it reliable and secure rate pair} 
        if $\exists 
         \{(\Phi_1^{(n)},\Phi_2^{(n)},$ $\Psi^{(n)})\}_{n \geq 1}$
        such that $\forall \gamma >0$,
        $\exists n_0=n_0(\gamma) \in \mathbb{N}$, 
	$\forall n\geq n_0$, we have 
	\begin{align*}
        &\frac{1}{n} \log |{\cal C}_i^{(n)}| 
          \leq  
          R_i+ \gamma,\: i=1,2,
\\  				
& p_{{\rm e}}^{(n)}(
\phi_1^{(n)},
\phi_2^{(n)},
\psi^{(n)}|{p}_{X_1X_2}^n)
\leq \errP,
\\
& I(C_1^{(n)}C_2^{(n)}; \rvcxone \rvcxtwo) 
\leq \secP.
\end{align*}
\end{definition}
\begin{definition}[Reliable and Secure Rate Region]
\label{eqn:DefTwo}
 Let $\mathcal{R}^{\ast}({\errP},$ ${\secP}|p_{X_1X_2},$ $p_{K_1K_2})$
denote the set of all $(R_1,R_2)$ such that $(R_1,$ $R_2)$ 
is an $({\errP},{\secP})$-reliable and secure 
rate pair. We call  
 $\mathcal{R}^{\ast}({\errP},{\secP}|p_{X_1X_2},$ 
 $p_{K_1K_2})$ \emph{\it the} 
 $({\errP},{\secP})$-\emph{\it reliable 
 and secure rate region}.     
Furthermore, set 
 \begin{align*}
&   \mathcal{R}^{\ast}(\delta| p_{X_1X_1},p_{K_1K_2}) \defeq  
    \bigcap_{\scs {\errP}\in (0,1)
            }
\mathcal{R}^{\ast}({\errP},{\secP}|
p_{X_1X_2},
p_{K_1K_2}),
\\
&   \mathcal{R}^{\ast}(p_{X_1X_2},p_{K_1K_2}) \defeq  
    \bigcap_{\scs ({\errP},{\secP}) \in (0,1)
                 \atop{\scs \times  [0,{\secP}_0]
                 }
            }
\mathcal{R}^{\ast}({\errP},{\secP}|
p_{X_1X_2}, p_{K_1K_2}).    
 \end{align*}      
We call $\mathcal{R}^{\ast}(\delta |
p_{X_1X_2},
p_{K_1K_2})$ 
   the \emph{\it reliable and $\delta$-secure rate region} 
and call $\mathcal{R}^{\ast}(
p_{X_1X_2},
p_{K_1K_2})$ 
   the \emph{\it reliable and secure rate region}. 
\end{definition}

\begin{definition}[
Reliable and Secure Rate Pair
in an Optimistic Sense]
We fix some positive constant ${\secP}_0$. 
    For a fixed pair $({\errP},{\secP}) 
        \in  (0,1)\times [0,{\secP}_0]$, 
        $(R_1,R_2)$ is said to be an 
        $({\errP},{\secP})$-{\it reliable and secure rate pair 
        in an optimistic sense} 
        if 
        $\exists\{(\Phi_1^{(n)},\Phi_2^{(n)},$ $\Psi^{(n)})\}_{n \geq 1}$
        and $\exists \{k_n\}_{n\geq 1}$,
        such that $\forall \gamma >0$,
        $\exists n_0=n_0(\gamma) \in \mathbb{N}$, 
	$\forall n\geq n_0$, we have 
    \begin{align*}
        &\frac{1}{k_n} \log |{\cal C}_i^{(k_n)}| 
          \leq  
          R_i+ \gamma,\: i=1,2,
\\  				
& p_{{\rm e}}^{(k_n)}(
\phi_1^{(k_n)},
\phi_2^{(k_n)},
\psi^{(k_n)}|{p}_{X_1X_2}^{k_n})
\leq \errP,
\\
& I(C_1^{(k_n)}C_2^{(k_n)}; \rvcxone \rvcxtwo) 
\leq \secP.
\end{align*}
Here in the above equations the random vectors $\rvcxi,$ $\rvcki,i=1,2$ have the length $k_n$. 
Let $\overline{\mathcal{R}}^{\ast}({\errP},$ ${\secP}|p_{X_1X_2},$ $p_{K_1K_2})$
denote the set of all $(R_1,R_2)$ such that $(R_1,$ $R_2)$ 
is an $({\errP},{\secP})$-reliable and secure 
rate pair in an optimistic sense. We call  
 $\overline{\mathcal{R}}^{\ast}
 ({\errP},{\secP}|p_{X_1X_2},$ 
 $p_{K_1K_2})$ \emph{\it the} 
 $({\errP},{\secP})$-\emph{\it reliable 
 and secure optimistic rate region}. 
\end{definition}

By definition it is obvious that
for every fixed pair $({\errP},{\secP}) 
        \in  (0,1)\times [0,{\secP}_0]$,
$$
{\mathcal{R}}^{\ast}
 ({\errP},{\secP}|p_{X_1X_2},p_{K_1K_2})
\subseteq 
\overline{\mathcal{R}}^{\ast}
 ({\errP},{\secP}|p_{X_1X_2},p_{K_1K_2}).
$$

\section{
Main Results 
}

In this section we state our main results.  
We first derive an explicit inner bound of 
$\mathcal{R}^{\ast}(
p_{X_1X_2},
p_{K_1K_2})$. 
This inner bound can easily be obtained by the previous 
works 
\cite{DBLP:conf/isit/SantosoO17}, 
\cite{santosoOhPEC:19}.   
We next derive explicit outer bounds of 
$\mathcal{R}^{\ast}({\errP},{\secP}|
p_{X_1X_2},
p_{K_1K_2})$ for $({\errP},{\secP}) 
            \in  (0,1) \times[0,{\secP}_0]$. 
Those outer bounds do not depend on 
$({\errP},{\secP}) 
\in  (0,1)\times [0,{\secP}_0]$ 
and coincide with the inner bound in several cases.
%

\subsection{
Inner Bound for the Distributed Source Encryption
}

In this subsection we derive an inner bound of 
$\mathcal{R}^{\ast}($ ${\errP},{\secP}| p_{X_1X_2}, p_{K_1K_2})$ 
for $({\errP},{\secP}) 
      \in(0,1) \times [0,{\secP}_0]$. 
Define the following two regions:
\begin{align*}
 \mathcal{R}_{\mathrm{sw}}(p_{X_1X_2})\defeq\{(R_1,R_2):
 \ &R_1 \geq H(X_1|X_2), 
 \\
& R_2 \geq H(X_2|X_1), 
\\
& R_1+R_2 \geq H(X_1X_2)\},
\\
\mathcal{R}_{\mathrm{key}}(p_{K_1K_2})
\defeq\{(R_1,R_2): 
\ &R_1 \leq H(K_1),R_2 \leq H(K_2),
\\
& R_1+R_2 \leq  H(K_1K_2)\}.
\end{align*}
Furthermore, we set
\begin{align*}
& {\cal R}(p_{X_1X_2},p_{K_1K_2})
  \defeq \{(R_1,R_2): R_i\geq \wt{R}_i,i=1,2,
\\
& \mbox{ for some } (\wt{R}_1, \wt{R}_2)\in 
\mathcal{R}_{\mathrm{key}}(p_{K_1K_2})
\cap 
\mathcal{R}_{\mathrm{sw}}(p_{X_1X_2})\}.
\end{align*}
According to the previous works 
\cite{DBLP:conf/isit/SantosoO17}, 
\cite{santosoOhPEC:19},
the bound 
$\mathcal{R}_{\mathrm{key}}^{\ast}($ $p_{K_1K_2})$
$\cap$ $\mathcal{R}_{\mathrm{sw}}(p_{X_1X_2})$
serves as  an inner bound of 
$\mathcal{R}^{\ast}($ $p_{X_1X_2}$, 
$p_{K_1K_2})$ in the case where the security criterion 
is measured by the mutual information 
$\Delta_{\rm MI}^{(n)}$.
By a simple observation we can see that 
if $(\wt{R}_1,\wt{R}_2)$ belongs to  
$\mathcal{R}^{\ast}$ 
$(p_{X_1X_2}, p_{K_1K_2})$, then 
every $({R}_1,{R}_2)$ satisfying 
$R_i\geq \wt{R}_i,i=1,2,$ also belongs to  
$\mathcal{R}^{\ast}(p_{X_1X_2}$, $p_{K_1K_2})$.
Hence we have the following theorem:  
\begin{theorem}
\label{th:directTh}
For each $({\errP},{\secP}) 
\in (0,1) \times (0,{\secP}_0]$, we have
\begin{align}
& \mathcal{R}(p_{X_1X_2},p_{K_1K_2})  
\subseteq \mathcal{R}^{\ast}(p_{X_1X_2},p_{K_1K_2})  
\notag\\
&\subseteq \mathcal{R}^{\ast}({\errP},{\secP}|
                      p_{X_1X_2},p_{K_1K_2}).
\end{align} 
\end{theorem}

For $\mathcal{R}(p_{X_1X_2},$ $p_{K_1K_2})$, 
we have several properties, which are listed 
in the following:
\begin{property}\label{pr:PrRegOne}
\begin{itemize}
\item[a)] We have that 
$\mathcal{R}(p_{X_1X_2},$ $p_{K_1K_2})\neq \emptyset$ 
if and only if
\begin{align}
& \mathcal{R}_{\mathrm{key}}(p_{K_1K_2})
\cap 
\mathcal{R}_{\mathrm{sw}}(p_{X_1X_2})\neq \emptyset.
\label{eqn:CondNonEmpty}
\end{align}
The above condition is equivalent to 
the following condition:
\begin{align}
& H(X_i)\leq H(K_i|K_{3-i}), i=1,2,
\\
& H(X_1X_2)\leq H(K_1K_2).
\end{align}
\item[b)] 
Define  
\begin{align*}
& {\cal S}(p_{X_1X_2},p_{K_1K_2})\defeq 
 \mathcal{R}(p_{X_1X_2}, p_{K_1K_2})
\\
& \cap \{(R_1,R_2): R_1+R_2=H(X_1X_2)\}.
\end{align*}
Let
\begin{align*}
& {\cal S}_{\rm sw}
(p_{X_1X_2})\defeq 
 \mathcal{R}_{\rm sw}(p_{X_1X_2})
\\
& \cap \{(R_1,R_2): R_1+R_2=H(X_1X_2)\}.
\end{align*}
Using ${\cal S}_{\rm sw}
(p_{X_1X_2})$, the region 
${\cal S}(p_{X_1X_2},p_{K_1K_2})$ 
is expressed as 
$$
 {\cal S}(p_{X_1X_2},p_{K_1K_2})
={\cal S}_{\rm sw}(p_{X_1X_2})
 \cap {\cal R}_{\rm key}(p_{K_1K_2}).
$$
The region ${\cal R}(p_{X_1X_2},p_{K_1K_2})$ 
has an expression using ${\cal S}(p_{X_1X_2},p_{K_1K_2})$. 
This expression is shown below:   
\begin{align*}
& {\cal R}(p_{X_1X_2},p_{K_1K_2})
  =\{(R_1,R_2): R_i\geq \wt{R}_i,i=1,2,
\\
& \mbox{ for some } (\wt{R}_1, \wt{R}_2)\in 
{\cal S}(p_{X_1X_2},p_{K_1K_2}).
\end{align*}
\item[c)] The region  
${\cal R}(p_{X_1X_2},p_{K_1K_2})$ coincides 
with the region 
${\cal R}_{\mathrm{sw}}($$p_{X_1X_2})$  
if and only if
$$
H(X_i)\leq H(K_i), i=1,2,\: H(X_1X_2)\leq H(K_1K_2).
$$
\end{itemize}
\end{property}

Proof of Property \ref{pr:PrRegOne} 
is easy. We omit the detail. 

\subsection{
Strong Converse for the Distributed Source Encryption
}

In this subsection we derive outer bounds of 
$\mathcal{R}^{\ast}({\errP},{\secP}|p_{X_1X_2},$\\$
p_{K_1K_2})$ for $({\errP},{\secP}) 
            \in  (0,1) \times [0,{\secP}_0]$.  
We first derive one outer bound by a simple 
observation based on previous works on the distributed 
source coding for correlated sources. 
From the communication 
scheme 
we can see that the common key cryptosystem 
can be regarded as the data compression system, where 
for each $i=1,2$, the encoder $\Phi_i^{(n)}$ and the decoder 
$\Psi^{(n)}$ can use the common side information $\rvcki$.
By the strong converse coding theorem for 
this data compression system \cite{oohama:94}, we have that if 
\begin{align*}
& R_1 < H(X_1|X_2K_1K_2)=H(X_1|X_2) \mbox{ or }
\\
& R_2 < H(X_2|X_1K_1K_2)=H(X_2|X_1) \mbox{ or }
\\
& R_1+ R_2 < H(X_1X_2|K_1K_2)=H(X_1X_2) 
\end{align*}
then $\forall \tau \in (0,1)$, $\forall \gamma >0$,
and $\forall\{(\phi_1^{(n)},\phi_2^{(n)},\psi^{(n)})\}_{n \geq 1}$, $\exists n_0=n_0(\tau,\gamma) \in \mathbb{N}$, 
       $\forall n\geq n_0$, we have the following:
	\begin{align*}
        & \frac{m}{n} \log |{\cal X}_i|\leq R_i + \gamma, i=1,2,\:
        \\
        & p_{{\rm e}}^{(n)}(\phi_1^{(n)},\phi_2^{(n)}, 
          \psi^{(n)}|p^n_{X_1X_2}
          )
        \geq 1-\tau.
	\end{align*} 
Hence we have the following theorem. 
\begin{theorem}
\label{th:SStConvTh}
For each $({\errP},{\secP}) 
\in(0,1)\times (0,{\secP}_0]$, we have
\begin{align}
&\mathcal{R}^{\ast}({\errP},{\secP}|p_{X_1X_2},p_{K_1K_2})
 \subseteq \overline{\mathcal{R}}^{\ast}({\errP},{\secP}|p_{X_1X_2},p_{K_1K_2})
\notag\\
& \subseteq \mathcal{R}_{\mathrm {sw}}(p_{X_1X_2}). 
\notag
\end{align} 
\end{theorem}

For the derivations of outer bounds 
we consider the following two cases:
\begin{itemize}
\item[$\:$] Case 1: $H(X_i)\leq H(K_i)$ for $i=1,2$.  
\item[$\:$] Case 2: $H(X_i)> H(K_i)$ for 
$i=1$ or $2$.
\end{itemize}

\noindent
\underline{\it Case 1:} \ We consider 
the case where $H(X_i)\leq H(K_i)$ 
for $i=1,2$. We define a region serving 
as an outer bound of 
$ 
\mathcal{R}^{\ast}({\errP},{\secP}|p_{X_1X_2},p_{K_1K_2})
$
for $({\errP},{\secP}) 
            \in  (0,1)\times [0,{\secP}_0]$.
Set 
\begin{align*}
&{\cal R}^{\rm (out)}(p_{X_1X_2}, p_{X_1X_2})
\\
&\defeq 
\left\{
\ba{cl}
\mathcal{R}_{\mathrm{sw}}(p_{X_1X_2})&
\mbox{ if } \mathcal{R}_{\mathrm{key}}(p_{K_1K_2})
\cap 
\mathcal{R}_{\mathrm{sw}}(p_{X_1X_2})\neq \emptyset,
\\
\emptyset & \mbox{ othrewise}
\ea
\right.
\end{align*}
Our main result is the following:
\begin{theorem}
\label{th:convTh}
For $({\errP},{\secP}) 
      \in  (0,1) \times [0,{\secP}_0]$,  
\begin{align*}
&\mathcal{R}^{\ast}({\errP},{\secP}|p_{X_1X_2},p_{K_1K_2})
\subseteq 
 \overline{\mathcal{R}}^{\ast}({\errP},{\secP}|p_{X_1X_2},p_{K_1K_2})
\\
&\subseteq 
{\cal R}^{\rm (out)}(p_{X_1X_2}, p_{K_1K_2}).
\end{align*}
\end{theorem}

Proof of Theorem \ref{th:convTh} is 
given in the next section.
By Property \ref{pr:PrRegOne} part c), 
we have that if  
$$
H(X_i)\leq H(K_i), i=1,2,\: H(X_1X_2)\leq H(K_1K_2),
$$
then 
${\cal R}^{\rm (out)}(p_{X_1X_2}, p_{K_1K_2})$ 
coincides with  ${\cal R}(p_{X_1X_2},$ $ p_{K_1K_2})$ 
serving as an inner bound of 
${\cal R}^{\ast}(p_{X_1X_2}, p_{K_1K_2})$. According 
to Property \ref{pr:PrRegOne} part a),  
the condition $H(X_1X_2)\leq H(K_1K_2)$ is included 
in the condition of 
$
\mathcal{R}_{\mathrm{key}}(p_{K_1K_2})
\cap 
\mathcal{R}_{\mathrm{sw}}(p_{X_1X_2})\neq \emptyset.
$
Hence the matching condition for 
${\cal R}^{\rm (out)}(p_{X_1X_2}, p_{K_1K_2})$ 
and ${\cal R}(p_{X_1X_2}, p_{K_1K_2})$ to match 
is given by  
$ H(X_i)\leq H(K_i), i=1,2.$
Summarizing the above argument we have the 
following corollary from Theorem \ref{th:convTh}. 
\begin{corollary} We assume that  
$H(X_i)\leq H(K_i), i=1,2$. 
Then we have that for  $({\errP},{\secP}) 
            \in (0,1)\times [0,{\secP}_0]$,
\begin{align*}
&{\cal R}(p_{X_1X_2}, p_{K_1K_2})
=\mathcal{R}^{\ast}(p_{X_1X_2}, p_{K_1K_2})
\\
&={\cal R}^{*}({\errP},{\secP}|p_{X_1X_2}, p_{K_1K_2})
 =\overline{\cal R}^{\ast}({\errP},{\secP}|p_{X_1X_2}, p_{K_1K_2})
\\ 
&={\cal R}^{\rm (out)}(p_{X_1X_2}, p_{K_1K_2}).
\end{align*}
\end{corollary}

The above equality implies that we have the
strong converse theorem for distributed 
source encryption in the case of 
$H(X_i)\leq H(K_i),i=1,2$.   

\noindent
\underline{\it Case 2:} \ We consider the case where
$H(X_i)> H(K_i)$, $i=1\mbox{ or }2$.     
In this case we obtain the following two results: 
\begin{itemize}
\item[a)] We prove that 
under a certain condition 
on $(\Phi_1^{(n)}, \Phi_2^{(n)},$ $\Phi^{(n)})$ stronger than
the Condition \ref{cond:CondOne}, 
the inner bound ${\cal R}(p_{X_1X_2},p_{K_1K_2})$ 
also serves as an outer bound of
$\mathcal{R}^{\ast}$$({\secP}|
                 p_{X_1X_2},p_{K_1K_2})$, 
${\secP}\in (0,{\secP}_0]$, 
thereby establishing the partial strong
converse theorem for the security parameter 
${\secP} \in (0,{\secP}_0]$. 
\item[b)]We prove that 
under a certain condition 
on $(\Phi_1^{(n)},\Phi_2^{(n)},$ $\Phi^{(n)})$ 
stronger than the condition stated in the part a), 
the inner bound ${\cal R}(p_{X_1X_2},p_{K_1K_2})$ 
also serves as an outer bound of
$\mathcal{R}^{\ast}$$({\errP},{\secP}|
                 p_{X_1X_2},p_{K_1K_2})$, 
$({\errP},{\secP})\in (0,1) \times (0,{\secP}_0]$, 
thereby establishing the strong converse theorem. 
\end{itemize}

We first examine some particular parts of 
the regions 
$
\mathcal{R}^{\ast}$$({\errP},{\secP}|
                 p_{X_1X_2},p_{K_1K_2})$,
for $({\errP},{\secP}) 
\in (0,1)\times (0,{\secP}_0] $
and $\mathcal{R}^{\ast}($ $p_{X_1X_2}, p_{K_1K_2})$.
For 
$({\errP},{\secP}) \in (0,1)\times (0,{\secP}_0]$, 
set  
\begin{align*}
& {\cal S}^{\ast}( 
{\errP},{\secP}| p_{X_1X_2},p_{K_1K_2})
\defeq 
 \mathcal{R}^{\ast}({\errP},{\secP} | p_{X_1X_2}, p_{K_1K_2})
\\
& \cap \{(R_1,R_2): R_1+R_2=H(X_1X_2)\}.
\end{align*}
Furthermore, set
\begin{align*}
& {\cal S}^{\ast}(p_{X_1X_2},p_{K_1K_2})
  \defeq 
 \mathcal{R}^{\ast}(p_{X_1X_2}, p_{K_1K_2})
\\
& \cap \{(R_1,R_2): R_1+R_2=H(X_1X_2)\}.
\end{align*}
In similar manners, we define $\overline{\cal S}^{\ast}( 
{\errP},{\secP}| p_{X_1X_2},p_{K_1K_2})$ 
and $\overline{\cal S}^{\ast}( 
p_{X_1X_2},p_{K_1K_2})$, 
respectively based on 
$\overline{\cal R}^{\ast}( 
{\errP},{\secP}|p_{X_1X_2}, p_{K_1}$ ${}_{K_2})$ 
and $\overline{\cal R}^{\ast}( 
p_{X_1X_2},p_{K_1K_2})$. 
By Theorems \ref{th:directTh} and \ref{th:convTh}, 
we have that for $({\errP},{\secP}) 
\in (0,1)\times (0,{\secP}_0]$,
\begin{align}
&  {\cal S}(p_{X_1X_2}, p_{K_1K_2}) 
      = {\cal S}_{\rm sw}(p_{X_1X_2})
   \cap {\cal R}_{\rm key}(p_{K_1K_2})
\notag\\
&  \subseteq
\mathcal{S}^{\ast}(p_{X_1X_2},
p_{K_1K_2})
\subseteq {\cal S}^{*}({\secP},
  {\errP}|p_{X_1X_2}, p_{K_1K_2})
\label{eqn:InclusionSets}\\
& \subseteq 
\overline{\cal S}({\secP},
  {\errP}|p_{X_1X_2}, p_{K_1K_2}) 
\subseteq {\cal S}_{\rm sw}(p_{X_1X_2}). 
\notag
\end{align}
We can show that 
the set ${\cal S}(p_{X_1X_2}, p_{K_1K_2})$
serving as an inner bound of 
the secure and reliable rate set
$\mathcal{S}^{\ast}(p_{X_1X_2}, p_{K_1K_2})$ 
in (\ref{eqn:InclusionSets}) also serves 
as an outer bound 
of the $({\errP},{\secP})$-rate 
set $
{\cal S}^{*}({\secP},
  {\errP}|p_{X_1X_2}, p_{K_1K_2})
$
for  $({\errP},{\secP}) 
            \in (0,1) \times [0,{\secP}_0]$. 
This result is presented in the following theorem.    
\begin{theorem}\label{th:ConvThTwo}
For $({\errP},{\secP}) 
            \in (0,1)\times[0,{\secP}_0]$,
\begin{align}
&  {\cal S}(p_{X_1X_2}, p_{K_1K_2}) 
      = {\cal S}_{\rm sw}(p_{X_1X_2})
   \cap {\cal R}_{\rm key}(p_{K_1K_2})
\notag\\
& = \mathcal{S}^{\ast}(p_{X_1X_2}, p_{K_1K_2})
  = {\cal S}^{*}({\errP},{\secP}|p_{X_1X_2}, p_{K_1K_2})
\label{eqn:SRegOne}  
\\
& =\overline{{\cal S}}^{*}({\errP},{\secP}|p_{X_1X_2}, p_{K_1K_2}).
\label{eqn:SRegTwo}  
\end{align} 
\end{theorem}

Proof of Theorem \ref{th:ConvThTwo} is 
given in the next section.
The above theorem implies that we have the strong
converse theorem for 
${\cal S}^{*}({\errP},{\secP}|
p_{X_1 X_2}, p_{K_1 K_2})$ and 
$\overline{{\cal S}}^{*}({\errP},{\secP}|p_{X_1X_2}, p_{K_1K_2})$, 
$({\errP},{\secP}) 
        \in (0,1) \times [0,{\secP}_0]$.
        
We next provide two propositions, each of which 
connects Theorem \ref{th:ConvThTwo} 
into arguments of proofs of converse coding 
theorems. Before stating those two 
propositions, we must provide three 
additional conditions imposed on  
$(\Phi_1^{(n)},
  \Phi_2^{(n)}, 
    \Psi^{(n)})$.  
We call the second, third, and fourth conditions in addition to the 
first condition of Condition \ref{cond:CondOne}, 
respectively, 
the Conditions \ref{cond:CondTwo}, 
               \ref{cond:CondThrA}, 
           and \ref{cond:CondThr}. 
To state the second condition, for each $i=1,2$, 
we let the representation of the symmetric group of   
permutations on ${\cal C}_i^{(n)}$ be denoted 
by $\mathscr{S}({\cal C}_i^{(n)})$. 
The second condition is as follows:  
\begin{condition}\label{cond:CondTwo}
We assume that ${\cal M}_i^{(n)}={\cal C}_i^{(n)}, i=1,2$.
For each distributed source encryption system 
$(\Phi_{1}^{(n)},\Phi_{2}^{(n)},\Psi^{(n)})$,  
there exists a distributed source coding system 
$(\phi_1^{(n)},$ $\phi_2^{(n)}$, 
$\psi^{(n)})$ such that  
for any $(\vckone, \vcktwo) \in \mathcal{X}_1^{n} 
\times \mathcal{X}_2^{n}$ and  
for some $\wt{\Phi}_{i,\lvckone}^{(n)}\in  
\mathscr{S}({\cal C}_i^{(n)}),i=1,2$, 
\begin{align*}
& \Phi_{i,\lvckone}^{(n)}
  =\wt{\Phi}_{i,\lvckone}^{(n)}\circ \phi_i^{(n)},\:i=1,2, 
\\
& \Psi_{\lvckone,\lvcktwo}^{(n)}(c_1,c_2)=\psi^{(n)} 
 ((\wt{\Phi}_{1,\lvckone}^{(n)})^{-1}(c_1),
  (\wt{\Phi}_{2,\lvckone}^{(n)})^{-1}(c_2)),\: 
\\
& \forall (c_1,c_2)\in  
  {\cal C}_1^{(n)}\times {\cal C}_2^{(n)}.
\end{align*}
\end{condition}

To state the third and forth conditions, we prepare 
some quantities, which are known as 
information spectrum quantities 
\cite{Han98InfSpec}. 
Let $\{Z^{(n)} \}_{n\geq 1}$ be a sequence of random
variables. For each $n\geq 1$, $Z^{(n)}$ takes values 
in a finite set ${\cal Z}^{(n)}$. we assume 
that $\sup_{n\geq1}(1/n)\log |{\cal Z}^{(n)}|<\infty$.
For $a\geq 0$, set
\begin{align*}
F_{Z^{(n)}}(a)
 &\defeq
\Pr\Biggl\{
  \frac{1}{n}
  \log\frac{1}{p_{{Z}^{(n)}}({Z}^{(n)})}  
\leq a\Biggr\}.
\end{align*}
The third condition is as follows:
\begin{condition}\label{cond:CondThrA}  
For $i=1\mbox{ or }2$ and for each $a\geq 0$, we have 
the following limit:   
\begin{align*}
\lim_{n\to\infty} 
 F_{M_i^{(n)}}(a) 
 =\lim_{n\to\infty}
 \Pr\Biggl\{
  \frac{1}{n}
  \log\frac{1}{p_{{M}_i^{(n)}}({M}_i^{(n)})}  
\leq a\Biggr\},
\end{align*}
which we denote by $F_{M_i^{(\infty)}}(a)$. 
\end{condition}

Let $\{A_n\}_{n=1}^{\infty}$ be a sequence of arbitrary 
real-valued random variables. We introduce 
the notion of the so-called {\it limit} 
{\it superior} {in probability} 
in the following.  
\begin{align*}
   \mbox{\rm p-}\limsup_{n\to\infty}A_n 
&\defeq  
\inf\{
\alpha : \lim_{n\to\infty}
\Pr\{A_n {>}\alpha\}=0\},
\\
 \mbox{\rm p-}\liminf_{n\to\infty}A_n 
&\defeq  
\sup\{
\alpha : \lim_{n\to\infty}
\Pr\{A_n {<}\alpha\}=0\}.  
\end{align*}
For each $i=1,2$, we define 
\begin{align*}
&\overline{H}
({M}_i^{(\infty)})
 \defeq 
 \mbox{\rm p-}\limsup_{n\to\infty}
 \frac{1}{n}\log\frac{1}{p_{{M}_i^{(n)}}  
   ({M}_i^{(n)})},
\\
&\underline{H}( {M}_i^{(\infty)})
 \defeq \mbox{\rm p-}\liminf_{n\to\infty}
 \frac{1}{n}\log\frac{1}{p_{{M}_i^{(n)}}
 ({M}_i^{(n)})}.
\end{align*}
The fourth condition is as follows:
\begin{condition}
\label{cond:CondThr}
For $i=1\mbox{ or }2$,   
 $\overline{H}( {M}_i^{(\infty)})=\underline{H}( {M}_i^{(\infty)}).$
\end{condition}

Note here that Condition \ref{cond:CondThrA} 
includes Condition \ref{cond:CondThr} as a special case. 
In fact $\overline{H}({M}_i^{(\infty)})=\underline{H}( {M}_i^{(\infty)})$
implies the following:
$$
F_{M_i^{(\infty)}}(a)=
\left\{ 
\ba{ll}
0& \mbox{ if }0 \leq a < \underline{H}
({M}_i^{(\infty)}), \\
1&  \mbox{ if } a \geq \underline{H}
({M}_i^{(\infty)}).\\
\ea
\right.
$$
We call the class of 
$\{(\Phi_1^{(n)},
    \Phi_2^{(n)},
    \Psi^{(n)})\}_{n \geq 1}$
 satisfying the Condition \ref{cond:CondOne} 
 the Class I.  
We call the class of 
$\{(\Phi_1^{(n)}, \Phi_2^{(n)},$ 
$ \Psi^{(n)})\}_{n \geq 1}$
 satisfying the Conditions 
 \ref{cond:CondOne}, 
 \ref{cond:CondTwo}, and 
 \ref{cond:CondThrA} the Class II. 
We call the class of 
$\{(\Phi_1^{(n)},
    \Phi_2^{(n)},
    \Psi^{(n)})\}_{n \geq 1}$
 satisfying the Conditions 
 \ref{cond:CondOne}, \ref{cond:CondTwo},
 and \ref{cond:CondThr} 
 the Class III. By the above definitions 
 it is obvious that the reliable and secure rate region 
$\mathcal{R}^{\ast}(p_{X_1X_2}, p_{K_1K_2})$,
the ${\secP}$-secure rate region 
${\cal R}^{*}({\secP}|$ $p_{X_1X_2}, p_{K_1K_2})$,  
${\secP} \in [0,\delta_0)$,
and the $({\errP},{\secP})$-reliable 
and secure rate region 
${\cal R}^{*}({\errP},{\secP}|p_{X_1X_2}, p_{K_1K_2})$, 
$({\errP},{\secP})\in (0,1)\times [0,\delta_0)$
are rate regions for the Class I.
To distinguish those to the rate regions for other 
coding classes we define the three rate regions for 
the Class II by  
$\mathcal{R}_{\rm II}^{\ast}(p_{X_1X_2}, p_{K_1K_2})$,
   ${\cal R}_{\rm II}^{*}({\secP}|p_{X_1X_2}, p_{K_1K_2})$,
${\secP}\in [0,\delta_0)$, 
and ${\cal R}_{\rm II}^{*}({\errP},{
\secP}|p_{X_1X_2}, p_{K_1K_2}),$
$({\errP},{\secP})\in (0,1)\times [0,\delta_0)$.
We similarly define three rate regions for 
the Class III. From the above definitions we have that
\begin{align*}
&\mathcal{R}^{\ast}(p_{X_1X_2}, p_{K_1K_2})
\supseteq
\mathcal{R}_{\rm II}^{\ast}(p_{X_1X_2}, p_{K_1K_2})
\\
& \supseteq
\mathcal{R}_{\rm III}^{\ast}(p_{X_1X_2}, p_{K_1K_2}),
\end{align*}
for 
${\secP}\in [0,\delta_0)$, 
\begin{align*}
&{\cal R}^{*}({\secP}|p_{X_1X_2}, p_{K_1K_2})
\supseteq 
\mathcal{R}_{\rm II}^{\ast}
({\secP}|p_{X_1X_2}, p_{K_1K_2})
\\
& \supseteq 
\mathcal{R}_{\rm III}^{\ast}
({\errP},{\secP}|p_{X_1X_2}, p_{K_1K_2}),
\end{align*}
and for $({\errP},{\secP})\in (0,1) \times [0,\delta_0)$, 
\begin{align*}
&{\cal R}^{*}({\errP},{\secP}|p_{X_1X_2}, p_{K_1K_2})
\supseteq 
\mathcal{R}_{\rm II}^{\ast}
({\errP},{\secP}|p_{X_1X_2}, p_{K_1K_2})
\\
& \supseteq 
\mathcal{R}_{\rm III}^{\ast}
({\errP},{\secP}|p_{X_1X_2}, p_{K_1K_2}).
\end{align*}

\begin{condition}[Ahlswede et al. \cite{ahlswede:76}]
 The joint distribution ${p_{X_1X_2}}$ of 
$({X_1}, {X_2})$ 
is {\it indecomposable} if there are no functions 
$f_1$ and $f_2$ with respective domains 
${\cal X}_1$ and ${\cal X}_2$ so that 
\begin{itemize}
\item[i)] $\Pr \left\{f_1({X}_1) = f_2({X_2})=1 \right\}$ 
and 
\item[ii)] $f_1(X_1)$ 
takes at least two values with non-zero probability. 
\end{itemize}
\end{condition}

We have the following proposition, 
which is a basis of the proofs of 
converse coding theorems for the Class II.

\begin{proposition}\label{pro:BaseProb}
We consider the case where 
$p_{X_1X_2}$ is indecomposable. 
Fix small $\kappa\in (0,1)$ arbitrary.
We assume that for ${\secP} 
   \in [0,{\secP}_0]$, 
   $(R_1,R_2)\in {\cal R}_{\rm II}^{\ast}({\secP}|$ $p_{X_1X_2},p_{K_1K_2})$.
Then we have the followings. 
\begin{itemize}
\item[a)]
$(R_1,R_2)$ must satisfy $(R_1,R_2)\in 
{\cal R}_{\rm sw}(p_{X_1X_2})$.
\item[b)]
$\exists (\wt{R}_1,\wt{R}_2)$ satisfying 
$$
\wt{R}_i\leq R_i,i=1,2,\mbox{ and }
(\wt{R}_1,\wt{R}_2)\in{\cal S}_{\rm sw}(p_{X_1X_2}),    
$$
and $\exists \omega \in (0,1)$, 
such that $\forall\tau\in (0,
\kappa(1-{\theTa})]$, 
we have  
$$
(\wt{R}_1,\wt{R}_2)\in
\overline{\cal S}^{\ast}(\tau+
{\theTa},
{\secP}|p_{X_1X_2},p_{K_1K_2}).
$$
\end{itemize}

\end{proposition}

We prove this proposition in 
Section \ref{sec:PrBasePro}. 
For the proof of Proposition 
\ref{pro:BaseProb}, we consider   
the distributed source coding problem
for the general correlated source specified with  
$\{(M_1^{(n)},M_2^{(n)})\}_{n=1}^{\infty}$. 
The distributed source coding for general correlated information sources was investigated by 
Han \cite{Han98InfSpec},\cite{han:1998}.
For the computation of transmission error 
indicating a reliability, we use the result 
of Ahlsewde and G\'acs \cite{ahlswede:76b}.  

Proposition \ref{pro:BaseProb} together with 
Theorem \ref{th:ConvThTwo} yields the 
following theorem:
\begin{theorem}\label{th:ConvThThrb}
We consider the case where 
$p_{X_1X_2}$ is indecomposable. 
For ${\secP} 
      \in [0,{\secP}_0] $,
\begin{align*}
&  {\cal R}(p_{X_1X_2}, p_{K_1K_2}) 
 = \mathcal{R}_{\rm II}^{\ast}
 (p_{X_1X_2}, p_{K_1K_2})
\\
&= {\cal R}^{*}_{\rm II}({\secP}|
p_{X_1X_2}, p_{K_1K_2}).
\end{align*} 
\end{theorem}

The above equality implies that we have the 
strong converse theorem for the region 
${\cal R}_{\rm II}^{*}({\secP}|p_{X_1X_2},$ $p_{K_1K_2}),$ 
${\secP}\in [0,{\secP}_0]$. 

       
We have the following proposition, 
which is a basis of the proofs of 
converse coding theorems for the Class III.        
\begin{proposition}\label{pro:BasePro}
Fix small $\kappa\in (0,1)$ arbitrary. 
We assume that for $({\errP},{\secP}) 
   \in  (0,1)\times [0,{\secP}_0]$, 
   $(R_1,R_2)\in {\cal R}_{\rm III}^{\ast}({\errP},{\secP}|$ $p_{X_1X_2},$ $p_{K_1K_2})$.
Then we have the followings. 
\begin{itemize}
\item[a)]
$(R_1,R_2)$ must satisfy $(R_1,R_2)\in 
{\cal R}_{\rm sw}(p_{X_1X_2})$.
\item[b)]
$\exists (\wt{R}_1,\wt{R}_2)$ satisfying 
$$
\wt{R}_i\leq R_i,i=1,2,\mbox{ and }
(\wt{R}_1,\wt{R}_2)\in{\cal S}_{\rm sw}(p_{X_1X_2}),    
$$
such that $\forall\tau\in (0,\kappa(1-{\errP})]$, 
we have  
$$
(\wt{R}_1,\wt{R}_2)\in
{\cal S}^{\ast}({\errP}+\tau,{\secP}|p_{X_1X_2},p_{K_1K_2}).
$$
\end{itemize}
\end{proposition}

We prove this proposition in 
Section \ref{sec:PrBasePro}.   
Similarly to the proof of Proposition 
\ref{pro:BaseProb}, we can prove 
Proposition \ref{pro:BasePro} by 
using the argument of the distributed source coding 
problem for the general correlated source specified 
with $\{(M_1^{(n)},M_2^{(n)})\}_{n=1}^{\infty}$, 
the general theory of which was developed by Han \cite{Han98InfSpec},\cite{han:1998}. 

Proposition \ref{pro:BasePro} together with 
Theorem \ref{th:ConvThTwo} yields the 
following theorem:

\begin{theorem}\label{th:ConvThThr}
For $({\errP},{\secP}) 
      \in (0,1)\times [0,{\secP}_0] $,
\begin{align*}
&  {\cal R}(p_{X_1X_2}, p_{K_1K_2}) 
 = \mathcal{R}_{\rm III}^{\ast}
 (p_{X_1X_2}, p_{K_1K_2})
\\
&= {\cal R}^{*}_{\rm III}({\errP},{\secP}|
p_{X_1X_2}, p_{K_1K_2}).
\end{align*} 
\end{theorem}

The above equality implies that we have the strong
converse theorem for 
the region ${\cal R}_{\rm III}^{*}({\errP},{\secP}|p_{X_1X_2},p_{K_1K_2}),$ 
$({\errP},{\secP}) 
        \in (0,1)\times [0,{\secP}_0]$.    

\section{Proofs of Theorems \ref{th:convTh} 
and \ref{th:ConvThTwo}} In this section 
we prove Theorems \ref{th:convTh} 
and \ref{th:ConvThTwo}. 
To prove  Theorems \ref{th:convTh} 
and \ref{th:ConvThTwo}, we present two propositions.
To describe those two propositions, we give several 
definitions. Define two subsets 
of ${\cal X}_1^n \times {\cal X}_2^n$ by  
\begin{align*}
&{{\cal A}}_{\gamma}^{(n)} \defeq 
 \biggl\{ (\vcxone,\vcxtwo):
\\
 &\:\:\:\left|\frac{1}{n}
 \log\frac{1}{p_{X_{i}|X_{3-i}}^{n}({\vcx}_i|\vcx_{3-i})}
- H(X_{i}|X_{3-i})\right|\leq \gamma,i=1,2\\
 &\:\:\left.\left| \frac{1}{n}\log\frac{1}{p_{X_{1}X_{2}}^{n}
  (\vcxone,\vcxtwo)}
 -H(X_{1}X_{2})\right|\leq \gamma\right\}, 
 \\
 &\wt{\mathcal{D}}_{\gamma}^{(n)}  
 \defeq {\cal A}_{\gamma}^{(n)}
\cap\mathcal{D}^{(n)}.
\end{align*}
Set  
$$
\nu_n(\gamma)
\defeq p_{X_{1}X_{2}}^{n}
\left(\left(
{\cal A}_{\gamma}^{(n)}
\right)^{c}\right),\:
{\nu}_{n}(\gamma,{\errP})
\defeq \nu_n(\gamma)+{\errP}.
$$
By the large deviation theory, we have that for fixed 
$\gamma>0$, $\nu_n(\gamma)$ decays exponentially
as $n\to\infty$.  On 
$\widetilde{\mathcal{D}}_{\gamma}^{(n)}$, we have 
the following bound:
\begin{align*}
& p_{X_{1}X_{2}}^{n}
 \left(\left(\wt{\mathcal{D}}_{\gamma}^{(n)}
     \right)^{c}\right)
\leq  p_{X_{1}X_{2}}^{n} \left(\left(
     {\mathcal{A}}_{\gamma}^{(n)}
     \right)^{c}\right)
\\
&\qquad\qquad  +p_{X_{1}X_{2}}^{n}
  \left(\left(\mathcal{D}^{(n)}\right)^{c}\right)
 \leq {\nu}_{n}(\gamma,\errP).
\end{align*}
We set 
$$
\zeta_n(\gamma,{\errP},{\secP})
\defeq \frac{1}{n}\left[
  \frac{{\secP}}
  {1-{\nu}_{n}(\gamma,{\errP})}
  +\log\frac{1}{1-{\nu}_{n}(\gamma,\errP)
  }\right].
$$
We have the following two propositions, which are 
key results to prove Theorems \ref{th:convTh} 
and \ref{th:ConvThTwo}.  
\begin{proposition}\label{pro:KeyProOne} 
Fix any $({\errP},{\secP}) 
\in (0,1)\times (0,{\secP}_0]$. 
We assume that 
${\cal R}^{\ast}({\errP},{\secP}|p_{X_1X_2},p_{K_1K_2})
  \neq \emptyset$. 
Then we must have that 
$\forall \gamma >0$,
        $\exists n_0=n_0(\gamma) \in \mathbb{N}$, 
$\forall n\geq n_0$, 
\begin{align}
\left.
\ba{rl}
H(X_{i}|X_{3-i})\leq & H(K_{i})+\gamma 
+\zeta_n(\gamma,{\errP},{\secP}),\: i=1,2,
\\
H(X_{1}X_{2}) \leq & H(K_{1}K_{2}) +\gamma
+\zeta_n(\gamma,{\errP},{\secP}).
\ea\right\}
\label{eqn:BdProThrOne}
\end{align}
We next assume that 
$\overline{\cal R}^{\ast}({\errP},{\secP}|p_{X_1X_2},p_{K_1K_2})
  \neq \emptyset$. 
Then we must have that 
$\exists\{k_n\}_{n\geq 1}$,
$\forall \gamma >0$,
        $\exists n_1=n_1(\gamma) \in \mathbb{N}$, 
$\forall n\geq n_1$, 
\begin{align}
\left.
\ba{rl}
H(X_{i}|X_{3-i})\leq & H(K_{i})+\gamma 
+\zeta_{k_n}(\gamma,{\errP},{\secP}),\: i=1,2,
\\
H(X_{1}X_{2}) \leq & H(K_{1}K_{2}) +\gamma
+\zeta_{k_n}(\gamma,{\errP},{\secP}).
\ea\right\}
\label{eqn:BdProThrTwo}
\end{align}

\end{proposition}
\begin{proposition}\label{pro:KeyProTwoA}
Fix any $({\errP},{\secP}) 
\in (0,1) \times (0,{\secP}_0] $. 
We assume that 
$(R_1,R_2)\in {\cal S}^{\ast}({\errP},{\secP}
               |p_{X_1X_2},p_{K_1K_2})$. 
Then we must have that 
$\forall \gamma >0$,
        $\exists n_0=n_0(\gamma) \in \mathbb{N}$, 
$\forall n\geq n_0$, 
\begin{align}
R_i\leq & H(K_{i}) 
+2\gamma +\zeta_n(\gamma,{\errP},{\secP}),\: i=1,2.
\label{eqn:BdProFourOne}
\end{align}
We next assume that 
$(R_1,R_2)\in 
\overline{\cal S}^{\ast}({\errP},{\secP}
               |p_{X_1X_2},p_{K_1K_2})$. 
Then we must have that $\exists\{k_n\}_{n\geq 1}$, 
$\forall \gamma >0$,
        $\exists n_1=n_1(\gamma) \in \mathbb{N}$, 
$\forall n \geq n_1$, 
\begin{align}
R_i\leq & H(K_{i}) 
+2\gamma +\zeta_{k_n}(\gamma,{\errP},{\secP}),\: i=1,2.
\label{eqn:BdProFourTwo}
\end{align}
\end{proposition}


Proofs of Propositions 
\ref{pro:KeyProOne} and 
\ref{pro:KeyProTwoA} are 
given in Section \ref{sec:PrKeyProOneTwo}. 
For the proof of the above two propositions 
we need several lemmas. We present those lemmas 
and prove them in 
Section 
\ref{ssec:PrfLemsFotTwoProp}. 
Proofs of two propositions are given in
Section 
\ref{ssec:PrfTwoProp}. 
Theorem \ref{th:convTh} immediately follows 
from Proposition \ref{pro:KeyProOne}. 
Theorem \ref{th:ConvThTwo} immediately follows 
from Proposition \ref{pro:KeyProTwoA}.

\begin{IEEEproof}[Proof of Theorem \ref{th:convTh}]
We assume that  
${\cal R}^{\ast}({\errP},{\secP}|p_{X_1X_2},$ $p_{K_1K_2})
  \neq \emptyset$. Then, 
  by Proposition \ref{pro:KeyProOne}, we have that 
  $\forall \gamma >0$,
 $\exists n_0=n_0(\gamma) \in \mathbb{N}$, 
$\forall n\geq n_0$, 
\begin{align}
H(X_{i}|X_{3-i})\leq & H(K_{i})+\gamma 
+\zeta_n(\gamma,{\errP},{\secP}),\: i=1,2,
\label{eqn:BoundOne}\\
H(X_{1}X_{2}) \leq & H(K_{1}K_{2}) +\gamma
+\zeta_n(\gamma,{\errP},{\secP}). 
\label{eqn:BoundTwo}
\end{align}
Letting $n\to \infty$ in 
(\ref{eqn:BoundOne}) and 
(\ref{eqn:BoundTwo}) and considering that 
$\gamma>0$ can be taken arbitrary small, 
we have that
\begin{align}
\left.
\ba{l}
H(X_i|X_{3-i})\leq H(K_i), i=1,2,
\vspace*{1mm}\\
H(X_1X_2)\leq H(K_1K_2).
\ea \right\}
\label{eqn:NonEmpCd}
\end{align}
According to Property \ref{pr:PrRegOne}, 
the condition (\ref{eqn:NonEmpCd}) 
is equivalent to the condition 
$\mathcal{R}(p_{X_1X_2},$ $p_{K_1K_2})
\neq \emptyset$. Furthermore, 
by Theorem \ref{th:SStConvTh}, we have that 
$\mathcal{R}_{\mathrm{sw}}(p_{X_1X_2})$ 
serves as an outer bound of 
$\mathcal{R}^{\ast}({\errP},{\secP}|p_{X_1X_2},$ $
p_{K_1K_2})$ for $({\errP},{\secP}) 
            \in  (0,1)\times [0,{\secP}_0]$.
Those imply that  
${\cal R}^{\rm (out)}(p_{X_1X_2},p_{K_1K_2})$ 
serves as an outer bound of 
$\mathcal{R}^{\ast}({\errP},{\secP}|p_{X_1X_2},$ 
$p_{K_1K_2})$    for $({\errP},{\secP}) 
            \in (0,1)\times [0,{\secP}_0]$.  

We next assume that  
$\overline{\cal R}^{\ast}({\errP},{\secP}|p_{X_1X_2},$ $p_{K_1K_2})
  \neq \emptyset$. Then, 
by Proposition \ref{pro:KeyProOne}, we have that 
  $\forall \gamma >0$,
 $\exists n_1=n_1(\gamma) \in \mathbb{N}$, 
$\forall n\geq n_1$, 
\begin{align}
H(X_{i}|X_{3-i})\leq & H(K_{i})+\gamma 
+\zeta_{k_n}(\gamma,{\errP},{\secP}),\: i=1,2,
\label{eqn:BoundOneB}\\
H(X_{1}X_{2}) \leq & H(K_{1}K_{2}) +\gamma
+\zeta_{k_n}(\gamma,{\errP},{\secP}). 
\label{eqn:BoundTwoB}
\end{align}
Letting $n\to \infty$ in 
(\ref{eqn:BoundOneB}) and 
(\ref{eqn:BoundTwoB}) and considering that 
$\gamma>0$ can be taken arbitrary small, we 
have the bound (\ref{eqn:NonEmpCd}),  
from which we know that ${\cal R}^{\rm (out)}(p_{X_1X_2},p_{K_1K_2})$ also
serves as an outer bound of 
$\overline{\mathcal{R}}^{\ast}({\errP},{\secP}|p_{X_1X_2},$ 
$p_{K_1K_2})$    for $({\errP},{\secP}) 
            \in (0,1)\times [0,{\secP}_0]$.           
\end{IEEEproof}
\begin{IEEEproof}[Proof of Theorem \ref{th:ConvThTwo}]
We assume that   
$(R_1,R_2)\in {\cal S}^{\ast}({\errP},{\secP}
               |p_{X_1X_2},p_{K_1K_2})$.
Then by Proposition \ref{pro:KeyProTwoA}, we have 
that $\forall \gamma >0$,
     $\exists n_0=n_0(\gamma) \in \mathbb{N}$, 
     $\forall n\geq n_0$,
\begin{align}
R_i\leq & H(K_{i}) 
+2\gamma +\zeta_n(\gamma,{\errP},{\secP}),\: i=1,2.
\label{eqn:BoundThr}
\end{align}
Letting $n \to \infty$ in 
(\ref{eqn:BoundThr}) and considering that 
$\gamma>0$ can be taken arbitrary small, we have
that $R_i\leq H(K_{i}),\: i=1,2$. This implies
that we must have $(R_1,R_2)
\in {\cal S}(p_{X_1X_2},p_{K_1K_2})$. 

We next assume that   
$(R_1,R_2)\in \overline{\cal S}^{\ast}({\errP},{\secP}
               |p_{X_1X_2},p_{K_1K_2})$.
Then by Proposition \ref{pro:KeyProTwoA}, we have 
that $\exists\{k_n\}_{n\geq 1}$, $\forall \gamma >0$,
     $\exists n_1=n_1(\gamma) \in \mathbb{N}$, 
     $\forall n\geq n_1$,
\begin{align}
R_i\leq & H(K_{i}) 
+2\gamma +\zeta_{k_n}(\gamma,{\errP},{\secP}),\: i=1,2.
\label{eqn:BoundThrB}
\end{align}
Letting $n \to \infty$ in 
(\ref{eqn:BoundThrB}) and considering that 
$\gamma>0$ can be taken arbitrary small, we have
that $R_i\leq H(K_{i}),\: i=1,2$. This implies
that we must have $(R_1,R_2)
\in {\cal S}(p_{X_1X_2},p_{K_1K_2})$. 
\end{IEEEproof}

\newcommand{\ProofBaseLm}{
}
\newcommand{\ProofBaseLmB}{
\subsection{Proof of Lemma \ref{lm:BaseLm}
}\label{apd:PrfBaseLm}}

\section{
Proof of Propositions 
\ref{pro:KeyProOne} and \ref{pro:KeyProTwoA}
}\label{sec:PrKeyProOneTwo}

In this section we prove of Propositions 
\ref{pro:KeyProOne} and 
\ref{pro:KeyProTwoA} we presented in Section \ref{sec:PrKeyProOneTwo}. 
For the proof of those two propositions 
we need several lemmas. 
We present those lemmas 
and prove them in Section 
\ref{ssec:PrfLemsFotTwoProp}. 
Proofs of two propositions are given  
in Section \ref{ssec:PrfTwoProp}. 

\subsection{
Several Preliminaries on Random Variables} 
\label{ssec:RVResults}

In this subsection we present 
several preliminary results 
on random variables necessary 
for the proof of Propositions 
            \ref{pro:KeyProOne} 
        and \ref{pro:KeyProTwoA}.
The preliminary results are also useful
for the proof of Proposition \ref{pro:BasePro}
developed in Section \ref{sec:PrBasePro}.

Define the random pair 
$(\rvctxone,\rvctxtwo)\in \wt{\cal D}_{\gamma}^{(n)}$ 
by 
\begin{align*}
p_{\lrvctxone\lrvctxtwo}(\vcxone,\vcxtwo)
=\left\{
\ba{cl}
\ds \frac{p_{X_1X_2}^{n}(\vcxone,\vcxtwo)}
{p_{X_1X_2}^n\left(\wt{\cal D}_{\gamma}^{(n)}
\right)}
 & \mbox{ if } (\vcxone,\vcxtwo)\in 
\wt{\cal D}_{\gamma}^{(n)},
\vspace*{2mm}\\
0& \mbox{ otherwise}.
\ea
\right.
\end{align*}
For each $i=1,2$, let $\wt{C}_i^{(n)}$ 
be a random variable induced by 
$\Phi_{i}^{(n)},\wt{\rvcx}_i ,$ and $\rvcki$ 
in the following way:

\begin{align}
\wt{C}_i^{(n)}\defeq&
\Phi_{i}^{(n)}\left(\rvcki,\wt{\rvcx}_i\right)
\notag\\
=&
\Phi_{i,\lrvcki}^{(n)}\left(\wt{\rvcx}_i\right)
 =\Phi_{i,\wt{\lrvcx}_i}^{(n)}\left(\rvcki\right).
\label{eqn:DeFtCtX}
\end{align}
Set 
\begin{align*}
 {\cal C}^{(n)}_{
 (\Phi_{1}^{(n)},\Phi_{2}^{(n)})
 (\wt{\cal D}_{\gamma}^{(n)})}
 \defeq & \{(c_1,c_2):c_i=\Phi^{(n)}_{i,\lvcki}(\vcxi),i=1,2
 \\
 & \mbox{ for some }
  (\vckone,\vcktwo) 
  \in {\cal X}_1^n \times {\cal X}_2^n 
\\  
 & \mbox{ and }
 (\vcxone,\vcxtwo) \in 
 \wt{\cal D}_{\gamma}^{(n)}
  \}.
\end{align*}
Fix any $(\vckone,\vcktwo) 
\in {\cal X}_1^n \times {\cal X}_2^n$.
Set 
\begin{align*}
 {\cal C}^{(n)}_{
 (\Phi_{1,\svckone}^{(n)},\Phi_{2,\svcktwo}^{(n)})
 (\wt{\cal D}_{\gamma}^{(n)})}
 \defeq & \{(c_1,c_2): c_i=\Phi^{(n)}_{i,\lvcki}(\vcxi),i=1,2
 \\
 & \mbox{ for some }(\vcxone,\vcxtwo) \in 
 \wt{\cal D}_{\gamma}^{(n)}\}.
\end{align*}
We have the following property.
\begin{property}\label{pr:PrtiRvAndRv}
For each 
$$
  (c_{1},c_{2},\vcxone,\vcxtwo) 
 \in {\cal C}^{(n)}_{
 (\Phi_{1}^{(n)},\Phi_{2}^{(n)})
 (\wt{\cal D}_{\gamma}^{(n)})}
 \times \wt{\cal D}_{\gamma}^{(n)},
$$
we have the following:   
\begin{align*}
& p_{\wt{C}_1^{(n)}\wt{C}_2^{(n)}
    \lrvctxone\lrvctxtwo}(c_{1},c_{2},
   \vcxone,\vcxtwo)
\\  
& =\Pr\Bigl\{(\wt{C}_{1}^{(n)},
              \wt{C}_{2}^{(n)},
   \rvctxone,\rvctxtwo)=(c_{1},c_{2},\vcxone,\vcxtwo)
\\
& =\Pr\Bigl\{(C_{1}^{(n)},C_{2}^{(n)},
   \rvcxone,\rvcxtwo)=(c_{1},c_{2},\vcxone,\vcxtwo)
\\
&  \qquad\quad
   \left| \left(\rvcxone,\rvcxtwo\right) \in  
   \wt{\DcSet}_{\gamma}^{(n)}
   \right.\Bigr\}.
\end{align*}

\end{property}

Proof of this property is given in 
Appendix \ref{apd:ProofPrtiRvAndRv}. 
This property is closely 
related to the proofs of Propositions 
\ref{pro:KeyProOne} and \ref{pro:KeyProTwoA}.

\newcommand{\ProofPrtiRvAndRv}{
\subsection{
Proof of Property \ref{pr:PrtiRvAndRv}
}\label{apd:ProofPrtiRvAndRv}

In this appendix we prove Property \ref{pr:PrtiRvAndRv}.
\begin{IEEEproof}
For each 
$$
 (c_{1},c_{2},\vcxone,\vcxtwo) 
 \in {\cal C}^{(n)}_{
 (\Phi_{1}^{(n)},\Phi_{2}^{(n)})
 (\wt{\cal D}_{\gamma}^{(n)})}
 \times \wt{\cal D}_{\gamma}^{(n)},
$$
we have the following chain of equalities:  
\begin{align}
& p_{\wt{C}_1^{(n)}\wt{C}_2^{(n)}
    \lrvctxone\lrvctxtwo}(c_{1},c_{2},
   \vcxone,\vcxtwo)
\notag\\ 
&\MEq{a}\Pr\Bigl\{
  (\Phi_{1,{\lrvctxone}}^{(n)}({\rvckone}),
   \Phi_{2,{\lrvctxtwo}}^{(n)}({\rvcktwo}),
   \rvctxone,\rvctxtwo)\Bigr\}
\notag\\
&\quad\quad 
 =(c_{1},c_{2},\vcxone,\vcxtwo)\Bigr\}
\notag\\
&=\Pr\Bigl\{
  (\Phi_{1,{\lvcxone}}^{(n)}({\rvckone}),
   \Phi_{2,{\lvcxtwo}}^{(n)}({\rvcktwo}),
   \rvctxone,\rvctxtwo)\Bigr\}
\notag\\
&\quad\quad 
 =(c_{1},c_{2},\vcxone,\vcxtwo)\Bigr\}
\notag\\
&\MEq{b}\Pr\Bigl\{
  (\Phi_{1,{\lvcxone}}^{(n)}({\rvckone}),
   \Phi_{2,{\lvcxtwo}}^{(n)}({\rvcktwo})
   =(c_{1},c_{2})\Bigr\}
\notag\\
&\quad\times\Pr\Bigl\{
 (\rvctxone,\rvctxtwo)=(\vcxone,\vcxtwo)\Bigr\}.
\label{eqn:PrCXOne} 
\end{align}
Step (a) follows from (\ref{eqn:DeFtCtX}). 
Step (b) follows from
$(\Phi_{1,{\lvcxone}}^{(n)}$ $({\rvckone}),
  \Phi_{2,{\lvcxtwo}}^{(n)}({\rvcktwo}))
  \perp (\rvctxone,\rvctxtwo).$
From (\ref{eqn:PrCXOne}), we continue to compute 
to obtain the following chain of equalities: 
\begin{align*}
&p_{\wt{C}_1^{(n)}\wt{C}_2^{(n)}
    \lrvctxone\lrvctxtwo}(c_{1},c_{2},
   \vcxone,\vcxtwo)
\notag\\ 
&\MEq{a}\Pr\Bigl\{
  (\Phi_{1,{\lvcxone}}^{(n)}({\rvckone}),
   \Phi_{2,{\lvcxtwo}}^{(n)}({\rvcktwo})
   =(c_{1},c_{2})\Bigr\}
\notag\\
&\quad\times\Pr\Bigl\{
 (\rvcxone,\rvcxtwo)=(\vcxone,\vcxtwo)
   \left| \left(\rvcxone,\rvcxtwo\right) \in  
   \wt{\DcSet}_{\gamma}^{(n)}
   \right.\Bigr\}
\\
&\MEq{b}\Pr\Bigl\{
  (\Phi_{1,{\lvcxone}}^{(n)}({\rvckone}),
   \Phi_{2,{\lvcxtwo}}^{(n)}({\rvcktwo}), 
  \rvcxone,\rvcxtwo)
\notag\\
& 
 \qquad\quad
  =(c_{1},c_{2},\vcxone,\vcxtwo)
   \left| \left(\rvcxone,\rvcxtwo\right) \in  
   \wt{\DcSet}_{\gamma}^{(n)}
   \right.\Bigr\}
\notag \\
&=\Pr\Bigl\{
  (\Phi_{1,{\lrvcxone}}^{(n)}({\rvckone}),
   \Phi_{2,{\lrvcxtwo}}^{(n)}({\rvcktwo}), 
  \rvcxone,\rvcxtwo)
\notag\\
& 
 \qquad\quad
  =(c_{1},c_{2},\vcxone,\vcxtwo)
   \left| \left(\rvcxone,\rvcxtwo\right) \in  
   \wt{\DcSet}_{\gamma}^{(n)}
   \right.\Bigr\}
\\   
&=\Pr\Bigl\{(C_{1}^{(n)},C_{2}^{(n)},
   \rvcxone,\rvcxtwo)=(c_{1},c_{2},\vcxone,\vcxtwo)
\notag\\
&  \qquad\quad
   \left| \left(\rvcxone,\rvcxtwo\right) \in  
   \wt{\DcSet}_{\gamma}^{(n)}
   \right.\Bigr\}.
\end{align*}
Step (a) follows from (\ref{eqn:PrCXOne}) and the definition 
of $(\rvctxone,\rvctxtwo)$. 
Step (b) follows from 
$(\Phi_{1,{\lvcxone}}^{(n)}({\rvckone}),
  \Phi_{2,{\lvcxtwo}}^{(n)}({\rvcktwo}))
  \perp (\rvcxone,\rvcxtwo)$.
\end{IEEEproof}
}

\newcommand{\OmiTTazz}{
The following joint or conditional 
distributions are important for later arguments.      
\begin{align*}
 & p_{\wt{C}_i^{(n)}\wt{\lrvcx}_{3-i} }(c_{i},\vcx_{3-i})
    =\Pr\Bigl\{   
   (C_{i}^{(n)},\rvcx_{3-i}) =(c_{i},\vcx_{3-i})  \\
 & \qquad \qquad \quad
   \left|\left(\rvcxone,\rvcxtwo\right)
    \in \widetilde{{\DcSet}}_{\gamma}^{(n)}\right.\Bigr\}, \\
 & p_{\wt{C}_i^{(n)}|\wt{\lrvcx}_{3-i}}
   (c_{i}|\vcx_{3-i})
     =\Pr\Bigl\{ C_{i}^{(n)}=c_{i} \\
 & \qquad  \qquad \quad
    \left|\rvcx_{3-i}
    =\vcx_{3-i},\left(\rvcxone,\rvcxtwo\right) 
   \in \widetilde{{\DcSet}}_{\gamma}^{(n)}\right.\Bigr\},
   i=1,2,
\\   
 & p_{\wt{C}_1^{(n)}\wt{C}_2^{(n)}|
     \lrvctxone\lrvctxtwo}
 (c_{1},c_{2}|\vcxone,\vcxtwo)
     =\Pr\Bigl\{ (C_{1}^{(n)},C_{2}^{(n)})
       =(c_{1},c_{2}) \\
 & \qquad  \qquad \quad
    \left|(\rvcxone,\rvcxtwo)
    =(\vcxone,\vcxtwo), \left(\rvcxone,\rvcxtwo\right) 
   \in \widetilde{{\DcSet}}_{\gamma}^{(n)}\right.\Bigr\}.
\end{align*}
}

\subsection{
Several Lemmas 
Necessary for the Proofs of  
Propositions \ref{pro:KeyProOne} 
         and \ref{pro:KeyProTwoA}}        
\label{ssec:PrfLemsFotTwoProp}
In this subsection we present 
several lemmas necessary for 
the proof of Propositions 
             \ref{pro:KeyProOne} 
         and \ref{pro:KeyProTwoA}.
We further prove those lemmas.
We first present several definitions. 
For each $i=1,2$, we set 
\begin{align}
(\wt{\DcSet}_{\gamma}^{(n)})_i
\defeq \{\vcxi:(\vcxone,\vcxtwo)
\in  \wt{\DcSet}_{\gamma}^{(n)}
\mbox{ for some }{\vc x}_{3-i}\}. 
\end{align}
For each $i=1,2$ and each $\vcxi\in 
(\wt{\DcSet}_{\gamma}^{(n)})_i$, 
we set 
$$
\wt{\DcSet}_{i|3-i,\gamma}^{(n)}(\vcxi|{\vc x}_{3-i})
\defeq \{\vcxi: (\vcxone,\vcxtwo)
\in \wt{\DcSet}_{\gamma}^{(n)}\}.
$$
Set
\begin{align*}
& Q_{12} \defeq 
p_{X_{1}X_{2}}^{n}
 \left(\widetilde{{\DcSet}}_{\gamma}^{(n)}
     \right)
= \Pr\left\{(\rvcxone,\rvcxtwo)
 \in\wt{{\DcSet}}_{\gamma}^{(n)}   
\right\}
\\
&=\sum_{(\lvcxone,\lvcxtwo)\in \wt{\DcSet}_{\gamma}^{(n)}}
    p_{\lrvcxone\lrvcxtwo}(\vcxone,\vcxtwo). 
\end{align*}
For each $i=1,2$, set
\begin{align*}
& Q_{i} \defeq 
\Pr\left\{\rvcxi\in \left(\wt{{\DcSet}}_{\gamma}^{(n)}
                    \right)_i \right\}
=\sum_{\lvcxi \in 
       \left(\wt{{\cal D}}_{\gamma}^{(n)}\right)_i} 
      p_{\lrvcxi}(\vcxi).  
\end{align*}
For each $i=1,2$ and $\vcx_{3-i}\in 
\left(\wt{{\DcSet}}_{\gamma}^{(n)}\right)_{3-i}$, set 
\begin{align*}
& Q_{i|3-i}(\vcx_{3-i})
\\
& \defeq 
 \Pr\left\{\left.\rvcxi\in\wt{{\DcSet}}_{i|3-i,\gamma}^{(n)}   
\left(\vcx_{3-i}\right) 
\right|\rvcx_{3-i}=\vcx_{3-i}\right\}
\\
&=\sum_{\lvcxi \in \wt{{\DcSet}}_{i|3-i,\gamma}^{(n)}
    \left(\lvcx_{3-i}\right)}
    p_{\lrvcxone|\lrvcxtwo}(\vcxi|\vcx_{3-i}).  
\end{align*}
For each $i=1,2$ and for any 
$(c_1,c_2)\in 
{\cal C}_1^{(n)}\times {\cal C}_2^{(n)}$ 
and any $\vcx_{3-i}\in 
\left(\wt{{\DcSet}}_{\gamma}^{(n)}\right)_{3-i}$, 
we have the followings: 
\begin{align}
 & p_{\wt{C}_i^{(n)}|\wt{\lrvcx}_{3-i}}
 (c_{i}|\vcx_{3-i})
   =[Q_{i|3-i}(\vcx_{3-i})]^{-1} \ds \sum_{\lvcxi\in
    \wt{{\DcSet}}_{\gamma}^{(n)}\left(\lvcx_{3-i}\right)}1
 \notag\\  
 & \times p_{C_{i}^{(n)}|\lrvcxi
                \lrvcx_{3-i}}(c_{i}|\vcxone,\vcxtwo)
          p_{\lrvcxi|\lrvcx_{3-i}}(\vcxi|\vcx_{3-i}),
\label{eqn:distqOne} \\
& p_{\wt{\lrvcx}_{3-i}
}(\vcx_{3-i})
\notag\\
& =\frac{Q_{i|3-i}(\vcx_{3-i})p_{\lrvcx_{3-i}}(\vcx_{3-i})}
  {\ds \sum_{\lvcxtwo\in
   \left(\widetilde{{\DcSet}}_{\gamma}^{(n)}\right)_{2}}
  Q_{i|3-i}(\vcx_{3-i})p_{\lrvcx_{3-i}}(\vcx_{3-i})},
\label{eqn:distqTwo} \\
 & p_{\wt{C}_1^{(n)}\wt{C}_2^{(n)}}(c_{1},c_{2})
   = [Q_{12}]^{-1}\sum_{(\lvcxone,\lvcxtwo)\in
    \wt{{\DcSet}}_{\gamma}^{(n)}}1
\notag\\
 & \times p_{C_{1}^{(n)}C_{2}^{(n)}|\lrvcxone \lrvcxtwo} 
          (c_{1},c_{2}|\vcxone,\vcxtwo)
          p_{\lrvcxone\lrvcxtwo}(\vcxone,\vcxtwo).
\label{eqn:distqThr}          
\end{align}
For the conditional distributions 
$p_{\wt{C}_i^{(n)}|\wt{\lrvcx}_{3-i}},i=1,2$ 
and the distribution 
$p_{\wt{C}_1^{(n)}\wt{C}_2^{(n)}}(c_{1},c_{2})$, 
we have the following lemma.
\begin{lemma}\label{lm:PrBdLem}
For any $(c_1,c_2)\in 
{\cal C}_1^{(n)}\times {\cal C}_2^{(n)}$ 
and any $\vcx_{3-i}\in 
\left(\wt{{\DcSet}}_{\gamma}^{(n)}\right)_{3-i}$, 
we have 
\begin{align}
& p_{\wt{C}_i^{(n)}|\wt{\lrvcx}_{3-i}}(c_{i}|\vcx_{3-i})
\notag\\
& \leq \left[Q_{i|3-i}(\vcx_{3-i})\right]^{-1}
   2^{-n[H(X_{i}|X_{3-i})-\gamma]}, i=1,2,
\label{lm:PrBdOne}\\
& p_{\wt{C}_1^{(n)}\wt{C}_2^{(n)}}(c_{1},c_{2})
\leq [Q_{12}]^{-1}2^{-n[H(X_{1}X_{2})-\gamma]}.
\label{lm:PrBdTwo}
\end{align}
\end{lemma}

\begin{IEEEproof}
We first prove (\ref{lm:PrBdOne}) for $i=1$.
By the definition $\wt{\DcSet}_{\gamma}^{(n)}
  =\wt{\cal A}_{\gamma}^{(n)}\cap\mathcal{D}^{(n)}$, we  
  have 
  $$ 
   \wt{\DcSet}_{1|2,\gamma}^{(n)}(\vcxtwo) 
   =\wt{\cal A}_{1|2,\gamma}^{(n)}(\vcxtwo)
   \cap\mathcal{D}_{1|2}^{(n)}(\vcxtwo),\: 
  \forall \vcxtwo 
  \in \left(\wt{{\DcSet}}_{\gamma}^{(n)}\right)_{2}, 
  $$
  implying that
  \begin{align}
  & p_{\lrvcxone|\lrvcxtwo}(\vcxone|\vcxtwo)
   \notag\\  
  &\leq 2^{-n[H(X_{1}|X_{2})-\gamma]}
    \mbox{ for }\vcxone\in
    \widetilde{{\DcSet}}_{1|2,\gamma}^{(n)}
  \left(\vcxtwo\right). 
   \label{eqn:probBd}
   \end{align}
For each $(c_1.\vcxtwo)\in {\cal C}_1^{(n)}
                \times 
                \left(\wt{\cal D}_{\gamma}^{(n)}\right)_2$, 
we compute 
$p_{\wt{C}_1^{(n)}|\lrvctxtwo}(c_{1}|$ $\vcxtwo)Q_{1|2}((\vcxtwo)$ 
to obtain as follows:
\begin{align*}
 & p_{\wt{C}_1^{(n)}|\lrvctxtwo}(c_{1}|\vcxtwo)
   Q_{1|2}(\vcxtwo)
  \\
 &\MEq{a}
 \sum_{\lvcxone \in 
    \wt{{\DcSet}}_{1|2,\gamma}^{(n)}(\lvcxtwo) }
  p_{C_{1}^{(n)}|\lrvcxone\lrvcxtwo}(c_{1}|\vcxone,\vcxtwo)
  p_{\lrvcxone|\lrvcxtwo}(\vcxone|\vcxtwo)
\\  
  & \MLeq{b}{\ds  2^{-n[H(X_{1}|X_{2})-\gamma]} 
   \sum_{\lvcxone\in\mathcal{D}_{1|2}^{(n)}(\lvcxtwo)}
   p_{C_{1}^{(n)}|\lrvcxone\lrvcxtwo}
   (c_{1}|\vcxone,\vcxtwo) }\\
 & \MLeq{c}2^{-n[H(X_{1}|X_{2})-\gamma]}.
\end{align*}
Step (a) follows from (\ref{eqn:distqOne}).
Step (b) follows from (\ref{eqn:probBd}).
Step (c) follows from Lemma \ref{lem:LemOne}.
In a similar manner, we can prove 
(\ref{lm:PrBdOne}) for $i=2$. 
We next prove (\ref{lm:PrBdTwo}). We 
first observe that 
By the definition 
${\wt{\DcSet}}_{\gamma}^{(n)}
={\wt{\cal A}}_{\gamma}^{(n)}
 \cap\mathcal{D}^{(n)}$, we have 
\begin{align}
  & p_{\lrvcxone\lrvcxtwo}(\vcxone,\vcxtwo)
   \notag\\  
  & \leq 2^{-n[H(X_{1}X_{2})-\gamma]}
    \mbox{ for }(\vcxone,\vcxtwo)\in
    \wt{{\DcSet}}_{\gamma}^{(n)}.
  \label{eqn:probBdTwo}
\end{align}
For each $(c_1.c_2)\in {\cal C}_1^{(n)}
                \times {\cal C}_2^{(n)}$, 
we compute $p_{\wt{C}_1^{(n)}\wt{C}_2^{(n)}}(c_{1},c_{2})$ $Q_{12}$ to obtain as follows:
\begin{align*}
 & p_{\wt{C}_1^{(n)}\wt{C}_2^{(n)}}(c_{1},c_{2})Q_{12}
   \MEq{a} \sum_{(\lvcxone,\lvcxtwo)\in
    \wt{{\DcSet}}_{\gamma}^{(n)}}1
\notag\\
 &\quad \times p_{C_{1}^{(n)}C_{2}^{(n)}|\lrvcxone \lrvcxtwo} 
          (c_{1},c_{2}|\vcxone,\vcxtwo)
          p_{\lrvcxone\lrvcxtwo}(\vcxone,\vcxtwo).
\\
 & \MLeq{b}\ds  2^{-n[H(X_{1}X_{2})-\gamma]} 
   \sum_{(\lvcxone,\lvcxtwo) \in {\DcSet}^{(n)}}1
\\   
 & \quad\times   
   p_{C_{1}^{(n)}C_{2}^{(n)}|\lrvcxone\lrvcxtwo}
   (c_{1},c_{2}|\vcxone,\vcxtwo) 
 \MLeq{c}2^{-n[H(X_{1}X_{2})-\gamma]}.
\end{align*}
Step (a) follows from (\ref{eqn:distqOne}).
Step (b) follows from (\ref{eqn:probBdTwo}).
Step (c) follows from Lemma \ref{lem:LemOne}.
\end{IEEEproof}

We next derive lower bounds of 
\begin{align*}
& H\left(C_{i}^{(n)}
\left|\rvcx_{3-i},(\rvcxone,\rvcxtwo)
 \in \widetilde{{\DcSet}}_{\gamma}^{(n)} \right.\right)
\\ 
&=H\left(\wt{C}_{i}^{(n)}\left|\wt{\rvcx}_{3-i}\right.\right),i=1,2,
\\
& H\left(C_{1}^{(n)}C_{2}^{(n)} \left| (\rvcxone,\rvcxtwo) 
   \in \wt{{\DcSet}}_{\gamma}^{(n)} \right.\right)
 =H\left(\wt{C}_{1}^{(n)}\wt{C}_{2}^{(n)}\right).
\end{align*}
On lower bounds of those three quantities 
we have the following lemma:
\begin{lemma}\label{lm:EntLow}
\begin{align}
& H\left(\left.\wt{C}_{i}^{(n)}\right|\wt{\rvcx}_{3-i}\right)
\notag\\
& \geq nH(X_i|X_{3-i})+\log Q_{12},\: i=1,2,
\label{eqn:entC1}
\\
&H\left(\wt{C}_{1}^{(n)}\wt{C}_{2}^{(n)}\right)
\geq nH(X_1X_{2}) + \log Q_{12}.
\label{eqn:entC2}
\end{align}
\end{lemma}

\begin{IEEEproof}
We first prove (\ref{eqn:entC1}). By a symmetrical 
structure of the problem it suffices to prove 
the bound for $i=1$.
We have the following chain of inequalities:
\begin{align}
& H\left(\left.\wt{C}_{1}^{(n)}\right|\wt{\rvcx}_2 
   \right)
\nonumber\\
& =\sum_{c_{1}\in\mathcal{M}_{1}^{(n)}}
   \sum_{\lvcxtwo \in
      \left(\widetilde{{\DcSet}}_{\gamma}^{(n)}\right)_{2}}
      p_{\wt{C}_1^{(n)}\lrvctxtwo}(c_{1},\vcxtwo)
  \log\frac{1}{p_{\wt{C}_1^{(n)}|\lrvctxtwo}(c_{1}|\vcxtwo)}
 \nonumber \\
 & \MGeq{a}\sum_{c_{1}\in\mathcal{M}_{1}^{(n)}}
   \sum_{\lvcxtwo \in
    \left(\widetilde{{\DcSet}}_{\gamma}^{(n)}\right)_{2}}
     p_{\wt{C}_1^{(n)}\lrvctxtwo}(c_{1},\vcxtwo)
 \nonumber\\
 & \qquad \times 
          \log\left[
          Q_{1|2}(\vcxtwo)2^{n[H(X_{1}|X_{2})-\gamma]}
          \right]
\nonumber \\
 & =\sum_{\lvcxtwo\in
     \left(\widetilde{{\DcSet}}_{\gamma}^{(n)}\right)_{2}}
    p_{\lrvctxtwo}(\vcxtwo)\log Q_{1|2}(\vcxtwo)
\nonumber\\
 &\quad  
 +n[H(X_{1}|X_{2})-\gamma].
\label{eqn:pxLogA}
\end{align}
Step (a) follows from (\ref{lm:PrBdOne}) in Lemma \ref{lm:PrBdLem}.
To derive a lower bound of 
$H\left(\left.\wt{C}_{1}^{(n)}\right|\wt{\rvcx}_2\right)$,
it suffices to estimate a lower bound of the first term
in the right member of  (\ref{eqn:pxLogA}).
We denote this quantity by $\Lambda$. On this quantity we have
\begin{align*}
\Lambda & =\frac{\ds\sum_{\lvcxtwo         \in\left(\widetilde{{\DcSet}}_{\gamma}^{(n)}\right)_{2}}
     p_{\lrvcxtwo}(\vcxtwo)Q_{1|2}(\vcxtwo)\log Q_{1|2}(\vcxtwo)}
       {\ds \sum_{\lvcxtwo
       \in \left(\wt{{\DcSet}}_{\gamma}^{(n)}\right)_{2}}
       p_{\lrvcxtwo}(\vcxtwo)Q_{1|2}(\vcxtwo)}.
\end{align*}
Define the probability distribution $\wt{p}_2$ on 
$\left(\wt{{\DcSet}}_{\gamma}^{(n)}\right)_{2}$ by
$$
\widetilde{p}_{2}(\vcxtwo)
\defeq
\frac{p_{\lrvcxtwo}(\vcxtwo)}
 {\ds  \sum_{\lvcxtwo \in 
   \left(\wt{{\DcSet}}_{\gamma}^{(n)}\right)_{2}
    }
   p_{\lrvcxtwo}(\vcxtwo)
 }
 =\frac{p_{\lrvcxtwo(\vcxtwo)}}   
   {p_{\lrvcxtwo}\left(\left(
        \wt{{\DcSet}}_{\gamma}^{(n)}
                 \right)_{2}\right)}.
$$
Set 
\begin{align*}
&\Lambda_1\defeq 
 \sum_{\lvcxtwo \in
    \left(\widetilde{{\DcSet}}_{\gamma}^{(n)}\right)_{2}}
     \wt{p}_{2}(\vcxtwo)Q_{1|2}(\vcxtwo)
     \log Q_{1|2}(\vcxtwo)
\\
&\Lambda_2 \defeq   \sum_{\lvcxtwo \in
    \left(\wt{{\DcSet}}_{\gamma}^{(n)}\right)_{2}}
     \wt{p}_{2}(\vcxtwo)Q_{1|2}(\vcxtwo)
\end{align*}
Then $\Lambda$ can be expressed as $\Lambda=\Lambda_1\Lambda_2^{-1}$.
Furthermore, by definition we have 
\begin{align}
  \Lambda_2=Q_{12}\left[p_{\lrvcxtwo}\left(\left(
        \wt{{\DcSet}}_{\gamma}^{(n)}
                 \right)_{2}\right)\right]^{-1}.
 \label{eqn:LambdaEq}                
\end{align} 
On lower bounds of $\Lambda$, we have the following chain 
of inequalities:
\begin{align}
&\Lambda=\frac{1}{\Lambda_2}{\ds\sum_{\lvcxtwo 
         \in\left(\wt{{\DcSet}}_{\gamma}^{(n)}\right)_{2}} 
    \wt{p}_{2}(\vcxtwo)Q_{1|2}(\vcxtwo)\log Q_{1|2}(\vcxtwo)}  
\notag\\
 & \MGeq{a}\frac{1}{\Lambda_2}\Lambda_2 \log \Lambda_2
 \MEq{b}\log \frac{Q_{12}}
              {p_{\lrvcxtwo}
              \left(\left(
                 \wt{{\DcSet}}_{\gamma}^{(n)}
                  \right)_{2}\right)}
  \geq\log Q_{12}. 
\label{eqn:lower-bound-H}
\end{align}
Step (a) follows from the convex property of 
$z\log z$ for $z>0$ and the Jensen's inequality.
Step (b) follows from (\ref{eqn:LambdaEq}).
\newcommand{\OTTS}{
 & =\log\left(\frac{\ds
 \sum_{\lvcxtwo\in\left(
   \widetilde{{\DcSet}}_{\gamma}^{(n)}
               \right)_{2}}
  p_{\lrvcxtwo(\vcxtwo)}   \sum_{\lvcxone
  \in\widetilde{{\DcSet}}_{\gamma}^{(n)}
  \left(\lvcxtwo\right)}
   p_{\lrvcxone|\lrvcxtwo}(\vcxone |\vcxtwo)}{p_{\lrvcxtwo}
   \left(\left(
   \widetilde{{\DcSet}}_{\gamma}^{(n)}
   \right)_{2}\right)}\right)
   \nonumber \\
   }
We next prove (\ref{eqn:entC2}).
We have the following chain of inequalities:
\begin{align*}
& H\left(\wt{C}_{1}^{(n)}\wt{C}_{2}^{(n)}\right)
  =\sum_{(c_{1},c_{2})
   \in {\cal C}_{1}^{(n)}\times {\cal C}_{2}^{(n)} }
      p_{\wt{C}_1^{(n)}\wt{C}_1^{(n)}}(c_{1},c_{2})
\nonumber\\
&\quad  \times 
  \log\frac{1}{q_{\wt{C}_1^{(n)}\wt{C}_1^{(n)}}(c_{1},c_{2})}
 \nonumber \\
 & \MGeq{a} 
 \sum_{(c_{1},c_{2})
   \in {\cal C}_{1}^{(n)}\times {\cal C}_{2}^{(n)} }
     p_{\wt{C}_1^{(n)},\wt{C}_2^{(n)}}(c_{1},c_{2})
 \nonumber\\
 & \qquad \times 
          \log\left[
          Q_{12}2^{n[H(X_{1}X_{2})-\gamma]}
          \right]
\nonumber \\
 & = \log Q_{12}+n[H(X_{1}X_{2})-\gamma].
\end{align*}
Step (a) follows from (\ref{lm:PrBdTwo}) in Lemma \ref{lm:PrBdLem}.
\end{IEEEproof}

We next evaluate the following quantities:
\begin{align*}
& H\left(C_{i}^{(n)}\left|\rvcxone,\rvcxtwo,
  (\rvcxone,\rvcxtwo)
   \in\wt{{\DcSet}}_{\gamma}^{(n)}\right.\right)
\\
&=H(\wt{C}_i^{(n)}|\wt{\rvcx}_{1}\wt{\rvcx}_{2})
  \leq H(\wt{C}_i^{(n)}|\wt{\rvcx}_{i}),\: i=1,2,
\\
& H\left(C_{1}^{(n)}C_{2}^{(n)}\left|\rvcxone,\rvcxtwo,
  (\rvcxone,\rvcxtwo)
  \in\wt{{\DcSet}}_{\gamma}^{(n)}\right.\right)
\\
&=H\left(\wt{C}_1^{(n)}\wt{C}_1^{(n)}
   \left|\wt{\rvcx}_{1}\wt{\rvcx}_{2}\right.\right). 
\end{align*}
We have the following lemma. 
\begin{lemma}\label{lm:EntUp}
We have the following:
\begin{align}
&H\left(C_{i}^{(n)}\left|\rvcxone\rvcxtwo, 
 (\rvcxone,\rvcxtwo)
  \in\widetilde{{\DcSet}}_{\gamma}^{(n)}\right.\right)
\notag\\
&\leq H\left(C_{i}^{(n)}\left|\rvcxi, 
 (\rvcxone,\rvcxtwo)
  \in\wt{{\DcSet}}_{\gamma}^{(n)}\right.\right)
\notag \\
& \leq nH(K_i),\:i=1,2,
\label{eqn:EntUpbOne}\\  
& H\left(C_{1}^{(n)}C_{2}^{(n)}\left|\rvcxone\rvcxtwo, 
 (\rvcxone,\rvcxtwo)
  \in\widetilde{{\DcSet}}_{\gamma}^{(n)}\right.\right)
\notag\\
& \leq nH(K_1K_2).
\label{eqn:EntUpbTwo}
\end{align}
\end{lemma}

\begin{IEEEproof}
We first prove the bound (\ref{eqn:EntUpbOne}) .
For each $i=1,2$, we have the following chain of 
inequalities:    
\begin{align}
 &  H\left(C_{i}^{(n)}\left|\rvcxone\rvcxtwo, 
 (\rvcxone,\rvcxtwo)
  \in\wt{{\DcSet}}_{\gamma}^{(n)}\right.\right)
  \nonumber \\
  &\leq  H\left(C_{i}^{(n)}\left|\rvcxi, 
  (\rvcxone,\rvcxtwo)
  \in\wt{{\DcSet}}_{\gamma}^{(n)}\right.\right)
 \nonumber\\
  &=
 H\left(\Phi_{\lrvcxi}^{(n)}(\rvcki)
 \left|\rvcxi,(\rvcxone,\rvcxtwo)
  \in\wt{{\DcSet}}_{\gamma}^{(n)}\right.\right)
\nonumber\\
 &\MLeq{a} H\left(\rvcki \left|\rvcxi,(\rvcxone,\rvcxtwo)
    \in\wt{{\DcSet}}_{\gamma}^{(n)}\right.\right)
  \MEq{b}nH(K_{i}).
 \notag
\end{align}
Step (a) follows from the data processing inequality.
Step (b) follows from $\rvcki \perp \rvcxone\rvcxtwo$ and 
the memoryless property of the key sources. 
We next prove (\ref{eqn:EntUpbTwo}). We have 
the following chain of inequalities:
\begin{align}
 &  H\left(C_{1}^{(n)}C_{2}^{(n)}\left|\rvcxone\rvcxtwo, 
 (\rvcxone,\rvcxtwo)
  \in\wt{{\DcSet}}_{\gamma}^{(n)}\right.\right)
 \nonumber \\
  &=H\left(
    \Phi_{\lrvcxone}^{(n)}(\rvckone)
    \Phi_{\lrvcxtwo}^{(n)}(\rvcktwo)
   \left|\rvcxone\rvcxtwo, (\rvcxone,\rvcxtwo)
  \in\wt{{\DcSet}}_{\gamma}^{(n)}\right.\right)
 \nonumber \\
 & \MLeq{a} H\left(\rvckone\rvcktwo    
  \left|\rvcxone\rvcxtwo, (\rvcxone,\rvcxtwo)
  \in \wt{{\DcSet}}_{\gamma}^{(n)}\right.\right)
\MEq{b}nH(K_{1}K_{2}).
 \notag
\end{align}
Step (a) follows from the data processing inequality.
Step (b) follows from $\rvckone\rvcktwo 
\perp \rvcxone\rvcxtwo$ and 
the memoryless property of the key sources. 
\end{IEEEproof}

The following lemma shows a relationship between 
security, reliability and mutual information. 
\begin{lemma}\label{lm:SecReMI}
We assume that 
$I(\rvcxone\rvcxtwo;C_{1}^{(n)}C_{2}^{(n)})
  \leq {\secP}$. Then 
 we have the followings.
\begin{align}
& Q_{12}^{-1}{\secP} \geq 
 I\left(\rvcxone\rvcxtwo;C_{1}^{(n)}C_{2}^{(n)}
     \left|(\rvcxone,\rvcxtwo)
      \in\widetilde{{\DcSet}}_{\gamma}^{(n)}
    \right.\right).
\label{eqn:MIUb}
\end{align}
\end{lemma}
\begin{IEEEproof}
Define
\begin{align*}
\xi(\rvcxone,\rvcxtwo) & \defeq
\begin{cases}
1 & \text{if }(\rvcxone,\rvcxtwo)
    \in\widetilde{{\DcSet}}_{\gamma}^{(n)},\\
0 & \text{otherwise}.
\end{cases}
\end{align*}
On lower bounds of 
$I (\rvcxone\rvcxtwo;$ $C_{1}^{(n)}C_{2}^{(n)})$, 
we have the following chain of inequalities: 
\begin{align*}
 & I(\rvcxone\rvcxtwo;C_{1}^{(n)}C_{2}^{(n)})
   =I(\rvcxone\rvcxtwo,\xi(\rvcxone,\rvcxtwo);
     C_{1}^{(n)}C_{2}^{(n)})
   \\
 & \ \geq I\left(\rvcxone\rvcxtwo; \left.
      C_{1}^{(n)}C_{2}^{(n)}
      \right|\xi(\rvcxone,\rvcxtwo)\right)
  \\    
 & \ \geq
     Q_{12}I\left(\rvcxone\rvcxtwo;C_{1}^{(n)}C_{2}^{(n)}
     \left|(\rvcxone,\rvcxtwo)
    \in\widetilde{{\DcSet}}_{\gamma}^{(n)}\right.\right),
\end{align*}
from which we have 
\begin{align*}
 & 
    Q_{12}^{-1}
      {I(\rvcxone\rvcxtwo;C_{1}^{(n)}C_{2}^{(n)})}
   \\
 & \ \geq
     I\left(\rvcxone\rvcxtwo;C_{1}^{(n)}C_{2}^{(n)}
     \left|(\rvcxone,\rvcxtwo)    \in\wt{{\DcSet}}_{\gamma}^{(n)}\right.\right).
\end{align*}
Since  
$I(\rvcxone\rvcxtwo;C_{1}^{(n)}C_{2}^{(n)})
  \leq {\secP}$, we have the bound (\ref{eqn:MIUb}).
\end{IEEEproof}

\subsection{Proofs of
Propositions \ref{pro:KeyProOne} 
         and \ref{pro:KeyProTwoA}
}\label{ssec:PrfTwoProp}
In this subsection we prove 
Propositions \ref{pro:KeyProOne} 
         and \ref{pro:KeyProTwoA}.

\begin{IEEEproof}[Proof of 
 Proposition \ref{pro:KeyProOne}]
We assume that 
${\cal R}^{\ast}({\errP},{\secP}| 
p_{X_1X_2},$ $p_{K_1X_2})\neq \emptyset$.
Then $\exists 
         \{(\Phi_1^{(n)}, \Phi_2^{(n)},$ $\Psi^{(n)})\}_{n \geq 1}$
        such that $\forall \gamma >0$,
        $\exists n_0=n_0(\gamma) \in \mathbb{N}$, 
 	$\forall n\geq n_0$, we have 
	\begin{align}
& p_{X_1X_2}^n
  ({\DcSet}^{(n)})=
p_{{\rm e}}^{(n)}(
\phi_1^{(n)},
\phi_2^{(n)},
\psi^{(n)}|{p}_{X_1X_2}^n)
\leq \errP,
\label{eqn:PrfPrTrOne}\\
& I(C_1^{(n)}C_2^{(n)}; \rvcxone \rvcxtwo) 
\leq \secP.
\label{eqn:PrfPrTrTwo}
\end{align}
By (\ref{eqn:PrfPrTrTwo}) and 
Lemma \ref{lm:SecReMI}, we have
\begin{align}
 Q_{12}^{-1}{\secP} \geq 
 I\left(\rvcxone\rvcxtwo;C_{1}^{(n)}C_{2}^{(n)}
     \left|(\rvcxone,\rvcxtwo)
      \in\widetilde{{\DcSet}}_{\gamma}^{(n)}
    \right.\right).
\label{eqn:MIUbb}   
\end{align}
On lower bounds of
$ I(\rvcxone\rvcxtwo;$ $C_{1}^{(n)}C_{2}^{(n)}|(\rvcxone,\rvcxtwo)
       \in\widetilde{{\DcSet}}_{\gamma}^{(n)})
$, we have the following three chains of inequalities: 
\begin{align}
  & I\left(\rvcxone\rvcxtwo;C_{1}^{(n)}C_{2}^{(n)}
     \left|(\rvcxone,\rvcxtwo)
      \in\widetilde{{\DcSet}}_{\gamma}^{(n)}
    \right.\right)
 \notag \\   
 &\geq I(\rvcxi,C_{i}^{(n)}|\rvcx_{3-i},(\rvcxone,\rvcxtwo)
       \in\widetilde{{\DcSet}}_{\gamma}^{(n)})
 \notag\\ 
 &=H\left(C_{i}^{(n)}\left|\rvcx_{3-i},(\rvcxone,\rvcxtwo)
       \in\wt{{\DcSet}}_{\gamma}^{(n)}\right.\right)
 \notag \\  
 & \quad\ -H\left(C_{i}^{(n)}\left|\rvcxone\rvcxtwo,(\rvcxone,\rvcxtwo)
      \in \wt{{\DcSet}}_{\gamma}^{(n)}\right.\right).
 \notag \\
 &\MGeq{a}
   H(\wt{C}_{i}^{(n)}|\wt{\rvcx}_{3-i})-nH(K_i)
 \notag \\
 &\MGeq{b}n[H(X_i|X_{3-i})-\gamma]+\log Q_{12}-nH(K_i).
\label{eqn:MILb} 
\end{align}
\begin{align}
  & I\left(\rvcxone\rvcxtwo;C_{1}^{(n)}C_{2}^{(n)}
     \left|(\rvcxone,\rvcxtwo)
      \in\widetilde{{\DcSet}}_{\gamma}^{(n)}
    \right.\right)
 \notag \\   
 &=H\left(C_{1}^{(n)}C_{2}^{(n)}\left|
       (\rvcxone,\rvcxtwo)
       \in\wt{{\DcSet}}_{\gamma}^{(n)}\right.\right)
 \notag \\  
 & \quad\ 
    -H\left(C_{1}^{(n)}C_{2}^{(n)}\left|
      \rvcxone\rvcxtwo.
      (\rvcxone,\rvcxtwo)
      \in \wt{{\DcSet}}_{\gamma}^{(n)}\right.\right)
 \notag \\
 &\MGeq{c}
   H(\wt{C}_{1}^{(n)}\wt{C}_{2}^{(n)})-nH(K_1K_2)
 \notag \\
 &\MGeq{d}n[H(X_1X_2)-\gamma]+\log Q_{12}
 -nH(K_1K_2).
\label{eqn:MILc} 
\end{align}
Step (a) follows from 
the bound (\ref{eqn:EntUpbOne}) 
in Lemma \ref{lm:EntUp}.
Step (b) follows from 
the bound (\ref{eqn:entC1}) 
in Lemma \ref{lm:EntLow}.
Step (c) follows from 
the bound (\ref{eqn:EntUpbTwo}) 
in Lemma \ref{lm:EntUp}.
Step d) follows from 
the bound (\ref{eqn:entC2}) 
in Lemma \ref{lm:EntLow}.
Combining (\ref{eqn:MIUbb}) with 
(\ref{eqn:MILb}) and (\ref{eqn:MILc}),
we have the following three bounds: 
\begin{align}
\left.
\ba{ll}
Q_{12}^{-1}{\secP} 
 \geq &n[H(X_i|X_{3-i})-\gamma]
\\
      &\:+\log Q_{12}-nH(K_i),\:i=1,2,
\\
Q_{12}^{-1}{\secP} 
\geq & n[H(X_1X_2)-\gamma]
\\
     &\: +\log Q_{12}-nH(K_1K_2).
\ea 
\right\}
\end{align}
Those are equivalent to
the followings:
\begin{align}
\left.
\ba{rl}
H(X_i|X_{3-i})\leq & H(K_i)+\gamma
\\
 & \:+\ds\frac{1}{n}\left[
   \frac{{\secP}}{Q_{12}}+\log\frac{1}{Q_{12}}
   \right],\: i=1,2
\vspace*{1.5mm}\\
H(X_1X_2)\leq & H(K_1K_2)+\gamma 
\\
& \:+\ds\frac{1}{n}\left[
  \frac{{\secP}}{Q_{12}}+\log\frac{1}{Q_{12}}
  \right].
\ea 
\right\}
\label{eqn:ProOneBd}
\end{align}
Here we note that by (\ref{eqn:PrfPrTrOne}), 
we have
\begin{align}
Q_{12}
&=p_{\lrvcxone\lrvcxtwo}
 \left(\wt{{\DcSet}}_{\gamma}^{(n)} \right)
=p_{X_1X_2}^n
\left(\wt{{\DcSet}}_{\gamma}^{(n)} \right)
\notag\\
&\geq 1-{\nu}_{n}(\gamma,{\errP}).
\label{eqn:LbQonetwo}
\end{align}
From (\ref{eqn:ProOneBd}) and 
     (\ref{eqn:LbQonetwo}), 
we have the bound the bound (\ref{eqn:BdProThrOne})
in Proposition \ref{pro:KeyProOne}.

We next assume that 
$\overline{\cal R}^{\ast}({\errP},{\secP}| 
p_{X_1X_2},$ $p_{K_1X_2})\neq \emptyset$.
Then $\exists 
         \{(\Phi_1^{(n)}, \Phi_2^{(n)},$ $\Psi^{(n)})\}_{n \geq 1}$, $\exists\{k_n\}_{n\geq 1}$    
        such that $\forall \gamma >0$,
        $\exists n_1=n_1(\gamma) \in \mathbb{N}$, 
 	$\forall n\geq n_1$, we have 
	\begin{align}
& p_{X_1X_2}^{k_n}
  ({\DcSet}^{(k_n)})
 \notag\\
&=p_{{\rm e}}^{(k_n)}(
\phi_1^{(k_n)},
\phi_2^{(k_n)},
\psi^{(n)}|{p}_{X_1X_2}^{k_n})
\leq \errP,
\label{eqn:PrfPrTrOneB}\\
& I(C_1^{(k_n)}C_2^{(k_n)}; \rvcxone \rvcxtwo) 
\leq \secP.
\label{eqn:PrfPrTrTwoB}
\end{align}
Then, using 
Lemmas \ref{lm:EntLow},
        \ref{lm:EntUp}, and 
        \ref{lm:SecReMI},
we can derive the following three bounds   
from (\ref{eqn:PrfPrTrTwoB}): 
\begin{align}
\left.
\ba{rl}
H(X_i|X_{3-i})\leq & H(K_i)+\gamma
\\
 & \:+\ds\frac{1}{k_n}\left[
   \frac{{\secP}}{Q_{12}}+\log\frac{1}{Q_{12}}
   \right],\: i=1,2
\vspace*{1.5mm}\\
H(X_1X_2)\leq & H(K_1K_2)+\gamma 
\\
& \:+\ds\frac{1}{k_n}\left[
  \frac{{\secP}}{Q_{12}}+\log\frac{1}{Q_{12}}
  \right].
\ea 
\right\}
\label{eqn:ProOneBdB}
\end{align}
Here we note that by (\ref{eqn:PrfPrTrOneB}), 
we have
\begin{align}
Q_{12}
&=p_{\lrvcxone\lrvcxtwo}
 \left(\wt{{\DcSet}}_{\gamma}^{(k_n)} \right)
=p_{X_1X_2}^{k_n}
\left(\wt{{\DcSet}}_{\gamma}^{(k_n)} \right)
\notag\\
&\geq 1-{\nu}_{k_n}(\gamma,{\errP}).
\label{eqn:LbQonetwoB}
\end{align}
From (\ref{eqn:ProOneBdB}) and 
     (\ref{eqn:LbQonetwoB}), 
we have the bound (\ref{eqn:BdProThrTwo}) 
in Proposition \ref{pro:KeyProOne}.
\end{IEEEproof}
\begin{IEEEproof}[Proof of 
 Proposition \ref{pro:KeyProTwoA}]
We assume that  
$(R_1,R_2)\in$ 
${\cal S}^{\ast}({\errP},{\secP}
          |p_{X_1X_2},p_{K_1K_2}).$
Then there exists a sequence 
 $\{(\Phi_1^{(n)},$ $\Phi_2^{(n)},$ $\Psi^{(n)})\}_{n \geq 1}$
 such that $\forall \gamma >0$,
 $\exists n_0=n_0(\gamma) \in \mathbb{N}$, 
 $\forall n\geq n_0$, we have 
\begin{align}
      &\frac{1}{n} \log |{\cal C}_i^{(n)}| 
        \leq  
         R_i+ \gamma,\: i=1,2,
\label{eqn:Rate} \\     
      & p_{X_1X_2}^{n}
  ({\DcSet}^{(n)})=
      p_{{\rm e}}(
        \phi_1^{(n)},
        \phi_2^{(n)},
        \psi^{(n)}|{p}_{X_1X_2}^n)
     \leq {\errP},
\label{eqn:Reliable} \\
& I(C_1^{(n)}C_2^{(n)}; \rvcxone \rvcxtwo) 
\leq {\secP}.
\label{eqn:Sec}
\end{align}
Fix $\gamma>0$ arbitrary. By Lemma \ref{lm:EntLow}, 
we have that 
\begin{align}
& H\left(\wt{C}_{1}^{(n)}\wt{C}_{2}^{(n)}\right)
 \geq n[H(X_1X_{2})-\gamma] + \log Q_{12}.
\label{eqn:EntCLb}
\end{align}
From (\ref{eqn:EntCLb}) and 
the assumption $R_1+R_2=H(X_1X_2)$, we have 
\begin{align}
& H\left(\wt{C}_{1}^{(n)}\wt{C}_{2}^{(n)}\right)
 \geq n[R_1+R_2-\gamma]+\log Q_{12}.
 \label{eqn:EntLba}
\end{align}
On the other hand, we have 
\begin{align}
&H\left(\wt{C}_{i}^{(n)}\right)
  \MLeq{a} \log |{\cal C}_i^{(n)}|
  \MLeq{b} n[R_i+\gamma], i=1,2.
\label{eqn:EntUba}  
\end{align}  
Step (a) follows from 
$\wt{C}_{i}^{(n)}\in {\cal C}_i^{(n)},i=1,2$.
Step (b) follows from (\ref{eqn:Rate}). 
From (\ref{eqn:EntLba}) and  (\ref{eqn:EntUba}),
we have 
\begin{align}
  H\left(\wt{C}_{i}^{(n)}\right)
  \geq n[R_i-2\gamma]+\log Q_{12},\:i=1,2. 
 \label{eqn:EntLbb} 
\end{align}
Then for each $i=1,2$, we have 
the following chain of inequalities:
\begin{align}
&Q_{12}^{-1}{\secP} \MGeq{a} 
 I\left(\rvcxone\rvcxtwo;C_{1}^{(n)}C_{2}^{(n)}
     \left|(\rvcxone,\rvcxtwo)
      \in\widetilde{{\DcSet}}_{\gamma}^{(n)}
    \right.\right)
\notag\\
& \geq I\left(\rvcxi;C_{i}^{(n)} 
     \left|(\rvcxone,\rvcxtwo)
      \in\widetilde{{\DcSet}}_{\gamma}^{(n)}
    \right.\right)
\notag\\
&=H\left(\wt{C}_{i}^{(n)}\right)
   -H\left(C_{i}^{(n)}\left|\rvcxi, 
 (\rvcxone,\rvcxtwo)
  \in\wt{{\DcSet}}_{\gamma}^{(n)}\right.\right)
\notag\\  
&\MGeq{b}
n[R_i-2\gamma]+\log Q_{12}-H(K_i).
\label{eqn:KeyAndRate}
\end{align}
Step (a) follows from Lemma  \ref{lm:SecReMI}.
Step (b) follows from (\ref{eqn:EntLbb}) and
Lemma \ref{lm:EntUp}. 
From (\ref{eqn:KeyAndRate}), we have 
\begin{align}
R_i\leq & H(K_i)+2\gamma
+\ds\frac{1}{n}\left[
   \frac{{\secP}}{Q_{12}}+\log\frac{1}{Q_{12}}
   \right],\: i=1,2.
   \label{eqn:ProTwoBd}
\end{align}
Here we note that by (\ref{eqn:Reliable}), we have 
\begin{align}
Q_{12}
&=p_{\lrvcxone\lrvcxtwo}
 \left(\wt{{\DcSet}}_{\gamma}^{(n)} \right)
=p_{X_1X_2}^n
\left(\wt{{\DcSet}}_{\gamma}^{(n)} \right)
\notag\\
&\geq 1-{\nu}_{n}(\gamma,{\errP}).
\label{eqn:LbQonetwoRe}
\end{align}
From (\ref{eqn:ProTwoBd}) and 
     (\ref{eqn:LbQonetwoRe}), 
we have the bound (\ref{eqn:BdProFourOne})
in Proposition \ref{pro:KeyProTwoA}. 

We next assume that  
$
(R_1,R_2)
\in \overline{\cal S}^{\ast}({\errP},{\secP}
              |p_{X_1X_2},p_{K_1K_2}).
$
Then  
 $\exists\{(\Phi_1^{(n)},\Phi_2^{(n)},$ $\Psi^{(n)})\}_{n \geq 1}$ and $\exists\{k_n\}_{n\geq 1}$, 
 such that $\forall \gamma >0$,
 $\exists n_1=n_1(\gamma) \in \mathbb{N}$, 
 $\forall n\geq n_1$, we have 
\begin{align}
      &\frac{1}{k_n} \log |{\cal C}_i^{(k_n)}| 
        \leq  
         R_i+ \gamma,\: i=1,2,
\label{eqn:RateB} \\     
 & p_{X_1X_2}^{k_n}
  ({\DcSet}^{(k_n)})
 \notag\\
& =p_{{\rm e}}(
        \phi_1^{(k_n)},
        \phi_2^{(k_n)},
        \psi^{(k_n)}|{p}_{X_1X_2}^{k_n})
     \leq {\errP},
\label{eqn:ReliableB} \\
& I(C_1^{(k_n)}C_2^{(k_n)}; \rvcxone \rvcxtwo) 
\leq {\secP}.
\label{eqn:SecB}
\end{align}
Then, using 
Lemmas \ref{lm:EntLow},
        \ref{lm:EntUp}, and 
        \ref{lm:SecReMI}, we can 
derive the following two bounds from 
(\ref{eqn:ReliableB}) and (\ref{eqn:SecB}):   
\begin{align}
R_i\leq & H(K_i)+2\gamma
+\ds\frac{1}{k_n}\left[
   \frac{{\secP}}{Q_{12}}+\log\frac{1}{Q_{12}}
   \right],\: i=1,2.
   \label{eqn:ProTwoBdB}
\end{align}
Here we note that by (\ref{eqn:ReliableB}), 
we have 
\begin{align}
Q_{12}
&=p_{\lrvcxone\lrvcxtwo}
 \left(\wt{{\DcSet}}_{\gamma}^{(k_n)} \right)
=p_{X_1X_2}^{k_n}
\left(\wt{{\DcSet}}_{\gamma}^{(k_n)} \right)
\notag\\
&\geq 1-{\nu}_{k_n}(\gamma,{\errP}).
\label{eqn:LbQonetwoReB}
\end{align}
From (\ref{eqn:ProTwoBdB}) and 
     (\ref{eqn:LbQonetwoReB}), 
we have the bound (\ref{eqn:BdProFourTwo})
in Proposition \ref{pro:KeyProTwoA}. 
\end{IEEEproof}

\section{Proof of Propositions \ref{pro:BaseProb} 
and \ref{pro:BasePro}}
\label{sec:PrBasePro}

\begin{figure}[t]
\centering
\includegraphics[width=0.47\textwidth]{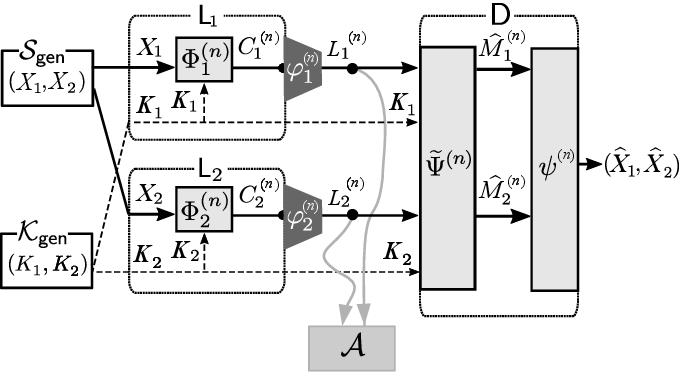}
\caption{
 The coding scheme  
 $(\varphi_1^{(n)},\varphi_2^{(n)},\wt{\psi}^{(n)})$
 internally connected with 
 $(\Phi_1^{(n)},\Phi_2^{(n)},{\Psi}^{(n)})$.}
\label{fig:NewEncScheme}
\vspace*{-5mm}
\end{figure}

In this section, we prove 
Propositions \ref{pro:BaseProb} and \ref{pro:BasePro}.
To prove those propositions we provide a new 
distributed coding scheme, which is internally 
connected with  
$(\Phi_1^{(n)},\Phi_2^{(n)},{\Psi}^{(n)})$.
 
\subsection{New Distributed Coding Scheme}
We first assume that $\{(\Phi_{1}^{(n)},
    \Phi_{2}^{(n)},\Psi^{(n)})\}_{n\geq 1}$ satisfies
belongs to the Class II.
\newcommand{\OmiTTzz}{
Fix a pair $({\errP},{\secP}) 
        \in  (0,1)\times [0,{\secP}_0]$, arbitrary.  
We start from the assumption that 
$(R_1,R_2)\in {\cal R}^{\ast}({\errP},{\secP}|
 p_{X_1X_2},p_{K_1K_2})$. 
Under this assumption we have a sequence 
  $\{(\Phi_1^{(n)},\Phi_2^{(n)},$ $\Psi^{(n)})\}_{n \geq 1}$
  such that $\forall \gamma >0$,
  $\exists n_0=n_0(\gamma) \in \mathbb{N}$, 
  $\forall n\geq n_0$, we have 
\begin{align*}
   &\frac{1}{n} \log |{\cal C}_i^{(n)}| 
          \leq  R_i+ \gamma,\: i=1,2,
\\  				
  & p_{{\rm e}}(
  \phi_1^{(n)},
  \phi_2^{(n)},
  \psi^{(n)}|{p}_{X_1X_2}^n) \leq \errP,
\\
   & I(C_1^{(n)}C_2^{(n)};\rvcxone\rvcxtwo)\leq \secP.
\end{align*}
}
We define a {\it new distributed encoding and joint 
decoding scheme} based on the above sequence 
$\{(\Phi_1^{(n)},\Phi_2^{(n)}$, 
$\Psi^{(n)})\}_{n \geq 1}$ attaining 
the reliable and secure rate pair $(R_1,R_2)$.  

Since 
$\{(\Phi_{1}^{(n)},
    \Phi_{2}^{(n)},\Psi^{(n)})\}_{n\geq 1}$ satisfies
the Condition \ref{cond:CondTwo}, we have     
${\cal M}_i^{(n)}={\cal C}_i^{(n)}, i=1,2$.
Furthermore, we have that for each distributed source encryption system 
$(\Phi_{1}^{(n)},\Phi_{2}^{(n)},\Psi^{(n)})$,  
there exists a distributed source coding system 
$(\phi_1^{(n)},$ $\phi_2^{(n)}$, 
$\psi^{(n)})$ such that  
for any $(\vckone, \vcktwo) \in \mathcal{X}_1^{n} 
\times \mathcal{X}_2^{n}$ and  
for some $\wt{\Phi}_{i,\lvckone}^{(n)}\in  
\mathscr{S}({\cal C}_i^{(n)}),i=1,2$, 
\begin{align}
& \Phi_{i,\lvcki}^{(n)}
  =\wt{\Phi}_{i,\lvckone}^{(n)}\circ \phi_i^{(n)},\:i=1,2, 
\label{eqn:CondOne}\\
& \Psi_{\lvckone,\lvcktwo}^{(n)}(c_1,c_2)=\psi^{(n)} 
 ((\wt{\Phi}_{1,\lvckone}^{(n)})^{-1}(c_1),
  (\wt{\Phi}_{2,\lvckone}^{(n)})^{-1}(c_2)),\: 
\notag\\
& \forall (c_1,c_2)\in  
  {\cal C}_1^{(n)}\times {\cal C}_2^{(n)}.
 \label{eqn:CondTwo} 
\end{align}
The data transmission scheme based on 
such $\{(\Phi_1^{(n)},\Phi_2^{(n)}$, 
$\Psi^{(n)})\}_{n \geq 1}$ 
is shown in Fig. \ref{fig:NewEncScheme}. 
For each $i=1,2$, we define 
$$  
\varphi_i^{(n)}:{\cal C}_i^{(n)}
\to {\cal L}_i^{(n)},\: 
r_i^{(n)} \defeq \frac{1}{n} \log |{\cal L}_i^{(n)}|,i=1,2.
$$
We further define 
$$
\wt{\Psi}^{(n)}:
       {\cal X}_1^{n} 
\times {\cal X}_2^{n}
\times {\cal L}_1^{(n)} 
\times {\cal L}_2^{(n)} 
   \to {\cal M}_1^{(n)} 
\times {\cal M}_2^{(n)}.
$$
Set 
\begin{align*}
L_i^{(n)}\defeq& \varphi_1^{(n)}\left(C_i^{(n)}
\right),\:i=1,2,
\\
 (\wh{M}_1^{(n)},\wh{M}_2^{(n)})
\defeq &\wt{\Psi}_{\lrvckone,\lrvcktwo}^{(n)}
\left(\varphi_1^{(n)}\left(C_1^{(n)}\right), 
\varphi_2^{(n)}\left(C_2^{(n)}\right)\right)    
\\
=&\wt{\Psi}^{(n)}_{\lrvckone,\lrvcktwo}
\left(L_1^{(n)},L_2^{(n)}\right).    
\end{align*}
Based on (\ref{eqn:CondOne}) 
     and (\ref{eqn:CondTwo}),              
we concretely construct 
$\varphi_i^{(n)},i=1,2$ and $\wt{\Psi}^{(n)}$. 
In the following argument on constructions 
of $\varphi_i^{(n)},i=1,2$ and  
 $\wt{\Psi}^{(n)}$, 
we fix $(\vckone, \vcktwo) 
\in    \mathcal{X}_1^{n} 
\times \mathcal{X}_2^{n}$ arbitrary.  

\noindent
\underline{\it Construction of $\varphi_i^{(n)},i=1,2$:} \ 
Let ${\cal J}_i^{(n)},i=1,2$ be finite sets. 
For each $i=1,2$, define $\wt{\phi}_i^{(n)}$ by 
$\wt{\phi}_i^{(n)}:{\cal M}_i^{(n)}\to {\cal J}_i^{(n)}$.
For each $i=1,2$, define $\varphi_i^{(n)}$ based on 
(\ref{eqn:CondTwo}) and $\wt{\phi}_i^{(n)}$. 
For each $i=1,2$, we introduce 
equivalence relation $\sim$ between
$c_i, c_i^{\prime} \in {\cal C}_i^{(n)}$ 
in the following  way:
\newcommand{\Mkdef}{
\stackrel{{\rm def}}{\Longleftrightarrow}}
\begin{align}
 c_i \sim c_i^{\prime}\Mkdef 
 \wt{\phi}_i\circ(\wt{\Phi}_{i,\lvcki}^{(n)})^{-1}(c_i)
=\wt{\phi}_i\circ(\wt{\Phi}_{i,\lvcki}^{(n)})^{-1}
(c_i^{\prime}).
\end{align}
We set ${\cal L}_i^{(n)}={\cal C}_i^{(n)}/\sim$ for $i=1,2,$.
Define 
$\varphi_i^{(n)},i=1,2$ by the following quotient map: 
\begin{align}
\varphi_i^{(n)}:{\cal C}_i^{(n)}
            \to {\cal C}_i^{(n)}/\sim.   
\end{align}
For each $i=1,2$, and $l_i\in{\cal L}_i^{(n)}$, define 
$$
(\varphi_i^{(n)})^{-1}(l_i)
\defeq \{ a_i:\varphi_i^{(n)}(a_i)=l_i\}.
$$
For each $i=1,2$, we fix any 
$l_i\in{\cal L}_i^{(n)}$. 
By the definition of $\varphi_i^{(n)},i=1,2,$ we 
have that for any 
$a_i \in (\varphi_i^{(n)})^{-1}(l_i)$,
the value of 
$\wt{\phi}_i^{(n)}\circ(\wt{\Phi}_{i,\lvcki}^{(n)})^{-1}(a_i)
\in {\cal J}_i^{(n)}$ does not depend on the choice of 
$a_i\in (\varphi_i^{(n)})^{-1}(l_i) $. 
We denote this quantity by $\wt{\Psi}_{i,\lvcki}^{(n)}(l_i)$. 
By definition we have that for each $i=1,2$, 
the map $\wt{\Psi}_{i,\lvcki}^{(n)}:
{\cal L}_i^{(n)} \to {\cal J}_i^{(n)}$ is a bijection.

\noindent
\underline{\it Construction of 
$\wt{\Psi}^{(n)}$:} \ 
Let 
$$
\wt{\psi}^{(n)} :
    {\cal J}_1^{(n)}\times{\cal J}_2^{(n)}
\to {\cal M}_1^{(n)}\times{\cal M}_2^{(n)}. 
$$
Using $\wt{\psi}^{(n)}$, we define 
$\wt{\Psi}^{(n)}$ by 
\begin{align*}
 \wt{\Psi}_{\lvckone,\lvcktwo}^{(n)}(l_1,l_2)=&
\wt{\psi}^{(n)}(
\wt{\Psi}_{1,\lvckone}^{(n)}(l_1),
\wt{\Psi}_{2,\lvcktwo}^{(n)}(l_2)),
\\
& \mbox{ for }(l_1,l_2)
\in {\cal L}_1^{(n)} \times {\cal L}_2^{(n)}. 
\end{align*}

\noindent
\underline{\it Computation 
of the  Reliability
:} \
For each $i=1,2$, set  
$$
J_i^{(n)}\defeq \wt{\phi}_i^{(n)}(M_i^{(n)})=
\wt{\phi}_i^{(n)}\circ{\phi}_i^{(n)}(\rvcx_i).
$$   
By (\ref{eqn:CondOne}), we have  
$$
C_i^{(n)}=\wt{\Phi}_{i,\lrvcki}
(M_i^{(n)})=\wt{\Phi}_{i,\lrvcki}\circ{\phi}_i^{(n)}(\rvcxi),\:
i=1,2.
$$
Furthermore, by the constructions of $\varphi_i^{(n)}$ 
and $\wt{\Psi}_{i,\lrvcki},$ $i=1,2$, we have that for 
$i=1,2$, 
\begin{align}
\wt{\Psi}_{i,\lrvcki}^{(n)}(L_i^{(n)})=J_i^{(n)}=\wt{\phi}_i^{(n)}(M_i^{(n)}).
\label{eqn:tildePsii}
\end{align}
Computing  $(\wh{M}_1^{(n)},\wh{M}_2^{(n)})$, 
we have the following chain of equalities:
\begin{align*}
&(\wh{M}_1^{(n)},\wh{M}_2^{(n)})
=\wt{\Psi}^{(n)}_{\lrvckone,\lrvcktwo}
\left(L_1^{(n)},L_2^{(n)}\right)
\\
&=\wt{\psi}^{(n)} \circ \wt{\Psi}^{(n)}_{\lrvckone,\lrvcktwo}
\left(L_1^{(n)},L_2^{(n)}\right)
\\
&\MEq{a}\wt{\psi}^{(n)}(
\wt{\Psi}_{1,\lrvckone}^{(n)}(L_1^{(n)}),
\wt{\Psi}_{2,\lrvcktwo}^{(n)}(L_2^{(n)}))
\\
&\MEq{b}\wt{\psi}^{(n)}(
\wt{\phi}_{1}^{(n)}(M_1^{(n)}),
\wt{\phi}_{2}^{(n)}(M_2^{(n)}))
\\
&=\wt{\psi}^{(n)}(
\wt{\phi}_{1}^{(n)}\circ{\phi}_{1}^{(n)}(\rvcxone),
\wt{\phi}_{2}^{(n)}\circ{\phi}_{2}^{(n)}(\rvcxtwo)).
\end{align*}
Step (a) follows from the definition of $\wt{\Psi}^{(n)}_{\lrvckone,\lrvcktwo}$.
Step (b) follows from (\ref{eqn:tildePsii}).
Computing the error probability of decoding, 
we have   
\begin{align*}
& \Pr\left\{\psi^{(n)}
(\wh{M}_1^{(n)},\wh{M}_2^{(n)})
\neq (\rvcxone,\rvcxtwo)\right\}
\\
&\MEq{a}\Pr\left\{ 
\psi^{(n)}\circ\wt{\psi}^{(n)}
(\wt{\phi}_1^{(n)}\circ\phi_1^{(n)}(\rvcxone),
 \wt{\phi}_2^{(n)}\circ\phi_2^{(n)}(\rvcxtwo)) 
\right.
\\
&\qquad \neq (\rvcxone,\rvcxtwo)
\Bigr\}.
\end{align*}
Since the above quantity depends only on 
$(\wt{\phi}_1^{(n)}\circ\phi_1^{(n)},
  \wt{\phi}_2^{(n)}\circ\phi_2^{(n)}, 
\psi^{(n)}\circ\wt{\psi}^{(n)})$,
we write the error probability 
$\wt{p}_{{\rm e}}^{(n)}$ of decoding as 
$$
 \wt{p}_{\rm e}^{(n)}=
 \wt{p}_{\rm e}^{(n)}\Bigl(
                \wt{\phi}_1^{(n)}\circ\phi_1^{(n)},
                \wt{\phi}_2^{(n)}\circ\phi_2^{(n)}, 
\psi^{(n)}\circ\wt{\psi}^{(n)}
   \Big|p_{X_1X_2}\Bigr).
$$
\underline{\it Computation 
of the Security
:} \ On the security we have
\begin{align}
 & I(L_1^{(n)}L_2^{(n)};\rvcxone\rvcxtwo)
\notag\\    
&= I(\varphi_1^{(n)}(C_1^{(n)})
     \varphi_2^{(n)}(C_2^{(n)});\rvcxone\rvcxtwo)  
\notag\\          
& \MLeq{a} I(C_1^{(n)}C_2^{(n)};\rvcxone\rvcxtwo)\leq \secP.
\label{eqn:SecBdFinO}
\end{align}
Step (a) follows from the data processing inequality.

From the above argument we can see that it suffices to 
evaluate upper bounds of $\wt{p}_{\rm e}^{(n)}$ to prove Propositions \ref{pro:BaseProb} and \ref{pro:BasePro}.
In the remaining part of this subsection, 
we present a result on an upper bound of 
$\wt{p}_{\rm e}^{(n)}$. To describe this lemma we prepare 
several quantities. For each $i=1,2$, let $\wt{M}_i^{(n)}$ 
$=\phi_{i}^{(n)}\left(\wt{\rvcx}_i\right)$ 
be a random variable induced by 
$\phi_{i}^{(n)},\wt{\rvcx}_i $.
Set 
\begin{align*}
 {\cal M}^{(n)}_{
 (\phi_{1}^{(n)},\phi_{2}^{(n)})
 (\wt{\cal D}_{\gamma}^{(n)})}
 \defeq & \{(m_1,m_2):m_i=\phi^{(n)}_{i}(\vcxi),i=1,2
 \\
 & \mbox{ for some }
  (\vcxone,\vcxtwo) \in 
 \wt{\cal D}_{\gamma}^{(n)}
  \}.
\end{align*}
For each
$
 (m_{1},m_{2}) \in {\cal M}^{(n)}_{
 (\phi_{1}^{(n)},\phi_{2}^{(n)})
 (\wt{\cal D}_{\gamma}^{(n)})},
$
we have
\begin{align*}
& p_{\wt{M}_1^{(n)}\wt{M}_2^{(n)}}(m_{1},m_{2})
 =\Pr\Bigl\{(\wt{M}_{1}^{(n)},\wt{M}_{2}^{(n)})
  =(m_{1},m_{2})\Bigr\}
\\   
& =\Pr\Bigl\{(M_{1}^{(n)},M_{2}^{(n)}
   )=(m_{1},m_{2})
   \left|\left(\rvcxone,\rvcxtwo\right) \in  
   \wt{\DcSet}_{\gamma}^{(n)}
   \right.\Bigr\}.
\end{align*}
For $i=1,2$, set
\begin{align*}
&{\cal U}_{i,\gamma}^{(n)}
\defeq
\biggl\{(m_1,m_2) \in 
 {\cal M}^{(n)}_{
 (\phi_{1}^{(n)},\phi_{2}^{(n)})
 (\wt{\cal D}_{\gamma}^{(n)})}:
\\
 &\:\left.\frac{1}{n}
 \log\frac{1}
  {p_{\wt{M}_i^{(n)}|\wt{M}_{3-i}^{(n)}}
  (m_i|m_{3-i})}
\geq {r}_i^{(n)}-\gamma\right\}. 
\end{align*}
Furthermore, set 
\begin{align*}
&{\cal U}_{3,\gamma}^{(n)} 
\defeq
\biggl\{(m_1,m_2)
\in
 {\cal M}^{(n)}_{
 (\phi_{1}^{(n)},\phi_{2}^{(n)})
 (\wt{\cal D}_{\gamma}^{(n)})}:
\\
&\:\left.
  \frac{1}{n}
     \log\frac{1}{p_{\wt{M}_1^{(n)}
      \wt{M}_2^{(n)}}(m_1,m_2)}
      \geq {r}_1^{(n)}+{r}_2^{(n)}-\gamma
   \right\}. 
\end{align*}
Set 
\begin{align}
&\Theta_n\left(\gamma,r_1^{(n)},r_2^{(n)}\right)
\notag\\
&= \Theta_n\left(\gamma,r_1^{(n)},r_2^{(n)}
\Bigl|
\phi_{1}^{(n)},\phi_{2}^{(n)},
\psi^{(n)}, 
p_{X_1X_2}\right)
\notag\\
&\defeq \Pr\Biggl\{ (\wt{M}_1^{(n)},\wt{M}_2^{(n)})\in
\bigcup_{i=1}^3
{\cal U}^{(n)}_{i,\gamma}\Biggr\}.
\label{eqn:ErBdTwozero}
\end{align}

Then we have the following lemma: 
\begin{lemma}\label{lm:LemRcBdB}
There exists at least one deterministic 
code $(\wt{\phi}_1^{(n)},
       \wt{\phi}_2^{(n)},$ $
       \wt{\psi}^{(n)})$ such that
\begin{align}
& 
 \wt{p}_{\rm e}^{(n)}\Bigl(
                \wt{\phi}_1^{(n)}\circ\phi_1^{(n)},
                \wt{\phi}_2^{(n)}\circ\phi_2^{(n)}, 
\psi^{(n)}\circ\wt{\psi}^{(n)}
   \Big|p_{X_1X_2}\Bigr)
\notag\\
&\leq 3\cdot{\rm 2}^{-n\gamma}+\nu_n(\gamma,\errP)
+\Theta_n\left(\gamma,r_1^{(n)},r_2^{(n)}\right).
\label{eqn:eRbdB}
\end{align}
\end{lemma}

Proof of Lemma \ref{lm:LemRcBdB} is given 
in Appendix \ref{apd:PrfLemRcBd}. In the 
next subsection we define several quantities
to evaluate upper bounds of $\Theta_n\left(\gamma,r_1^{(n)},r_2^{(n)}\right)$.
We further present some results describing 
their properties. Those results are useful 
for deriving upper bounds of 
this quantity. 
\newcommand{\PrfLemRcBd}{
\subsection{Proof of Lemma \ref{lm:LemRcBdB}}
\label{apd:PrfLemRcBd}

In this appendix we prove Lemma \ref{lm:LemRcBdB}.
For each $i=1,2$, define
\begin{align*}
&{\cal T}_{i,\gamma}^{(n)}
\defeq
\biggl\{(m_1,m_2)
\in {\cal M}_1^{(n)}\times{\cal M}_2^{(n)} :
\\
 &\:\left.\frac{1}{n}
 \log\frac{1}
  {p_{M_i^{(n)}|M_{3-i}^{(n)}}(m_i|m_{3-i})}
\leq {r}_i^{(n)}-\gamma\right\}.
\end{align*}
We further define 
\begin{align*}
&{\cal T}_{3,\gamma}^{(n)} 
\defeq
\biggl\{ (m_1,m_2)
\in {\cal M}_1^{(n)}\times{\cal M}_2^{(n)} :
\\
&\:\left.
 \frac{1}{n}
   \log\frac{1}{p_{M_1^{(n)}M_2^{(n)}}(m_1,m_2)}
\leq {r}_1^{(n)}+{r}_2^{(n)}-\gamma
 \right\},
 \\
& {\mathcal{T}}_{\gamma}^{(n)} 
\defeq \bigcap_{i=1}^3
{\mathcal{T}}_{i,\gamma}^{(n)}. 
\end{align*}
Then, we have the following lemma.

\begin{lemma}\label{lm:LemRcBd}
There exists at least one deterministic 
code $(\wt{\phi}_1^{(n)},
       \wt{\phi}_2^{(n)},$ $
       \wt{\psi}^{(n)})$ such that
\begin{align}
& 
 \wt{p}_{\rm e}^{(n)}\Bigl(
                \wt{\phi}_1^{(n)}\circ\phi_1^{(n)},
                \wt{\phi}_2^{(n)}\circ\phi_2^{(n)}, 
\psi^{(n)}\circ\wt{\psi}^{(n)}
   \Big|p_{X_1X_2}\Bigr)
\notag\\
&\leq \Pr\biggl\{ 
(\rvcxone,\rvcxtwo) \notin {\cal D}^{(n)}
\mbox{ or }(M_1^{(n)},M_2^{(n)})\notin 
{\cal T}^{(n)}_{\gamma} 
\biggr\}
\notag\\
&\quad \qquad
+3\cdot{\rm 2}^{-n\gamma}.
\label{eqn:eRbd}
\end{align}
\end{lemma}

To prove Lemma \ref{lm:LemRcBd}, we give 
some definitions. We further 
present a lemma useful for deriving 
the error probability bound 
in Lemma \ref{lm:LemRcBd}. 
We first present several definitions. 

For each $i=1,2$, we set 
\begin{align*}
({\cal T}_{\gamma}^{(n)})_i
\defeq \{m_i:(m_1, m_2)
\in {\cal T}_{\gamma}^{(n)}
\mbox{ for some }{m}_{3-i}\}. 
\end{align*}
For each $i=1,2$ and each 
$m_i\in ({\cal T}_{\gamma}^{(n)})_i$, 
we set 
$$ 
{\cal T}_{i|3-i,\gamma}^{(n)}({c}_{3-i})
\defeq \{m_i: (m_1,m_2)
\in {\cal T}_{\gamma}^{(n)}\}.
$$
The following lemma is useful to derive the 
error provability bound in Lemma \ref{lm:LemRcBd}.

\begin{lemma}\label{lm:LemTgCarBd}
We have the following:
\begin{align}
&\left| {\cal T}_{i|3-i,\gamma}^{(n)}({m}_{3-i})\right|
 \leq |{\cal L}_i^{(n)}|\cdot 2^{-n\gamma}, i=1,2,
\label{eqn:CarBdOneTwo}\\
&\left| {\cal T}_{\gamma}^{(n)}\right|
 \leq |{\cal L}_1^{(n)}||{\cal L}_2^{(n)}|
 \cdot 2^{-n\gamma}. 
 \label{eqn:CarBdThr}
\end{align}  
\end{lemma}
\begin{IEEEproof}
We prove (\ref{eqn:CarBdOneTwo}).
For each $i=1,2$, we have the following chain of 
inequalities:
\begin{align}
 1&\geq \Pr\left\{M_i^{(n)}\in  
{\cal T}_{i|3-i,\gamma}^{(n)}({m}_{3-i})
\Bigl|M_{3-i}^{(n)}={m}_{3-i}\right\}
\notag\\
&
=\sum_{m_i\in {\cal T}_{\gamma}^{(n)}({m}_{3-i})}
 p_{M_i^{(n)}|M_{3-i}^{(n)}}(m_i|m_{3-i})
\notag\\
& \MGeq{a}  \frac{2^{n\gamma}}{|{\cal L}_i^{(n)}|}
\sum_{m_i\in {\cal T}_{i|3-i,\gamma}^{(n)}({m}_{3-i})}1
\notag\\           
&
= \frac{2^{n\gamma}}{|{\cal L}_i^{(n)}|} 
\left|{\cal T}_{i|3-i,\gamma}^{(n)}
({m}_{3-i})\right|.
\label{eqn:PrfCaBd}
\end{align}
Step (a) follows from that for 
$m_i\in {\cal T}_{i|3-i,\gamma
           }^{(n)}({m}_{3-i})$
$$
 p_{M_i^{(n)}|M_{3-i}^{(n)}}(m_i|m_{3-i})
 \geq \frac{2^{n\gamma}}{|{\cal L}_i^{(n)}|}.
$$
From (\ref{eqn:PrfCaBd}),  
we have the bound (\ref{eqn:CarBdOneTwo}) in 
Lemma \ref{lm:LemTgCarBd}. In a similar manner 
we can prove (\ref{eqn:CarBdThr}). 
\end{IEEEproof}

\begin{IEEEproof}[Proof of Lemma \ref{lm:LemRcBd}]
We prove this lemma by using information 
spectrum method.  

\noindent
\underline{\it Random Coding:} \ For each 
$m_1\in {\cal M}_1^{(n)}$, we generate 
$l_1 \in $\\ ${\cal L}_1^{(n)}$ randomly according to the 
uniform distribution over ${\cal L}_1^{(n)}$ and 
define $\wt{\phi}_1^{(n)}(m_1)=l_1$. 
Similarly, for each $m_2 \in {\cal M}_2^{(n)}$, 
we generate $l_2\in {\cal L}_2^{(n)}$ 
randomly according to the 
uniform distribution over ${\cal L}_2^{(n)}$ 
and define $\wt{\phi}_2^{(n)}(m_2)=l_2$. 

\noindent 
\underline{\it Decoding:} \ 
Suppose that
a decoder $\wt{\psi}^{(n)}$ receives a 
pair of the outputs 
$(l_1,l_2)\in {\cal L}^{(n)}$
from the two encoders
$\wt{\phi}_1^{(n)}$ and $\wt{\phi}_2^{(n)}$. 

The decoding process consists of the 
two steps shown below.   
\begin{itemize}
 \item [1)]We first define the decoder 
$\wt{\psi}^{(n)}:
    {\cal L}_1^{(n)}\times {\cal L}_2^{(n)}
\to {\cal M}_1^{(n)}\times {\cal M}_2^{(n)}$ 
in the following way. If there exists a {\it unique} 
$(\wh{m}_1,\wh{m}_2)$  satisfying 
$(\wh{m}_1,\wh{m}_2)
\in {\cal T}^{(n)}_{\gamma}$, 
we define the decoder by 
$\wt{\psi}^{(n)}(l_1,l_2)=(\wh{m}_1,\wh{m}_2)$
for such $(\wh{m}_1,\wh{m}_2)$. 
If there exists no such $(\wh{m}_1,\wh{m}_2)$ 
or exist more than one such $(\wh{m}_1,\wh{m}_2)$, 
we define
$\wt{\psi}^{(n)}(l_1,l_2)$ 
as an arbitrary specified element 
in ${\cal M}_1^{(n)}\times {\cal M}_2^{(n)}$.   
 \item [2)] For $(\wh{m}_1,\wh{m}_2)$, 
 we decode $(\wh{\vcx}_1,\wh{\vcx}_2)
 =\psi^{(n)}(\wh{m}_1,\wh{m}_2)$,
 using the decoder function $\psi^{(n)}$.   
\end{itemize}
 
\noindent
\underline{\it Evaluation of 
the Error probability:} \ We set 
\begin{align*}
& \wt{\varepsilon}_n \defeq \wt{p}_{\rm e}^{(n)}
\\
&= \wt{p}_{\rm e}^{(n)}(\wt{\phi}_1^{(n)}\circ \phi_1^{(n)} ,
                 \wt{\phi}_2^{(n)}\circ \phi_2^{(n)},
\psi^{(n)}\circ\wt{\psi}^{(n)}|
p_{X_1X_2}).
\notag\\
\end{align*}
On upper bound of $\wt{\varepsilon}_n$, we 
have the following chain of inequalities: 
\begin{align}
& \wt{\varepsilon}_n
=\Pr\left\{(\rvcxone,\rvcxtwo)
\notin {\cal D}^{(n)} \mbox{ or }
(M_1^{(n)},M_2^{(n)})\notin 
  {\cal T}^{(n)}_{\gamma}
\right.
\notag\\
& \qquad\qquad \left.\mbox{ or }
(\wh{M}_1^{(n)},\wh{M}_1^{(n)})\neq 
({M}_1^{(n)},{M}_2^{(n)})
\right\}
\notag\\
&\leq \Pr\left\{
(\rvcxone,\rvcxtwo)
\notin {\cal D}^{(n)} \mbox{ or }
(M_1^{(n)},M_2^{(n)})\notin 
  {\cal T}^{(n)}_{\gamma}
  \right\}
\notag  \\
&\quad+\Pr\left\{(M_1^{(n)},M_2^{(n)})\in 
  {\cal T}^{(n)}_{\gamma}
  \right.
\notag  \\
& \qquad \quad \mbox{ and } (\wh{M}_1^{(n)},\wh{M}_1^{(n)})\neq 
 ({M}_1^{(n)},{M}_2^{(n)})\Bigr\}.
\label{eqn:ErBdTwoa}
\end{align}
Note that the first term in the right member 
of (\ref{eqn:ErBdTwoa}) is constant under 
the random choice of 
$(\wt{\phi}_1^{(n)},\wt{\phi}_2^{(n)})$. Let 
$\Xi(\wt{\phi}_1^{(n)},\wt{\phi}_2^{(n)})$ denote 
the second term in the right members 
of (\ref{eqn:ErBdTwoa}).  
In the following argument we evaluate upper bounds 
of the term $\Xi(\varphi_1^{(n)},\varphi_2^{(n)})$.
For this evaluation we consider the following 
three failure events:   
\begin{align*}
{\cal E}_i \defeq& \Bigl\{
  \exists\wh{m}_i\neq M_i^{(n)}, 
   \wt{\phi}_i^{(n)}(\wh{m}_i)
  =\wt{\phi}_i^{(n)}( M_i^{(n)})
 \\
& \mbox{ and } \wh{m}_i
   \in {\cal T}^{(n)}_{i|3-i,\gamma} (M_{3-i}^{(n)})
   \Bigr\} \mbox{ for }i=1,2,  
 \\
{\cal E}_3\defeq &\Bigl\{
 \exists\wh{m}_i\neq M_i^{(n)}, 
  \wt{\phi}_i^{(n)}(\wh{m}_i)=\wt{\phi}_i^{(n)}(M_i^{(n)}), i=1,2,
 \\
& \mbox{ and } 
   (\wh{m}_1,\wh{m}_2)\in {\cal T}^{(n)}_{\gamma}\Bigr\}. 
\end{align*}
Then, we have the following:
\begin{align}
\Xi (\wt{\phi}_1^{(n)},\wt{\phi}_2^{(n)})
=\Pr\left\{\bigcup_{i=1}^3{\cal E}_i \right\}
\leq \sum_{i=1}^3\Pr\left\{{\cal E}_i\right\}.
\label{eqn:ErBdThrb}
\end{align}
We denote the probability measure and the expectation 
based on the randomness of the choice of 
$(\wt{\phi}_1^{(n)},\wt{\phi}_2^{(n)})$ by 
$\mathbb{P}(\cdot)$ and 
$\mathbb{E}[\cdot]$, respectively to distinguish 
them with those for other random variables. 
From (\ref{eqn:ErBdThrb}), we have 
\begin{align}
& \mathbb{E}\left[\Xi (\wt{\phi}_1^{(n)},\wt{\phi}_2^{(n)})\right]
\leq \sum_{i=1}^3\mathbb{E}\left[\Pr\left\{{\cal E}_i\right\}\right].
\label{eqn:ErBdFou}
\end{align}
For each $i=1,2,3$, an exact form of 
$\mathbb{E}\left[\Pr\left\{{\cal E}_i\right\}\right]$ 
is given by  
\begin{align}
&\mathbb{E}\left[\Pr\left\{{\cal E}_i\right\}\right]
 =\sum_{\scs (m_1,m_2)
          {\in{\cal T}^{(n)}_{\gamma}
          }
       }                  
p_{\wt{M}_1^{(n)}\wt{M}_2^{(n)}} 
    (m_1,m_2) \mathbb{P}_i.
 \label{eqn:ErBdFiv}  
\end{align}
Here for $i=1,2$, we have 
\begin{align*}
\mathbb{P}_i&=\mathbb{P}
   \left(\bigvee_{\scs 
         \wh{m}_i\neq m_i,
          \atop{ \scs \wh{m}_i
          \in  {\cal T}^{(n)}_{i|3-i,
          \gamma}(m_{3-i})}}
          \{\wt{\phi}_i^{(n)}(\wh{m}_i)
          =\wt{\phi}_i^{(n)} (m_i) \}\right).
\end{align*}
For $i=3$, we have 
\begin{align*}
\mathbb{P}_3&=\mathbb{P} 
   \left(\bigvee_{\scs 
         \wh{m}_1\neq m_1, \wh{m}_2\neq m_2,
          \atop{ \scs (\wh{c}_1,\wh{c}_2)
          \in  {\cal T}^{(n)}_{\gamma}}}
    \bigwedge_{i=1,2}\{\wt{\phi}_i^{(n)}(\wh{m}_i)
       =\wt{\phi}_i^{(n)} (m_i) \}\right).
\end{align*}
On upper bounds of $\mathbb{P}_i$, $i=1,2$, we have the following:
\begin{align}
\mathbb{P}_i\leq & 
\sum_{\scs \wh{m}_i\neq m_i,
  \atop{\scs \wh{m}_i
        \in  {\cal T}^{(n)}_{i|3-i,\gamma}(m_{3-i})}}
  \mathbb{P}\left(\wt{\phi}_i^{(n)}(\wh{m}_i)
      =\wt{\phi}_i^{(n)}(m_i)\right)
\notag\\
=& \frac{
\left|{\cal T}^{(n)}_{i|3-i,\gamma}(m_{3-i})\right|
}{\left|{\cal L}_i^{(n)}\right|}\MLeq{a}2^{-n\gamma}.
\label{eqn:ErBdSix} 
\end{align}
Step (a) follows from (\ref{eqn:CarBdOneTwo}) 
in Lemma \ref{lm:LemTgCarBd}.
On upper bounds of $\mathbb{P}_3$, we have the following:
\begin{align}
\mathbb{P}_3\leq & 
\sum_{\scs 
\wh{m}_1\neq m_1, \wh{m}_2\neq m_2,
          \atop{ \scs (\wh{m}_1,\wh{m}_2)
          \in  {\cal T}^{(n)}_{\gamma}}
          }
 \prod_{i=1,2}\mathbb{P}\left(\wt{\phi}_i^{(n)}(\wh{c}_i)
      =\wt{\phi}_i^{(n)}(c_i)\right)          
\notag\\
\leq& \frac{
\left|{\cal T}^{(n)}_{\gamma}\right|
}{\left|{\cal L}_1^{(n)}\right|
  \left|{\cal L}_2^{(n)}\right|}\MLeq{a}2^{-n\gamma}.
 \label{eqn:ErBdSev} 
\end{align}
Step (a) follows from (\ref{eqn:CarBdThr}) 
in Lemma \ref{lm:LemTgCarBd}.
Combining (\ref{eqn:ErBdFou}), (\ref{eqn:ErBdFiv}), (\ref{eqn:ErBdSix}), and (\ref{eqn:ErBdSev}) together, 
we obtain 
$\mathbb{E}\left[\Xi (\wt{\phi}_1^{(n)},\wt{\phi}_2^{(n)})\right]\
\leq$ $ 3\cdot 2^{-n\gamma}.$ Hence there exists at least one 
pair $(\wt{\phi}_1^{(n)},\wt{\phi}_2^{(n)})$ 
of deterministic functions such that we have    
$
\Xi (\wt{\phi}_1^{(n)},\wt{\phi}_2^{(n)})
\leq$ $ 3\cdot2^{-n\gamma}, 
$
which together with (\ref{eqn:ErBdTwoa}) yields 
the bound (\ref{eqn:eRbd}) in Lemma \ref{lm:LemRcBd}. 
\end{IEEEproof}

\begin{IEEEproof}[Proof of Lemma \ref{lm:LemRcBdB}]
On the first term in the right members of (\ref{eqn:eRbd})
in Lemma \ref{lm:LemRcBd}, we have the following chain of inequalities:
\begin{align}
& \Pr\biggl\{(M_1^{(n)},M_2^{(n)})\notin 
{\cal T}^{(n)}_{\gamma} 
\mbox{ or }(\rvcxone,\rvcxtwo)
\notin{\cal D}^{(n)} \biggr\}
\notag\\
&\MLeq{a} \Pr\biggl\{(M_1^{(n)},M_2^{(n)})\notin 
{\cal T}^{(n)}_{\gamma} 
\mbox{ or }(\rvcxone,\rvcxtwo)
\notin\wt{\DcSet}_{\gamma}^{(n)} \biggr\}
\notag\\
&=\Pr\left\{ (\rvcxone,\rvcxtwo)
\notin \wt{\DcSet}_{\gamma}^{(n)}
\right\}
+\Pr\Biggl\{(M_1^{(n)},M_2^{(n)})
\notag\\
&\qquad \quad \notin 
\bigcap_{i=1}^3 
{\cal T}^{(n)}_{i,\gamma}
\mbox{ and }(\rvcxone,\rvcxtwo)
\in \wt{\DcSet}_{\gamma}^{(n)}
\Biggr\}
\notag\\
& \leq \nu_n(\gamma,\errP)
+ \Pr\Biggl\{ (M_1^{(n)},M_2^{(n)})\notin
\bigcap_{i=1}^3{\cal T}^{(n)}_{i,\gamma}
 \Biggl|
\notag\\
&\qquad \qquad \qquad\qquad
(\rvcxone,\rvcxtwo)
\in  \wt{\DcSet}_{\gamma}^{(n)}\Biggr\} 
\notag\\
& \MEq{b} \nu_n(\gamma,\errP) 
+\Pr\Biggl\{ (\wt{M}_1^{(n)},\wt{M}_2^{(n)})\in
\bigcup_{i=1}^3
{\cal U}^{(n)}_{i,\gamma}\Biggr\}.
\label{eqn:ErBdTwo}
\end{align}
Step (a) follows from 
         $\wt{\DcSet}_{\gamma}^{(n)}
   \subseteq {\DcSet}^{(n)}$. 
Step (b) follows from the definition of  
$(\wt{M}_1^{(n)},\wt{M}_2^{(n)})$.  
From (\ref{eqn:ErBdTwo}) and Lemma \ref{lm:LemRcBd},
we have the bound (\ref{eqn:eRbdB}) in 
Lemma \ref{lm:LemRcBdB}.
\end{IEEEproof}

}

\subsection{Information Spectrum Method}

We define 
\begin{align*}
&\overline{H}(\wt{X}_1^{\infty}\wt{X}_2^{\infty})
 \defeq 
 \mbox{\rm p-}\limsup_{n\to\infty}
 \frac{1}{n}\log\frac{1}
{p_{\wt{\lrvcx}_1 \wt{\lrvcx}_2} 
   (\wt{\rvcx}_{1},\wt{\rvcx}_{2})},
\\
&\underline{H}(\wt{X}_1^{\infty}\wt{X}_2^{\infty})
 \defeq 
 \mbox{\rm p-}\liminf_{n\to\infty}
 \frac{1}{n}\log\frac{1}
{p_{\wt{\lrvcx}_1 \wt{\lrvcx}_2} 
   (\wt{\rvcx}_{1},\wt{\rvcx}_{2})},
\\
&\overline{H}
(\wt{M}_1^{(\infty)}\wt{M}_2^{(\infty)})
\\
& \defeq 
 \mbox{\rm p-}\limsup_{n\to\infty}
 \frac{1}{n}\log\frac{1}
{p_{\wt{M}_1^{(n)}\wt{M}_2^{(n)}}  
   (\wt{M}_1^{(n)},\wt{M}_2^{(n)})},
\\
&\underline{H}(\wt{M}_1^{(\infty)}\wt{M}_2^{(\infty)}
)
\\
& \defeq 
 \mbox{\rm p-}\liminf_{n\to\infty}
 \frac{1}{n}\log\frac{1}
{p_{\wt{M}_1^{(n)}\wt{M}_2^{(n)}
}(
\wt{M}_1^{(n)},
\wt{M}_2^{(n)})}.
\end{align*}
For $i=1,2$, we define 
$$
\overline{H}(\wt{X}_i^\infty), 
\underline{H}(\wt{X}_i^\infty),
\overline{H}(\wt{M}_i^{(\infty)}),
\underline{H}(\wt{M}_i^{(\infty)})
$$           
in a similar manner. Furthermore, we define
\begin{align*}
&\underline{I}
(\wt{M}_1^{(\infty)};\wt{M}_2^{(\infty)})
\\
& \defeq 
 \mbox{\rm p-}\liminf_{n\to\infty}
 \frac{1}{n}\log \frac
{p_{\wt{M}_1^{(n)}|\wt{M}_2^{(n)}} 
  (\wt{M}_1^{(n)}|\wt{M}_2^{(n)})}
{p_{\wt{M}_1^{(n)}}
 (\wt{M}_1^{(n)})}.
\end{align*}
In the following argument for simplicity of notation
we set
\begin{align*}
  \overline{H}_{12} &\defeq 
  \overline{H}(\wt{M}_1^{(\infty)}\wt{M}_2^{(\infty)}),\:
  \underline{H}_{12} \defeq 
  \underline{H}(\wt{M}_1^{(\infty)}\wt{M}_2^{(\infty)}),
\\
 \overline{H}_{i} &\defeq 
  \overline{H}(\wt{M}_i^{(\infty)}),\: 
  \underline{H}_i  \defeq \underline{H}(\wt{M}_i^{(\infty)}),\:i=1,2,
\\
 \underline{I}&\defeq\underline{I} 
   (\wt{M}_1^{(\infty)};\wt{M}_2^{(\infty)}).
\end{align*}
We present two results
on the above information spectrum quantities.
To describe the first result we define
\begin{align}
&  \eta_{n}(\gamma)
\notag\\
&
\defeq \Pr\left\{
\left|\frac{1}{n}
   \log\frac{1}{p_{\wt{\lrvcx}_1\wt{\lrvcx}_2}(
     \wt{\rvcx}_1 ,\wt{\rvcx}_2)}
-H(X_1X_2)\right|\geq {\ts \frac{3}{2}}\gamma \right\}.  
\notag
\end{align}
For $i=1,2$, define
\begin{align}
&\eta_{i,n}(\gamma)
\defeq \Pr\left\{
\left|\frac{1}{n}
   \log\frac{1}{p_{\wt{\lrvcx}_i}
    (\wt{\rvcx}_i)}
  -H(X_i)\right|\geq {3}\gamma 
 \right\}.
 \notag
\end{align}
Furthermore, for $i=1,2$, we define 
\begin{align*}
& \theta_{i,n}(\gamma)
 \defeq
\Pr\left\{
    \frac{1}{n}
    \log\frac{1}{p_{\wt{M}_i^{(n)}}(\wt{M}_i^{(n)})}
    \leq\underline{H}_i-{\ts \frac{1}{2}}\gamma 
\right\}.
\end{align*}
We have the following property 
on $\overline{H}_{12},
    \underline{H}_{12}$
   $\overline{H}_i,
  \underline{H}_i,i=1,2$. 
\begin{property}\label{pr:InfSpecLm}
\begin{itemize}
\item[a)]
For each 
$
(m_1,m_2) \in {\cal M}^{(n)}_{
 (\phi_1^{(n)},\phi_2^{(n)})
 (\wt{\cal D}_{\gamma}^{(n)})}, 
$
there exists a unique $(\vcxone,\vcxtwo)
\in \wt{\cal D}_{\gamma}^{(n)}$ such that 
$\vcxi=\phi_{i}^{(n)}$ $(m_i),i=1,2$. 
Furthermore, we have the following:
\begin{align}
& p_{\wt{M}_1^{(n)}\wt{M}_2^{(n)}}
     (m_{1},m_{2})  
=\Pr\Bigl\{(\wt{M}_{1}^{(n)},\wt{M}_{2}^{(n)})
 =(m_{1},m_{2})\Bigr\}
\notag\\
&= p_{\lrvctxone\lrvctxtwo}(\vcxone,\vcxtwo),
\notag
\end{align}
implying that
\begin{align}
 & p_{\wt{M}_1^{(n)}\wt{M}_2^{(n)}}
     \Bigl(\wt{M}_{1}^{(n)},\wt{M}_{2}^{(n)}\Bigr)
=p_{\lrvctxone\lrvctxtwo}
\left(\wt{\rvcx}_1,\wt{\rvcx}_2\right).
\label{eqn:PrCKtoPrXb}
\end{align}
From (\ref{eqn:PrCKtoPrXb}), we have the following:  
\begin{align*}
&\underline{H}
(\wt{M}_1^{(\infty)}\wt{M}_2^{(\infty)})= 
     \underline H(\wt{X}_1^{\infty}
                  \wt{X}_2^{\infty}),
\\
& 
\overline{H}(\wt{M}_1^{(\infty)}\wt{M}_2^{(\infty)}
             )= 
             \overline H(\wt{X}_1^{\infty}
                          \wt{X}_2^{\infty}).
\end{align*}
\item[b)] For any $\gamma>0$, we have the following:
\begin{align}
0 \leq \eta_n(\gamma)\leq \frac{4}{n\gamma}
\log\left(\frac{{\rm e}^{2{\rm e}^{-1}}}
    {1-\nu_n(\gamma,\varepsilon)}\right).
\label{eqn:EfBdB}    
\end{align}
For each fixed $\gamma>0$, the right member of (\ref{eqn:EfBdB}) vanish when $n\to\infty$, implying the following:  
\begin{align*}
& \underline H(\wt{X}_1^{\infty}
                          \wt{X}_2^{\infty})
  = \overline H(\wt{X}_1^{\infty}
                          \wt{X}_2^{\infty}) 
  =  H(X_1X_2).                       
\end{align*}
For each $i=1,2$ and for any $\gamma>0$, 
\begin{align}
0 \leq \eta_{i,n}(\gamma)\leq \frac{3}{n\gamma}
\log\left(\frac{{\rm e}^{2{\rm e}^{-1}}}
    {1-\nu_n(\gamma,\varepsilon)}\right).
\label{eqn:EfBdBc}    
\end{align}
For each fixed $\gamma>0$, the right member 
of (\ref{eqn:EfBdBc}) vanish when $n\to\infty$, implying 
the following:  
\begin{align*}
& \underline H(\wt{X}_i^{\infty})                        
  = \overline H(\wt{X}_i^{\infty}) 
  =  H(X_i),\:i=1,2.                       
\end{align*}
\item[c)]For each $i=1,2$, we have 
\begin{equation}
\left.\ba{l}
 \overline{H}(\wt{M}_i^{(\infty)})
=\overline{H}({M}_i^{(\infty)}),
\\
\underline{H}(\wt{M}_i^{(\infty)}) 
=\underline{H}({M}_i^{(\infty)}).
\ea\right\}
\label{eqn:PrCEq}
\end{equation}
\item[d)] 
For each fixed $\gamma>0$, 
\begin{align}
\lim_{n\to \infty}\theta_{i,n}(\gamma)=0\mbox{ for }i=1,2.
\label{eqn:ThetaZero}
\end{align} 
For each fixed $i=1,2$, we have 
\begin{align}
& 0 \leq    \underline{H}_i \leq  
            \overline{H}_i \leq 
        \min\{R_i, H(X_i)\}\mbox{ for }i=1,2.
\label{eqn:UbHi}       
\end{align}
\end{itemize}  
\end{property}

Proof of Property 
\ref{pr:InfSpecLm} is given 
in Appendix \ref{apd:ProofPrInfSpec}. 
\newcommand{\ProofPrInfSpec}{
\subsection{
Proof of Property \ref{pr:InfSpecLm}
}\label{apd:ProofPrInfSpec}

In this appendix we prove 
Property \ref{pr:InfSpecLm}.

We first prove the part a). 
\begin{IEEEproof}
[Proof of Property \ref{pr:InfSpecLm} part a)]
We first observe that for 
$
(m_1,m_2) \in {\cal M}^{(n)}_{
 (\phi_{1}^{(n)},\phi_{2}^{(n)})
 (\wt{\cal D}_{\gamma}^{(n)})}, 
$
\begin{align}
& \Pr\Bigl\{(\wt{M}_{1}^{(n)},\wt{M}_{2}^{(n)})
    =(m_{1},m_{2})\Bigr\}
\notag\\  
&=\Pr\Bigl\{(M_{1}^{(n)},M_{2}^{(n)})
   = (m_{1},m_{2})\Bigl|
\notag\\
&\quad\qquad \left(\rvcxone,\rvcxtwo\right) \in  
   \wt{\DcSet}_{\gamma}^{(n)}\Bigr\}.
\label{eqn:tiCandCbb}   
\end{align}
For each 
$
(m_1,m_2) \in {\cal M}^{(n)}_{
 (\phi_{1}^{(n)},\phi_{2}^{(n)})
 (\wt{\cal D}_{\gamma}^{(n)})}, 
$
there exists a unique $(\vcxone,\vcxtwo)
\in \wt{\cal D}_{\gamma}^{(n)}$ 
such that 
$\vcxi=\phi_{i}^{(n)}(m_i),i=1,2$. 
Furthermore we have the following:
\begin{align}
& p_{\wt{M}_1^{(n)}\wt{M}_2^{(n)}} (m_{1},m_{2})  
=\Pr\Bigl\{(\wt{M}_{1}^{(n)},\wt{M}_{2}^{(n)})=(m_{1},m_{2})
    \Bigr\}
\notag\\  
& \MEq{a}\Pr\Bigl\{(M_{1}^{(n)},M_{2}^{(n)})=(m_{1},m_{2})
   \Bigl| 
\left(\rvcxone,\rvcxtwo\right) \in  
   \wt{\DcSet}_{\gamma}^{(n)}
  \Bigr\}
\notag\\
& =\Pr\Bigl\{
\left(\rvcxone,\rvcxtwo\right)=(\vcxone,\vcxtwo)
   \Bigl| 
 \left(\rvcxone,\rvcxtwo\right) \in  
   \wt{\DcSet}_{\gamma}^{(n)}\Bigr\} 
\notag\\
&= p_{\lrvctxone\lrvctxtwo}(\vcxone,\vcxtwo).
\notag
\end{align}
Step (a) follows from (\ref{eqn:tiCandCbb}). 
\end{IEEEproof}

We next prove the part b). 
We use the following lemma. 
\begin{lemma}\label{lm:DivLm}
Let $Z_1,Z_2$ be arbitrary random
variable taking values in a finite
set ${\cal Z}$. Then we have      
\begin{align}
 {\rm E}\left[\left|\log
 \frac{p_{Z_1}(Z_1)}{p_{Z_2}(Z_1)}
\right|\right]
\leq D(p_{Z_1}||p_{Z_2}) 
+{\ts 2{\rm e}^{-1}}\log {\rm e}.
 \label{eqn:DvibIeqZero} 
\end{align}
\end{lemma}
\begin{IEEEproof}
Set 
$
\omega(z) \defeq 
{p_{Z_1}(z)}({p_{Z_2}(z)})^{-1}. 
$
Furthermore set 
\begin{align*}
& D^{(+)}\defeq {\rm E} \left[
{\vc 1}\left[ 
       \omega(Z_1)\geq 1 \right]
       \log \omega (Z_1) 
       \right],\\  
& D^{(-)}\defeq {\rm E}\left[
 {\vc 1}\left[
  0<\omega(Z_1)\leq 1 
 \right]
 (-1)\log \omega(Z_1) 
\right].
\end{align*}
Then we have 
\begin{align}
\left.
\ba{rcl}
D(p_{Z_1}||p_{Z_2}) &=&D^{(+)}-D^{(-)},
\vspace{1mm}\\
{\rm E}\left[\ds \left|\log\frac{p_{Z_1}(Z_1)}{p_{Z_2}(Z_1)}
  \right|\right] &=& D^{(+)}+D^{(-)}.    
\ea
\right\}
\label{eqn:DivEqOne}
\end{align}
From (\ref{eqn:DivEqOne}), we have 
\begin{align}
{\rm E}\left[\ds \left|\log\frac{p_{Z_1}(Z_1)}{p_{Z_2}(Z_1)}
  \right|\right] &= D(p_{Z_1}||p_{Z_2}) +2D^{(-)}.    
\label{eqn:DivEqTwo}
\end{align}
On upper bounds of $D^{(-)}$, we have the following chain 
of inequalities:
\begin{align}
& D^{(-)}=
  {\rm E}\left[
  {\vc 1}\left[ 0<\omega(Z_1)\leq 1 \right]
   (-1)\log{\omega(Z_1)}\right]
\notag\\  
&={\rm E}\left[
  {\vc 1}\left[ 0<\omega(Z_2)\leq 1 \right]
   (-\omega(Z_2))\log{\omega(Z_2)}\right]
\notag\\   
&
\MLeq{a}({\rm e}^{-1}\log {\rm e})
{\rm E}\left[
  {\vc 1}\left[0<\omega(Z_2)\leq 1 \right]\right]
  \leq {\rm e}^{-1}\log {\rm e}.  
\label{eqn:DivIeqThr}
\end{align}
Step (a) follows from that 
$$
(-\omega)\log \omega \leq {\rm e}^{-1}\log {\rm e}
\mbox{ for }0\leq \omega \leq 1.
$$  
From (\ref{eqn:DivEqTwo}) and (\ref{eqn:DivIeqThr}), 
we have the bound (\ref{eqn:DvibIeqZero}) 
in Lemma \ref{lm:DivLm}.
\end{IEEEproof}

\begin{IEEEproof}
[Proof of Property \ref{pr:InfSpecLm} part b)]
We first prove (\ref{eqn:EfBdB}).
On upper bound of $\eta_n(\gamma)$, we have the following 
chain of inequalities:
\begin{align}
 &  \eta_{n}(\gamma)
=\Pr\left\{
\left|\frac{1}{n}
   \log\frac{p_{{\lrvcx}_1{\lrvcx}_2}(
                \wt{\rvcx}_1 ,\wt{\rvcx}_2)}
            {p_{\wt{\lrvcx}_1\wt{\lrvcx}_2}(
                \wt{\rvcx}_1 ,\wt{\rvcx}_2)}              
  \right.\right.
\notag\\
& \left.\left. 
\quad  +\frac{1}{n}
    \log\frac{1}{p_{{\lrvcx}_1{\lrvcx}_2}(
    \wt{\rvcx}_1 ,\wt{\rvcx}_2)} 
-H(X_1X_2)\right| \geq {\ts\frac{3}{2}}\gamma \right\}
\notag\\
&\leq\Pr\left\{
\left|\frac{1}{n}
   \log\frac{p_{{\lrvcx}_1{\lrvcx}_2}(
                \wt{\rvcx}_1 ,\wt{\rvcx}_2)}
            {p_{\wt{\lrvcx}_1\wt{\lrvcx}_2}(
                \wt{\rvcx}_1 ,\wt{\rvcx}_2)}              
  \right|\right.
\notag\\
& \left. 
 \quad+\left|\frac{1}{n}
    \log\frac{1}{p_{{\lrvcx}_1{\lrvcx}_2}(
    \wt{\rvcx}_1 ,\wt{\rvcx}_2)} 
-H(X_1X_2)\right| \geq 
 {\ts\frac{3}{2}}\gamma \right\}.
\label{eqn:etaUbZero}
\end{align}
From (\ref{eqn:etaUbZero}), we further continue to 
evaluate upper bound of $\eta_n(\gamma)$ to obtain 
the following: 
\begin{align}
&\eta_n(\gamma)\leq \Pr\left\{
\left|\frac{1}{n}
   \log\frac{p_{\lrvctxone\lrvctxtwo}(
                \wt{\rvcx}_1 ,\wt{\rvcx}_2)}
            {p_{{\lrvcx}_1{\lrvcx}_2}(
                \wt{\rvcx}_1 ,\wt{\rvcx}_2)}              
  \right|\geq {\ts \frac{1}{4}}\gamma \right\}
\notag\\
& \quad +\Pr\left\{ 
     \left|\frac{1}{n}
     \log\frac{1}{p_{{\lrvcx}_1{\lrvcx}_2}(
    \wt{\rvcx}_1 ,\wt{\rvcx}_2)} 
-H(X_1X_2)\right| \geq {\ts \frac{5}{4}}\gamma \right\}
\notag\\
& \MLeq{a} \frac{4}{n\gamma}{\rm E}
\left[
\left|\log\frac{p_{\lrvctxone\lrvctxtwo}(
                 \wt{\rvcx}_1 ,\wt{\rvcx}_2)}
                {p_{{\lrvcx}_1{\lrvcx}_2}(
                \wt{\rvcx}_1 ,\wt{\rvcx}_2)}              
  \right|
\right]
+p_{\lrvctxone\lrvctxtwo}\left(
\left({\cal A}_{\frac{6}{5}\gamma}^{(n)}\right)^{\rm c}
\right)
\notag\\
&\MEq{b}
 \frac{4}{n\gamma}{\rm E}
\left[\left|\log\frac{p_{\lrvctxone\lrvctxtwo}(
                 \wt{\rvcx}_1 ,\wt{\rvcx}_2)}
                {p_{{\lrvcx}_1{\lrvcx}_2}(
                \wt{\rvcx}_1 ,\wt{\rvcx}_2)} 
\right|\right].
\label{eqn:EtaGamUBOne}
\end{align}
Step (a) follows from the Markov inequality and 
$\frac{5}{4}\gamma>\frac{6}{5}\gamma>0$.
Step (b) follows from the following:
\begin{align*}
 & p_{\lrvctxone\lrvctxtwo}\left(
\left({\cal A}_{\frac{6}{5}\gamma}^{(n)}\right)^{\rm c}
\right)
=p_{\lrvctxone\lrvctxtwo}\left(
\wt{\cal D}_{\gamma}^{(n)}
\bigcap
\left({\cal A}_{\frac{6}{5}\gamma}^{(n)}\right)^{\rm c}
\right)
\notag\\
&=p_{\lrvctxone\lrvctxtwo}\left(
{\cal D}^{(n)}
\bigcap
{\cal A}_{\gamma}^{(n)}
\bigcap
\left({\cal A}_{\frac{6}{5}\gamma}^{(n)}\right)^{\rm c}
\right)=0.
\end{align*}
The upper bound (\ref{eqn:EtaGamUBOne})
of $\eta_n(\gamma)$
together with Lemma \ref{lm:DivLm} yields 
the following:
\begin{align*} 
&\eta_n(\gamma)\leq\frac{4}{n\gamma}\left[
   D(p_{\lrvctxone\lrvctxtwo}||
     p_{\lrvcxone\lrvcxtwo})
           +{\ts 2{\rm e}^{-1}}\log {\rm e} \right]
\notag\\
&=\frac{4}{n\gamma}
\log\left(
   \frac{{\rm e}^{2{\rm e}^{-1}}}
   { 
       p_{\lrvcxone\lrvcxtwo}\left(
      \wt{\cal D}_{\gamma }^{(n)}\right) 
   }\right)
\leq \frac{4}{n\gamma}
\log\left(\frac{{\rm e}^{2{\rm e}^{-1}}}
    {1-\nu_n(\gamma,\varepsilon) }\right).
\end{align*} 
We next prove (\ref{eqn:EfBdBc}). 
For $\tau>0$, we set 
\begin{align*}
&\wt{{\cal A}}_{\tau}^{(n)} \defeq 
 \biggl\{ (\vcxone,\vcxtwo):
\left|\frac{1}{n}
 \log\frac{1}{p_{X_{i}}^{n}({\vcx}_i)}
- H(X_{i})\right|\leq \tau,i=1,2\\
 &\:\:\left.\left| \frac{1}{n}\log\frac{1}{p_{X_{1}X_{2}}^{n}
  (\vcxone,\vcxtwo)}
 -H(X_{1}X_{2})\right|\leq \tau\right\}. 
 \end{align*}
Note that for any $\tau>0$, we have   
$
              {{\cal A}}_{\frac{1}{2}\tau}^{(n)}
\subseteq  \wt{{\cal A}}_{\tau}^{(n)}.    
$
In a manner quite similar to the derivation of (\ref{eqn:etaUbZero}), we evaluate upper bounds  
of $\eta_{i,n}(\gamma), i=1,2$ to obtain the
following:  
\begin{align}
&  \eta_{i,n}(\gamma)
\leq\Pr\left\{
\left|\frac{1}{n}
   \log\frac{p_{{\lrvcx}_i }(\wt{\rvcx}_i)}
            {p_{\wt{\lrvcx}_i}(\wt{\rvcx}_i)}              
  \right|\right.
\notag\\
& \left. 
 \quad+\left|\frac{1}{n}
    \log\frac{1}{p_{{\lrvcx}_i}(
    \wt{\rvcx}_i)} 
-H(X_i)\right| \geq 
  3\gamma \right\}.
\label{eqn:etaUbZerob}
\end{align}
From (\ref{eqn:etaUbZerob}), 
we have the following: 
\begin{align}
&\eta_{i,n}(\gamma)
\leq \Pr\left\{
\left|\frac{1}{n}
   \log\frac{p_{\wt{\lrvcx}_i}(\wt{\rvcx}_i)}
            {p_{{\lrvcx}_i}(\wt{\rvcx}_i )}              
  \right|\geq {\ts \frac{1}{3}}\gamma \right\}
\notag\\
& \quad +\Pr\left\{ 
     \left|\frac{1}{n}
     \log\frac{1}{p_{{\lrvcx}_i}(\wt{\rvcx}_i)} 
   -H(X_i)\right| \geq {\ts \frac{8}{3}}\gamma \right\}
\notag\\
& \MLeq{a} \frac{3}{n\gamma}{\rm E}
\left[
\left|\log\frac{p_{\wt{\lrvcx}_i }(\wt{\rvcx}_i)}
               {p_{{\lrvcx}_i }(\wt{\rvcx}_i )}              
 \right| \right]
+p_{\lrvctxone\lrvctxtwo}\left(
\left(\wt{\cal A}_{\frac{9}{4}\gamma}^{(n)}\right)^{\rm c}
\right)
\notag\\
& \MLeq{b} \frac{3}{n\gamma}{\rm E}
\left[
\left|\log\frac{p_{\wt{\lrvcx}_i }(\wt{\rvcx}_i)}
               {p_{{\lrvcx}_i }(\wt{\rvcx}_i )}              
 \right| \right]
+p_{\lrvctxone\lrvctxtwo}\left(
\left({\cal A}_{\frac{9}{8}\gamma}^{(n)}\right)^{\rm c}
\right)
\notag\\
& \MEq{c}
\frac{3}{n\gamma}{\rm E}
\left[
\left|\log\frac{p_{\wt{\lrvcx}_i }(\wt{\rvcx}_i)}
               {p_{{\lrvcx}_i }(\wt{\rvcx}_i )}              
 \right|\right].
\label{eqn:EtaiGamUb}
\end{align}
Step (a) follows from the Markov inequality 
and $\frac{8}{3}\gamma>\frac{9}{4}\gamma>0$. 
Step (b) follows from  
$\wt{\cal A}_{\frac{9}{4}\gamma}^{(n)}
 \supseteq{\cal A}_{\frac{9}{8}\gamma}^{(n)}$.
Step (c) follows from the following:
\begin{align*}
 & p_{\lrvctxone\lrvctxtwo}\left(
\left({\cal A}_{\frac{9}{8}\gamma}^{(n)}\right)^{\rm c}
\right)
=p_{\lrvctxone\lrvctxtwo}\left(
\wt{\cal D}_{\gamma}^{(n)}
\bigcap
\left({\cal A}_{\frac{9}{8}\gamma}^{(n)}\right)^{\rm c}
\right)
\notag\\
&=p_{\lrvctxone\lrvctxtwo}\left(
{\cal D}^{(n)}
\bigcap
{\cal A}_{\gamma}^{(n)}
\bigcap
\left({\cal A}_{\frac{9}{8}\gamma}^{(n)}\right)^{\rm c}
\right)=0.
\end{align*}
For each $i=1,2$, the upper bound (\ref{eqn:EtaiGamUb}) of $\eta_{i,n}(\gamma)$
together with  Lemma \ref{lm:DivLm} yields 
the following: 
\begin{align}
& \eta_{i,n}(\gamma) \leq\frac{3}{n\gamma}\left[
   D(p_{\wt{\lrvcx}_i}||p_{{\lrvcx}_i} )
           +{\ts 2{\rm e}^{-1}}
           \log {\rm e} \right]
\notag\\
& \leq\frac{3}{n\gamma}\left[   D(p_{\wt{\lrvcx}_1\wt{\lrvcx}_2}||p_{{\lrvcx}_1{\lrvcx}_2})
           +{\ts 2{\rm e}^{-1}}\log {\rm e} \right]           
\notag\\
& \leq \frac{3}{n\gamma}
\log\left(\frac{{\rm e}^{2{\rm e}^{-1}}}
    {1-\nu_n(\gamma,\varepsilon)}\right),
\notag    
\end{align} 
completing the proof.
\end{IEEEproof}

Let $\{Z^{(n)}\in {\cal Z}^{(n)}\}_{n\geq 1}$ 
be a sequence of random variables with 
$\sup_{n\geq 1}\log |{\cal Z}^{(n)}| < \infty$.
For $a\geq 0$, set
\begin{align*}
G_{Z^{(n)}}(a)
 &\defeq
\Pr\Biggl\{
  \frac{1}{n}
  \log\frac{1}{p_{{Z}^{(n)}}({Z}^{(n)})}  
\geq a\Biggr\}.
\end{align*}
To prove the part c), we use the following lemma.
\begin{lemma}  \label{lm:IfSpLmb}
For each $i=1,2$ and $\forall\tau>0$, we have 
\begin{align}
G_{\wt{M_i}^{(n)}}(a)
&\geq G_{M_i^{(n)}}(a+\tau) -\nu_n(\gamma,\vep)
\notag\\
&\quad -\frac{1}{n\tau}
 \log\left(\frac{{\rm e}^{2{\rm e}^{-1}}}
    {1-\nu_n(\gamma,\varepsilon)}\right),
\label{eqn:IfSpPrPartCOne}\\    
G_{\wt{M_i}^{(n)}}(a)&\leq 
G_{M_i^{(n)}}(a-\tau)[1-\nu_n(\gamma,\vep)]^{-1}
\notag\\ 
&\quad+\frac{1}{n\tau}
 \log\left(\frac{{\rm e}^{2{\rm e}^{-1}}}
    {1-\nu_n(\gamma,\varepsilon)}\right).
\label{eqn:IfSpPrPartCThr}
\end{align}
Furthermore, for each $i=1,2$, we have 
\begin{align}
F_{\wt{M}_i^{(n)}}(a)
&\geq F_{M_i^{(n)}}(a-\tau) -\nu_n(\gamma,\vep)
\notag\\ 
&\quad -\frac{1}{n\tau}
 \log\left(\frac{{\rm e}^{2{\rm e}^{-1}}}
    {1-\nu_n(\gamma,\varepsilon)}\right),
\label{eqn:IfSpPrPartCbOne}\\    
F_{\wt{M}_i^{(n)}}(a)&\leq F_{M_i^{(n)}}(a+\tau)
 {[1-\nu_n(\gamma,\vep)]^{-1}}
\notag\\ 
&\quad +\frac{1}{n\tau}
 \log\left(\frac{{\rm e}^{2{\rm e}^{-1}}}
    {1-\nu_n(\gamma,\varepsilon)}\right).
\label{eqn:IfSpPrPartCbThr}
\end{align}
\end{lemma}
\begin{IEEEproof}
We prove the first two bounds described 
 in (\ref{eqn:IfSpPrPartCOne}) 
 and (\ref{eqn:IfSpPrPartCThr}).   
Proofs of the second two bounds described 
in  (\ref{eqn:IfSpPrPartCbOne}) 
and (\ref{eqn:IfSpPrPartCbThr})   
are quite similar to those of the first two bounds.
We omit the detail of the proofs. For each $i=1,2$, we set
\begin{align}
P_{1,i}&\defeq\Pr\left\{
\frac{1}{n}\left|\log\frac{p_{\wt{M}_i^{(n)}}(\wt{M}_i^{(n)})}
{p_{{M}_i^{(n)}}(\wt{M}_i^{(n)})}\right|
  \geq \tau \right\}.
\label{eqn:DefPonEi}
\end{align}
On upper bounds of $P_{1,i},i=1,2$, we have the following 
chain of inequalities:
\begin{align}
& P_{1,i} \MLeq{a} \frac{1}{n\tau}{\rm E}\left[
\left|\log\frac{p_{\wt{M}_i^{(n)}}(\wt{M}_i^{(n)})}
{p_{{M}_i^{(n)}}(\wt{M}_i^{(n)})}\right|
\right]
\label{eqn:UpbPzero}\\
& 
\MLeq{b}
\frac{1}{n\tau}\left[D(p_{\wt{M}_i^{(n)}}||p_{{M}_i^{(n)}})
+2{\rm e}^{-1}\log {\rm e}\right]
\notag\\
&=\frac{1}{n\tau}\left[
D(p_{\phi_i^{(n)}(\wt{\lrvcxi})}||
  p_{\phi_i^{(n)}(   {\lrvcxi})})
+2{\rm e}^{-1}\log {\rm e}\right]
\notag\\
& \leq 
\frac{1}{n\tau}\left[
D(p_{\wt{\lrvcx}_1\wt{\lrvcx}_2}||p_{{\lrvcx}_1{\lrvcx}_2})
+{2{\rm e}^{-1}}\log {\rm e}\right]
\notag\\
& \leq \frac{1}{n\tau}
 \log\left(\frac{{\rm e}^{2{\rm e}^{-1}}}
    {1-\nu_n(\gamma,\varepsilon)}\right).
\label{eqn:UpbPone}    
\end{align}
Step (a) follows from the Markov inequality.
Step (b) follows from Lemma \ref{lm:DivLm}.
On upper bounds of $G_{\wt{M_i}^{(n)}}(a),i=1,2$, we have the following 
chain of inequalities:  
\begin{align}
&G_{\wt{M_i}^{(n)}}(a)\leq \Pr\left\{
\frac{1}{n}\log\frac{1}
{p_{{M}_i^{(n)}} 
      (\wt{M}_i^{(n)})}
  \geq a-\tau \right\}  
\notag\\
&\quad  +\Pr\left\{
\frac{1}{n} \log\frac{p_{{M}_i^{(n)}}(\wt{M}_i^{(n)})}
{p_{\wt{M}_i^{(n)}}(\wt{M}_i^{(n)})}
 \geq \tau \right\} 
\notag\\
&\leq
\Pr\left\{\left.
\frac{1}{n}\log\frac{1}
{p_{{M}_i^{(n)}} 
   ({M}_i^{(n)})}
  \geq a-\tau
\right| (\rvcxone,\rvcxtwo)\in
        \wt{\cal D}_{\gamma }^{(n)}\right\}
\notag\\ 
&\quad + P_{1,i}.
%
%
\label{eqn:PrLmbOne} 
\end{align}
On the first term in (\ref{eqn:PrLmbOne}), 
we have the following:
\begin{align}
&  \Pr\left\{\left.
\frac{1}{n}\log\frac{1}
{p_{{M}_i^{(n)}} 
   ({M}_i^{(n)})}
  \geq a-\tau
\right| (\rvcxone,\rvcxtwo)\in
        \wt{\cal D}_{\gamma }^{(n)}\right\}
\notag\\
&\leq \frac{G_{{M_i}^{(n)}}(a)}{p_{\lrvcxone\lrvcxtwo}\left(
      \wt{\cal D}_{\gamma }^{(n)}\right)}  
\leq G_{{M_i}^{(n)}}(a)[1-\nu_n(\gamma,\vep)]^{-1}.
\label{eqn:PrLmbTwo}  
\end{align}
Combining (\ref{eqn:UpbPone}),
          (\ref{eqn:PrLmbOne}),
     and  (\ref{eqn:PrLmbTwo}),       
we obtain the bound 
(\ref{eqn:IfSpPrPartCThr}) in Lemma \ref{lm:IfSpLmb}.  
On lower bounds of $G_{\wt{M_i}^{(n)}}(a),i=1,2$, we have the following 
chain of inequalities:  
\begin{align}
&G_{\wt{M_i}^{(n)}}(a)\geq \Pr\left\{
\frac{1}{n}\log\frac{1}
{p_{{M}_i^{(n)}} 
      (\wt{M}_i^{(n)})}
  \geq a+\tau \right\}  
\notag\\
&\quad  -\Pr\left\{
\frac{1}{n} \log\frac{p_{{M}_i^{(n)}}(\wt{M}_i^{(n)})}
{p_{\wt{M}_i^{(n)}}(\wt{M}_i^{(n)})}
 \leq -\tau \right\} 
\notag\\
&\geq
\Pr\left\{\left.
\frac{1}{n}\log\frac{1}
{p_{{M}_i^{(n)}} 
   ({M}_i^{(n)})}
  \geq a+\tau
\right| (\rvcxone,\rvcxtwo)\in
        \wt{\cal D}_{\gamma }^{(n)}\right\}
\notag\\ 
&\quad - P_{1,i}.
\label{eqn:PrLmbThr} 
\end{align}
On the first term in (\ref{eqn:PrLmbThr}), 
we have the following:  
\begin{align}
& \Pr\left\{\left.
\frac{1}{n}\log\frac{1}
{p_{{M}_i^{(n)}} 
   ({M}_i^{(n)})}
  \geq a+\tau
\right| (\rvcxone,\rvcxtwo)\in
        \wt{\cal D}_{\gamma }^{(n)}\right\}
\notag\\
&\geq
\Pr\left\{
\frac{1}{n}\log\frac{1}
{p_{{M}_i^{(n)}} 
   ({M}_i^{(n)} )}
  \geq a+\tau,  (\rvcxone,\rvcxtwo)\in
        \wt{\cal D}_{\gamma }^{(n)} 
 \right\}
\notag\\ 
&\geq G_{{M_i}^{(n)}}(a+\tau)
+p_{\lrvcxone\lrvcxtwo}\left(
      \wt{\cal D}_{\gamma }^{(n)}\right)-1
\notag\\      
&\geq 
G_{{M_i}^{(n)}}(a+\tau)-\nu_n(\gamma,\vep)
\label{eqn:PrLmbFou}.
\end{align}
Combining (\ref{eqn:UpbPone}),
          (\ref{eqn:PrLmbThr}),
     and  (\ref{eqn:PrLmbFou}),       
we obtain the bound  
(\ref{eqn:IfSpPrPartCOne})
in Lemma \ref{lm:IfSpLmb}.
\end{IEEEproof}

\begin{IEEEproof}
[Proof of Property \ref{pr:InfSpecLm} part c)]
In the bound described in 
(\ref{eqn:IfSpPrPartCOne}) 
in Lemma \ref{lm:IfSpLmb}, 
we choose $a=\overline{H}(\wt{M}_i^{(\infty)})+\tau$ 
and let $n\to \infty$ in this bound. Then we have 
\begin{align} 
&\lim_{n\to\infty}
\Pr\left\{
\frac{1}{n}\log\frac{1}
{p_{{M}_i^{(n)}} 
   ({M}_i^{(n)} )}
  \geq \overline{H}(\wt{M}_i^{(\infty)})+2\tau 
 \right\}
\notag\\ 
&\leq \lim_{n\to\infty}\Pr\left\{
\frac{1}{n}\log\frac{1}
{p_{\wt{M}_i^{(n)}} 
   (\wt{M}_i^{(n)})}
  \geq \overline{H}(\wt{M}_i^{(\infty)})+\tau \right\}
\notag\\
&\MEq{a}0.
\label{eqn:PrCProofOne}
\end{align}
Step (a) follows from the definition of  
$\overline{H}(\wt{M}_i^{(\infty)})$. 
The bound (\ref{eqn:PrCProofOne}) implies that
$\overline{H}(   {M}_i^{(\infty)})\leq 
 \overline{H}(\wt{M}_i^{(\infty)})+2\tau$.
Similarly, in the bound described in 
(\ref{eqn:IfSpPrPartCThr}) in Lemma \ref{lm:IfSpLmb}, 
we choose $a=\overline{H}({M}_i^{(\infty)})+2\tau$ 
and let $n\to \infty$ in this bound. Then we have 
\begin{align} 
&\lim_{n\to\infty}
\Pr\left\{
\frac{1}{n}\log\frac{1}
{p_{\wt{M}_i^{(n)}} 
   (\wt{M}_i^{(n)} )}
  \geq \overline{H}({M}_i^{(\infty)})+2\tau 
 \right\}
\notag\\ 
&\leq \lim_{n\to\infty}\Pr\left\{
\frac{1}{n}\log\frac{1}
{p_{{M}_i^{(n)}} 
   ({M}_i^{(n)})}
  \geq \overline{H}({M}_i^{(\infty)})+\tau \right\}
\notag\\
&\MEq{a}0.
\label{eqn:PrCProofTwo}
\end{align}
Step (a) follows from the definition of  
$\overline{H}({M}_i^{(\infty)})$. 
The bound (\ref{eqn:PrCProofTwo}) implies that
$\overline{H}(\wt{M}_i^{(\infty)})\leq 
 \overline{H}({M}_i^{(\infty)})+2\tau$.
Hence we have the bound
$$
\overline{H}(   {M}_i^{(\infty)})-2\tau \leq
\overline{H}(\wt{M}_i^{(\infty)})       \leq 
\overline{H}(   {M}_i^{(\infty)})+2\tau.
$$
Since $\tau>0$ can be taken arbitrary small, we have 
$   \overline{H}(\wt{M}_i^{(\infty)})
$ $=\overline{H}(   {M}_i^{(\infty)})$. 
Using the second two bounds described 
in  (\ref{eqn:IfSpPrPartCbOne}) 
and (\ref{eqn:IfSpPrPartCbThr}) 
in Lemma \ref{lm:IfSpLmb}, we can prove 
the second inequality in (\ref{eqn:PrCEq}). 
The proof is quite parallel with that of 
the first equality in (\ref{eqn:PrCEq}). We 
omit the detail. 
\end{IEEEproof}

To prove the part d), we use the following lemma. 
\begin{lemma}\label{lm:IfSpLm}
Let $(Z_1,Z_2)$ be an arbitrary 
correlated random pair taking values in ${\cal Z}_1 \times {\cal Z}_2$. 
Then, for any $\tau>0$, we have      
\begin{align}
&\Pr\left\{
\log \frac{1}{p_{Z_1|{Z}_2}(Z_1|Z_2)}
\geq \log |{\cal Z}_1| +\tau
\right\}
\leq 2^{-\tau}.
\label{eqn:InfSpecIeq} 
\end{align}
\end{lemma}
  
\begin{IEEEproof}
We have the following:
\begin{align*}
&\Pr\left\{
\log \frac{1}{p_{Z_1|Z_2}(Z_1|Z_2)}
\geq \log |{\cal Z}_1| +\tau
     \right\}
\\     
&= \sum_{\scs 
(z_1,z_2)\in{\cal Z}_1\times {\cal Z}_2:
\atop{\scs 
{p_{Z_1|Z_2}(z_1|z_2)}
 \leq \frac{2^{-\tau}}{|{\cal Z}_1|}}
}p_{Z_1|Z_2}(z_1|z_2)p_{Z_2}(z_2)
\\
& \leq \frac{2^{-\tau}}{|{\cal Z}_1|}
\sum_{\scs 
(z_1,z_2)\in{\cal Z}_1\times {\cal Z}_2:
\atop{\scs 
{p_{Z_1|Z_2}(z_1|z_2)}
 \leq \frac{2^{-\tau}}{|{\cal Z}_1|}}
}p_{Z_2}(z_2)
\\
& \leq \frac{2^{-\tau}}{|{\cal Z}_1|}\sum_{\scs 
(z_1,z_2)\in{\cal Z}_1\times {\cal Z}_2
}p_{Z_2}(z_2)
=2^{-\tau},
\end{align*}
completing the proof.
\end{IEEEproof}

We finally prove the part d).
\begin{IEEEproof}
[Proof of Property \ref{pr:InfSpecLm} part d)]
The bound (\ref{eqn:ThetaZero}) is obvious 
from the definition of $\underline{H}_i,i=1,2$. 
We prove the bound (\ref{eqn:UbHi}). We first prove 
$
\underline{H}_i\leq \overline{H}_i\leq R_i,i=1,2.
$
The bounds 
$\underline{H}_i\leq \overline{H}_i, i=1,2$ are 
obvious by definition. We prove 
$\overline{H}_i\leq R_i,i=1,2.$
Since $(R_1,R_2)$ is a $(\errP,\secP)$-reliable and 
secure rate pair we have that $\forall\gamma >0$, 
$\exists n_0(\gamma)$, $\forall n\geq n_0$, we have   
\begin{align}
   \frac{1}{n}\log \left|{\cal M}_i^{(n)}\right|
\leq R_i+\gamma,\: i=1,2. 
\label{eqn:RateCond}
\end{align}
Under this condition, we have the following chain 
of inequalities:  
\begin{align}
&\Pr\left\{
\frac{1}{n}\log\frac{1}
{p_{\wt{M}_i^{(n)}} 
   (\wt{M}_i^{(n)})}
  \geq R_i+\gamma +\tau 
 \right\}  
\notag\\
&\MLeq{a}
\Pr\left\{
\log\frac{1}
{p_{\wt{M}_i^{(n)}} 
   (\wt{M}_i^{(n)} )}
  \geq \log |{\cal M}_i^{(n)}| +n\tau 
 \right\}
\notag\\ 
& \MLeq{b} 2^{-n\tau}.
\label{eqn:IfSpecPrPartcOne}
\end{align}
Step (a) follows from (\ref{eqn:RateCond}).
Step (b) follows from Lemma \ref{lm:IfSpLm}.
The right member of (\ref{eqn:IfSpecPrPartcOne}) 
tends to zero as $n\to\infty$. Then by the definition of 
$\overline{H}_i$, we have  
$\overline{H}_i\leq R_i+\gamma+\tau,$
$i=1,2$. 
Since $\gamma>0$ and $\tau>0$ can arbitrary be small, 
we conclude that $\overline{H}_i\leq R_i, i=1,2$.
We next prove $\overline{H}_i\leq H(X_i),i=1,2$.
For each $i=1,2$, we have the following 
chain of inequalities:
\begin{align}
& 0\leq \underline{H}(\wt{X}_i^{\infty}|
  \wt{M}_i^{\infty}) 
= \mbox{\rm p-}\liminf_{n\to\infty}
 \frac{1}{n}\log\frac{1}
{p_{\wt{\lrvcx}_i|\wt{M}_i^{(n)}}
(\wt{\rvcx}_i|\wt{M}_i^{(n)})}
\notag\\
&=\mbox{\rm p-}\liminf_{n\to\infty}
\hugebl
\frac{1}{n}\log\frac{1}
{p_{\wt{\lrvcx}_i}(\wt{\rvcx}_i)}
-\frac{1}{n}\log\frac{1}
{p_{\wt{M}_i^{(n)} }(
\wt{M}_i^{(n)})}
\hugebr
\notag\\
&\leq \mbox{\rm p-}\limsup_{n\to\infty}
\frac{1}{n}\log\frac{1} {p_{\wt{\lrvcx}_i}
(\wt{\rvcx}_i)} 
\notag\\
&\quad 
+\mbox{\rm p-}\liminf_{n\to\infty}
\hugebl
-\frac{1}{n}\log\frac{1}
{p_{\wt{M}_i^{(n)}}(\wt{M}_i^{(n)})}
\hugebr
\notag\\
&=\overline{H}(\wt{X}_i^{\infty})-\overline{H}_i
\MEq{a}H(X_i)-\overline{H}_i.
\label{eqn:infHiBd}
\end{align}
Step (a) follows from the part b).
From (\ref{eqn:infHiBd}), we conclude that
$\overline{H}_i\leq H(X_i),i=1,2$.  
\end{IEEEproof}
}
To present the second result we set 
\begin{align*}
&\underline{H(\wt{M}_1^{(\infty)}) 
  +{H}(\wt{M}_2^{(\infty)})}
\notag\\
&\defeq\mbox{\rm p-}\liminf_{n\to\infty}
\hugebl
\sum_{i=1,2}\frac{1}{n}\log 
\frac{1}{p_{\wt{M}_i^{(n)}}(\wt{M}_i^{(n)})}
\hugebr.
\end{align*}
For simplicity of notation we set
$$
\underline{H_1+H_2}
=
\underline{H(\wt{M}_1^{(\infty)}) 
  +{H}(\wt{M}_2^{(\infty)})}.
$$
In general we have 
\beq
\underline{H}_1+\underline{H_2}
\leq \underline{H_1+H_2}\leq 
\underline{H}_1+\overline{H_2}.
\label{eqn:EntInfIeq}
\eeq
We have the following property, which is related 
to the quantities $\underline{H_1+H_2}$ and $\underline{I}$.  
\begin{property}\label{pr:InfSpecLmb}$\quad$
\begin{itemize}
\item[a)] We have the following:  
\begin{align}
&\underline{I}=
\underline{I}(\wt{M}_1^{(\infty)};
              \wt{M}_2^{(\infty)})
\notag\\
&=\underline{H_1+H_2}
 -H(X_1X_2)\geq 0.
\label{eqn:MuInfoOne}
\end{align}

\item[b)]We consider the case where     
$
 \underline{H}({M}_i^{(\infty)}) 
=\overline{H}({M}_i^{(\infty)})$ 
for $i=1\mbox{ or }2.$
In this case by Property \ref{pr:InfSpecLm} part c),
we have that for $i=1\mbox{ or }2,$
\begin{align}
\overline{H}_i&=\overline{H}(\wt{M}_i^{(\infty)})
 =\overline{H}({M}_i^{(\infty)}) 
\notag\\
&=\underline{H}({M}_i^{(\infty)})
 =\underline{H}(\wt{M}_i^{(\infty)})
 =\underline{H}_i.
\label{eqn:PrCEqB} 
\end{align}
From (\ref{eqn:EntInfIeq}) and (\ref{eqn:PrCEqB}), we have
$\underline{H_1+H_2}
 =\underline{H}_1+\underline{H_2}, 
$
implying that   
\begin{align}
&\underline{I}=\underline{I}(\wt{M}_1^{(\infty)};
   \wt{M}_2^{(\infty)})
\notag\\
& =\underline{H}_1+\underline{H}_2-H(X_1X_2)\geq 0.
\label{eqn:MuInfoThr}
\end{align}
\item[c)]We have the following:
\begin{align}
 \underline{H_{1}+H_{2}}&=
\underline{H(\wt{M}_1^{(\infty)}) 
         +{H}(\wt{M}_2^{(\infty)})}
\notag\\
&=\underline{H({M}_1^{(\infty)}) 
  +{H}({M}_2^{(\infty)})}.
 \label{eqn:SumEntSpEq} 
\end{align}
\end{itemize}
\end{property}

Proof of Property \ref{pr:InfSpecLmb} 
parts a) and c) is given in Appendix \ref{apd:ProofPrInfSpecb}. 
\newcommand{\ProofPrInfSpecB}{
\subsection{
Proof of Property \ref{pr:InfSpecLmb}
}\label{apd:ProofPrInfSpecb}

In this appendix we prove Property 
\ref{pr:InfSpecLmb}. 
\begin{IEEEproof}
[Proof of Property \ref{pr:InfSpecLmb} part a)]
On lower and upper bounds of $\underline{I}$, we 
have
\begin{align}
&0\MLeq{a}\underline{I}=\underline{I}(
\wt{M}_1^{(\infty)};
\wt{M}_2^{(\infty)})
\notag\\
&=\mbox{\rm p-}\liminf_{n\to\infty}
\hugebl
\sum_{i=1,2}
\frac{1}{n}\log 
\frac{1}
{p_{\wt{M}_i^{(n)}}(\wt{M}_i^{(n)} )}
\notag\\
& \qquad -  
\frac{1}{n}\log 
 \frac{1}
 {p_{\wt{M}_1^{(n)}\wt{M}_2^{(n)}} 
    (\wt{M}_1^{(n)}\wt{M}_2^{(n)})}
\hugebr
\notag\\
&\MEq{b}\mbox{\rm p-}\liminf_{n\to\infty}
\hugebl
\sum_{i=1,2}
\frac{1}{n}\log \frac{1}
{p_{\wt{M}_i^{(n)}}(\wt{M}_i^{(n)})}
\notag\\
& \qquad -  
\frac{1}{n}\log 
 \frac{1} {p_{\lrvctxone\lrvctxtwo} 
    (\wt{\rvcx}_1,\wt{\rvcx}_2)}
\hugebr.
\label{eqn:infIOne}
\end{align}
Step (a) follows from a well known 
result on $\underline{I}$, e.g. \cite{Han98InfSpec}.
Step (b) follows from 
Property \ref{pr:InfSpecLm} part a).
From (\ref{eqn:infIOne}), we have the following 
two bounds:
\begin{align}
 0<&\underline{I} 
   \leq \underline{H_1+H_2}-
    \underline{H}(\wt{X}_1^{\infty}\wt{X}_2^{\infty}),
\label{eqn:UpBinfI}\\
 &\underline{I}\geq 
   \underline{H_1+H_2}-
   \overline{H}(\wt{X}_1^{\infty}\wt{X}_2^{\infty}).
\label{eqn:LoBinfI}
 \end{align}
 On the other hand, by Property \ref{pr:InfSpecLm} 
 part b), we have
 \begin{align}
    \overline{H}(\wt{X}_1^{\infty}\wt{X}_2^{\infty})
  =\underline{H}(\wt{X}_1^{\infty}\wt{X}_2^{\infty})
  =H(X_1X_2).
 \label{eqn:HOneTwo}
 \end{align}
From 
(\ref{eqn:UpBinfI}), (\ref{eqn:LoBinfI}),
and (\ref{eqn:HOneTwo}), we have 
the bound (\ref{eqn:MuInfoOne}) in the part a).
\end{IEEEproof}

We next prove the part c). We have 
the following lemma, which is a generalization 
of Lemma \ref{lm:IfSpLmb}.  
\begin{lemma} \label{lm:IfSpLmc}
For each $i=1,2$, we have 
\begin{align}
&\quad
\Pr\left\{
\sum_{i=1}^2\frac{1}{n}
\log\frac{1}
{p_{{M}_i^{(n)}} 
   ({M}_i^{(n)} )}
  \leq a-\tau 
 \right\}-\nu_n(\gamma,\vep)
\notag\\ 
&\quad -\frac{2}{n\tau}
 \log\left(\frac{{\rm e}^{2{\rm e}^{-1}}}
    {1-\nu_n(\gamma,\varepsilon)}\right)
\notag\\    
&\leq \Pr\left\{
\sum_{i=1}^2\frac{1}{n}\log\frac{1}
{p_{\wt{M}_i^{(n)}} 
   (\wt{M}_i^{(n)})}
  \leq a \right\}  
\label{eqn:IfSpPrPartCTwob}\\
&\leq
\Pr\left\{\sum_{i=1}^2
\frac{1}{n}\log\frac{1}
{p_{{M}_i^{(n)}} 
   ({M}_i^{(n)} )}
  \leq a+\tau 
 \right\}
\notag\\ 
&\quad \times [1-\nu_n(\gamma,\vep)]^{-1}+\frac{2}{n\tau}
 \log\left(\frac{{\rm e}^{2{\rm e}^{-1}}}
    {1-\nu_n(\gamma,\varepsilon)}\right).
\label{eqn:IfSpPrPartCThrb}
\end{align}
\end{lemma}
\begin{IEEEproof}
We set
\begin{align*}
Q_1\defeq\Pr\left\{
\sum_{i=1}^2
\frac{1}{n}\left|\log\frac{p_{\wt{M}_i^{(n)}}(\wt{M}_i^{(n)})}
{p_{{M}_i^{(n)}}(\wt{M}_i^{(n)})}\right|
  \geq \tau \right\}, 
\notag\\
%
Q_2\defeq \Pr\left\{
\sum_{i=1}^2\frac{1}{n}\log\frac{1}
{p_{\wt{M}_i^{(n)}} 
   (\wt{M}_i^{(n)})}
  \leq a \right\}.
\end{align*}
On upper bounds of $Q_1$, we have the following 
chain of inequalities:
\begin{align}
& Q_1 \MLeq{a} 
\sum_{i=1}^2\frac{1}{n\tau}{\rm E}\left[
\left|\log\frac{p_{\wt{M}_i^{(n)}}(\wt{M}_i^{(n)})}
{p_{{M}_i^{(n)}}(\wt{M}_i^{(n)})}\right|
\right]
\notag\\
&\MLeq{b}
\frac{2}{n\tau}
 \log\left(\frac{{\rm e}^{2{\rm e}^{-1}}}
    {1-\nu_n(\gamma,\varepsilon)}\right).
\label{eqn:UpbPoneb}    
\end{align}
Step (a) follows from the Markov inequality.
Step (b) follows from (\ref{eqn:UpbPzero}) and (\ref{eqn:UpbPone})
in the proof of Lemma \ref{lm:IfSpLmb}. 
On upper bounds of $Q_2$, we have the following 
chain of inequalities:  
\begin{align}
&Q_2\leq \Pr\left\{
\sum_{i=1}^2\frac{1}{n}\log\frac{1}
{p_{{M}_i^{(n)}} 
      (\wt{M}_i^{(n)})}
  \leq a+\tau \right\}  
\notag\\
&\quad  +\Pr\left\{\sum_{i=1}^2
\frac{1}{n} \log\frac{p_{{M}_i^{(n)}}(\wt{M}_i^{(n)})}
{p_{\wt{M}_i^{(n)}}(\wt{M}_i^{(n)})}
 \leq -\tau \right\} 
\notag\\
&\leq
\Pr\Biggl\{
\sum_{i=1}^2
\frac{1}{n}\log\frac{1}
{p_{{M}_i^{(n)}} 
   ({M}_i^{(n)})}
\notag\\ 
&\qquad  \quad 
 \leq a+\tau
\Biggr| (\rvcxone,\rvcxtwo)
\in \wt{\cal D}_{\gamma }^{(n)}\Biggr\}
+Q_1.
\label{eqn:PrLmbOneb} 
\end{align}
On the first term in (\ref{eqn:PrLmbOneb}), 
we have the following:
\begin{align}
&  \Pr\left\{\left.
\sum_{i=1}^2\frac{1}{n}\log\frac{1}
{p_{{M}_i^{(n)}} 
   ({M}_i^{(n)})}
  \leq a+\tau
\right| (\rvcxone,\rvcxtwo)\in
        \wt{\cal D}_{\gamma }^{(n)}\right\}
\notag\\
& \leq
\Pr\left\{
\sum_{i=1}^2\frac{1}{n}\log\frac{1}
{p_{{M}_i^{(n)}} 
   ({M}_i^{(n)} )}
  \leq a+\tau 
 \right\}
 \notag\\
& \quad \times [1-\nu_n(\gamma,\vep)]^{-1}.
\label{eqn:PrLmbTwob}  
\end{align}
Combining (\ref{eqn:UpbPoneb}),
          (\ref{eqn:PrLmbOneb}),
     and  (\ref{eqn:PrLmbTwob}),       
we obtain 
the bound 
(\ref{eqn:IfSpPrPartCThrb}) in Lemma \ref{lm:IfSpLmc}.  
On lower bounds of $Q_2$, we have the following 
chain of inequalities:  
\begin{align}
&Q_2\geq \Pr\left\{
\sum_{i=1}^2 \frac{1}{n}\log\frac{1}
{p_{{M}_i^{(n)}} 
      (\wt{M}_i^{(n)})}
  \leq a-\tau \right\}  
\notag\\
&\quad  -\Pr\left\{
\sum_{i=1}^2\frac{1}{n} \log\frac{p_{{M}_i^{(n)}}(\wt{M}_i^{(n)})}
{p_{\wt{M}_i^{(n)}}(\wt{M}_i^{(n)})}
 \geq \tau \right\} 
\notag\\
&\geq
\Pr\Biggl\{
\sum_{i=1}^2 \frac{1}{n}\log\frac{1}
{p_{{M}_i^{(n)}} 
   ({M}_i^{(n)})}
\notag\\
&\qquad\quad 
  \geq a+\tau
\Biggr| (\rvcxone,\rvcxtwo)\in
        \wt{\cal D}_{\gamma }^{(n)}\Biggr\}
-Q_1.
\label{eqn:PrLmbThrb} 
\end{align}
On the first term in (\ref{eqn:PrLmbThrb}), 
we have the following:  
\begin{align}
& \Pr\left\{\left.
\sum_{i=1}^2\frac{1}{n}\log\frac{1}
{p_{{M}_i^{(n)}} 
   ({M}_i^{(n)})}
  \leq a-\tau
\right| (\rvcxone,\rvcxtwo)\in
        \wt{\cal D}_{\gamma }^{(n)}\right\}
\notag\\
&\geq \Pr\left\{
\sum_{i=1}^2\frac{1}{n}\log\frac{1}
{p_{{M}_i^{(n)}} 
   ({M}_i^{(n)} )}
  \leq a-\tau\right\}
\notag\\
&\quad 
-\nu_n(\gamma,\vep)
\label{eqn:PrLmbFoub}.
\end{align}
Combining (\ref{eqn:UpbPoneb}),
          (\ref{eqn:PrLmbThrb}),
     and  (\ref{eqn:PrLmbFoub}),       
we obtain 
the bound  
(\ref{eqn:IfSpPrPartCTwob}) 
in Lemma \ref{lm:IfSpLmc}.
\end{IEEEproof}

\begin{IEEEproof}
[Proof of Property \ref{pr:InfSpecLmb} part c)]
For simplicity of notation we 
set
\begin{align*}
        A &\defeq
\underline{H({M}_1^{(\infty)}) 
  +{H}({M}_2^{(\infty)})},
\\
\wt{A} &\defeq
\underline{H(\wt{M}_1^{(\infty)}) 
         +{H}(\wt{M}_2^{(\infty)})}.
\end{align*}
In the bound 
(\ref{eqn:IfSpPrPartCTwob}) in Lemma \ref{lm:IfSpLmc}, 
we choose $a=\wt{A}-\tau$ 
and let $n\to \infty$ in this bound. Then we have 
\begin{align} 
&\lim_{n\to\infty}
\Pr\left\{
\sum_{i=1}^2\frac{1}{n}\log\frac{1}
{p_{{M}_i^{(n)}} 
   ({M}_i^{(n)} )}
  \leq \wt{A} -2\tau 
 \right\}
\notag\\ 
&\leq \lim_{n\to\infty}\Pr\left\{
\sum_{i=1}^2\frac{1}{n}\log\frac{1}
{p_{\wt{M}_i^{(n)}} 
   (\wt{M}_i^{(n)})}
  \leq \wt{A}-\tau \right\}
\notag\\
& \MEq{a}0.
\label{eqn:PrCProofOneb}
\end{align}
Step (a) follows from the definition of $\wt{A}$.  
The bound (\ref{eqn:PrCProofOneb}) implies that
${A}-2\tau \leq \wt{A}$. Similarly, in the bound 
(\ref{eqn:IfSpPrPartCThrb}) in Lemma \ref{lm:IfSpLmc}, 
we choose $a=A-2\tau$ 
and let $n\to \infty$ in this bound. Then we have 
\begin{align} 
&\lim_{n\to\infty}
\Pr\left\{
\sum_{i=1}^2\frac{1}{n}\log\frac{1}
{p_{\wt{M}_i^{(n)}} 
   (\wt{M}_i^{(n)} )}
  \leq A -2\tau 
 \right\}
\notag\\ 
&\leq \lim_{n\to\infty}\Pr\left\{
\sum_{i=1}^2\frac{1}{n}\log\frac{1}
{p_{{M}_i^{(n)}} 
   ({M}_i^{(n)})}
  \leq A-\tau \right\}
\notag\\
&\MEq{a}0.
\label{eqn:PrCProofTwob}
\end{align}
Step (a) follows from the definition of  
$A$. 
The bound (\ref{eqn:PrCProofTwob}) implies that
$\wt{A}-2\tau \leq A.$ 
Hence we have the bound
$
A-2\tau \leq
\wt{A}\leq A+2\tau. 
$
Since $\tau>0$ can be taken arbitrary small, 
we have $A=\wt{A}$. 
\end{IEEEproof}
}
We set 
$$
2\sigma \defeq \underline{H_1+H_2}-
(\underline{H}_1+\underline{H}_2).
$$
Since 
$\underline{H_1+H_2}\geq \underline{H}_1+\underline{H}_2$,
$\sigma$ is nonnegative. For each $i=1,2$, we set 
$\underline{H}_{i,\sigma}\defeq \underline{H}_i+\sigma$. 
Note here that  
\beq
\underline{H}_1+\underline{H}_2\leq \underline{H_1+H_2}
\leq 
\min_{i=1,2}\left\{\overline{H}_i+\underline{H}_{3-i}
\right\}.
\label{eqn:HoneHtwoBouds}
\eeq
From (\ref{eqn:HoneHtwoBouds}), we 
have that for $\sigma>0$ and for each $i=1,2$,
\beq
\underline{H}_i< 
\underline{H}_{i,\sigma} < 
\underline{H}_{i,\sigma}+\sigma \leq \overline{H}_i.
\label{eqn:sigma}
\eeq 
Set  
\begin{align*}
 \theta_{n}(\sigma,\gamma)
 &\defeq
\Pr\Biggl\{
  \frac{1}{n}
    \log\frac{1}{p_{\wt{M}_i^{(n)}}(\wt{M}_i^{(n)})}
    \leq\underline{H}_{i,\sigma}
     -{\ts \frac{1}{2}}\gamma,
\\
& \qquad \quad \mbox{ for }i=1\mbox{ or }2    
\Biggr\}.
\end{align*}
When $\sigma=0$, we write $\theta_n(0,\gamma)$ 
as $\theta_n(\gamma)$. For the above quantity 
we have the following lemma.
\begin{lemma}\label{lm:LimTheta}
When $\sigma=0$, we have 
\begin{align}
  \lim_{n\to\infty} \theta_n(\gamma)=0.
\label{eqn:limthetaOne}  
\end{align}
When $\sigma>0$, we assume Condition \ref{cond:CondThrA}.
We further assume that $p_{X_1X_2}$ is indecomposable. 
Under those assumptions,
$\exists \{k_n\}_{n\geq 1}$ and 
$\exists \omega \in (0,1)$ such that 
$\forall \gamma\in (0,2\sigma]$, 
\begin{align}
& \limsup_{n\to\infty} \theta_{k_n}
(\sigma,\gamma)\leq {\theTa}
\in (0,1).
\label{eqn:LimThetaUpBd}
\end{align}
\end{lemma}

Proof of Lemma \ref{lm:LimTheta} is given 
in Appendix \ref{apd:PrfLimTheta}. 
\newcommand{\PrfLimTheta}{
\subsection{
Proof of Lemma \ref{lm:LimTheta}
}\label{apd:PrfLimTheta}

In this appendix we prove Lemma \ref{lm:LimTheta}.
We first provide a lemma necessary for the proof.

\begin{lemma}\label{lm:LmforCompTheta}
We have the following: 
\begin{align}  
&\Pr\left\{
\bigcup_{i=1}^2\left\{\frac{1}{n}\log\frac{1}
{p_{\wt{M}_i^{(n)}} 
   (\wt{M}_i^{(n)})}
  \leq a \right\}\right\}  
\notag\\
&\leq
\Pr\left\{\bigcup_{i=1}^2\left\{
\frac{1}{n}
\log\frac{1}
{p_{{M}_i^{(n)}} 
   ({M}_i^{(n)} )}
  \leq a+\tau 
 \right\}\right\}
\notag\\ 
&\quad \times {[1-\nu_n(\gamma,\vep)]^{-1}} +\frac{2}{n\tau}
 \log\left(\frac{{\rm e}^{2{\rm e}^{-1}}}
    {1-\nu_n(\gamma,\varepsilon)}\right).
\label{eqn:IfSpBdTwo}
\end{align}
\end{lemma}

\begin{IEEEproof}
We set
\begin{align*}
\wt{Q}&\defeq \Pr\left\{
\bigcup_{i=1}^2\left\{\frac{1}{n}\log\frac{1}
{p_{\wt{M}_i^{(n)}} 
   (\wt{M}_i^{(n)})}
  \leq a \right\} \right\}.
\end{align*}
On upper bounds of $\wt{Q}$, we have the following 
chain of inequalities:
\begin{align}
&\wt{Q}\leq \Pr\left\{
\bigcup_{i=1}^2\left\{
\frac{1}{n}\log\frac{1}
{p_{{M}_i^{(n)}} 
      (\wt{M}_i^{(n)})}
  \leq a+\tau \right\}\right\}  
\notag\\
&\qquad +\sum_{i=1}^2\Pr\left\{
\frac{1}{n} \log\frac{p_{{M}_i^{(n)}}(\wt{M}_i^{(n)})}
{p_{\wt{M}_i^{(n)}}(\wt{M}_i^{(n)})}
 \leq -\tau \right\} 
\notag\\
&\leq
\Pr\Biggl\{
\bigcup_{i=1}^2\Biggl\{
\frac{1}{n}\log\frac{1}
{p_{{M}_i^{(n)}} 
   ({M}_i^{(n)})}
\notag\\ 
&\qquad  \qquad 
 \leq a+\tau
\Biggr\}\Biggr| (\rvcxone,\rvcxtwo)
\in \wt{\cal D}_{\gamma }^{(n)}\Biggr\}
+\sum_{i=1}^2{P}_{1,i}.
\label{eqn:PrLmbOneC} 
\end{align}
Here ${P}_{1,i},i=1,2$ are the quantities 
defined by (\ref{eqn:DefPonEi}) 
in the proof of Lemma \ref{lm:IfSpLmb}. 
The quantities ${P}_{1,i},i=1,2$ have upper bounds given by (\ref{eqn:UpbPone}) in the proof of Lemma \ref{lm:IfSpLmb}.
On the first term in (\ref{eqn:PrLmbOneC}), 
we have the following:
\begin{align}
&\Pr\Biggl\{
\bigcup_{i=1}^2\Biggl\{
\frac{1}{n}\log\frac{1}
{p_{{M}_i^{(n)}} 
   ({M}_i^{(n)})}
\notag\\ 
&\qquad  \qquad 
 \leq a+\tau
\Biggr\}\Biggr| (\rvcxone,\rvcxtwo)
\in \wt{\cal D}_{\gamma }^{(n)}\Biggr\}
\notag\\
& \leq
\Pr\left\{
\bigcup_{i=1}^2\left\{\frac{1}{n}\log\frac{1}
{p_{{M}_i^{(n)}} 
   ({M}_i^{(n)} )}
  \leq a+\tau 
 \right\}\right\}
 \notag\\
& \quad \times [1-\nu_n(\gamma,\vep)]^{-1}.
\label{eqn:PrLmbTwoC}  
\end{align}
Combining (\ref{eqn:UpbPone}),
          (\ref{eqn:PrLmbOneC}),
     and  (\ref{eqn:PrLmbTwoC}),       
we obtain the bound (\ref{eqn:IfSpBdTwo}) 
in Lemma \ref{lm:LmforCompTheta}.  
\end{IEEEproof}

We next define several quantities necessary 
for the proof of Lemma 
\ref{lm:LimTheta}.
Set  
\begin{align*}
 \vartheta_{n}(\sigma,\gamma)
 &\defeq
\Pr\Biggl\{
  \frac{1}{n}
  \log\frac{1}{p_{{M}_i^{(n)}}({M}_i^{(n)})}  
\leq \underline{H}_{i,\sigma}-{\ts \frac{1}{4}}\gamma,
\\
& \qquad \quad \mbox{ for }i=1\mbox{ or }2    
\Biggr\}.
\end{align*}
For each $i=1,2$, we set 
\begin{align*}
&{\cal F}_i \defeq 
\left\{\vcxi: \frac{1}{n}\log 
\frac{1}{p_{\phi_i^{(n)}({\lrvcxi)}}(\phi_i^{(n)}(\vcxi))}
\leq \underline{H}_{i,\sigma}-{\ts\frac{1}{4}}\gamma
\right\},
\\
& \vartheta_{i,n}=\vartheta_{i,n}(\sigma,\gamma)
\defeq 
\Pr\left\{ \rvcxi \in {\cal F}_i \right\}.
\end{align*}
Furthermore, we define
\begin{align*}
&{\vartheta}_{12,n}(\gamma)
\\
&\defeq \Pr\left\{
\sum_{i=1}^2
\frac{1}{n}
    \log\frac{1}{p_{{M}_i^{(n)}}({M}_i^{(n)})}
\leq \underline{H_1+H_2}-{\ts \frac{1}{2}}\gamma
\right\}. 
\end{align*}
On $\vartheta_{n}$, $\vartheta_{12,n}$, and $\vartheta_{i,n},i=1,2$, we 
have the following lemma:
\begin{lemma}\label{lm:LimThetaIOneTwo}
We have the followings:
\begin{itemize}
\item[a)]For any $\sigma>0$ and $\gamma > 0$, 
\begin{align}
\limsup_{n\to\infty}\theta_n(\sigma,\gamma)
 &\leq \limsup_{n\to\infty}\vartheta_n(\sigma,\gamma).
\label{eqn:LimVarTheta}
\\
\lim_{n\to\infty}{\vartheta}_{12,n}(\gamma)&=0. 
\label{eqn:LimtrThetaUpdTwo}
\end{align} 
\item[b)] We assume that 
$F_{M_1^{(\infty)}}(a)$ 
exists for $a\geq 0$. Then $\exists 
\varsigma\in $ $(0,\frac{1}{2}]$ such that 
$\forall \sigma>0$ and 
$\forall \gamma\in (0,2\sigma]$, we have  
 $\vartheta_{1,n}(\sigma,\gamma)\in 
          [\varsigma,
 \overline{\varsigma}]$ for infinitely many $n$.
\end{itemize}
\end{lemma}

\begin{IEEEproof} 
We first prove the part a). By Lemma \ref{lm:LmforCompTheta}, we have
\begin{align}
 \theta_n(\sigma,\gamma)
&\leq \vartheta_n(\sigma,\gamma)
 [1-\nu_n(\gamma,\vep)]^{-1}
\notag\\ 
&\quad 
+\frac{8}{n\gamma}
 \log\left(\frac{{\rm e}^{2{\rm e}^{-1}}}
    {1-\nu_n(\gamma,\varepsilon)}\right).
\label{eqn:VarthetaBd}   
\end{align}
By letting $n\to\infty $ in both sides of (\ref{eqn:VarthetaBd}), we have the bound (\ref{eqn:LimVarTheta}) in the part a). 
Since by Property \ref{pr:InfSpecLmb} part c), 
$$ 
\underline{H({M}_1^{(\infty)}) 
  +{H}({M}_2^{(\infty)})}=\underline{H_1+H_2},
$$
we have (\ref{eqn:LimtrThetaUpdTwo}) 
in the part a). 

We next prove the part b). 
We first recall the following definition 
of $\vartheta_{1,n}=\vartheta_{1,n}(\sigma,\gamma)$:   
\begin{align*}
 &\vartheta_{1,n}=\vartheta_{1,n}(\sigma,\gamma)
 ={\Pr}\{\rvcxone \in {\cal F}_1\}
 \\
 &=
\Pr\Biggl\{
\frac{1}{n}
 \log\frac{1}{p_{{M}_1^{(n)}}({M}_1^{(n)})} 
\leq \underline{H}_{1}+\sigma-{\ts \frac{1}{4}}\gamma 
\Biggr\},
\end{align*}
from which we have 
\begin{align}
& \vartheta_{1,n} 
\geq  
\Pr\Biggl\{
\frac{1}{n}\log \frac{1}{p_{{M}_1^{(n)}}({M}_1^{(n)})} 
\leq \underline{H}_{1}+{\ts \frac{1}{2}}\sigma
\Biggr\},
\label{eqn:varThetaOneLb}\\
& \overline{\vartheta}_{1,n} \geq 
\Pr\Biggl\{
\frac{1}{n}
 \log\frac{1}{p_{{M}_1^{(n)}}({M}_1^{(n)})} 
> \underline{H}_{1} +\sigma \Biggr\}.
\label{eqn:AvarThetaOneLb}
\end{align}
We further note the following: 
\begin{align}
&\underline{H}(M_1^{(\infty)})=\underline{H}_1,
 \overline{H}(M_1^{(\infty)})=\overline{H}_1.
\label{eqn:LemSevenCondTwo}
\end{align}
We prove the part b) by deriving a contradiction 
under the following assumption:
\begin{align}
&\lim_{n \to \infty}\vartheta_{1,n}
 \overline{\vartheta}_{1,n}=0.
\label{eqn:limVarThetaAa}
\end{align}
Under Condition \ref{cond:CondThrA}, i.e.,
the existence of $F_{M_1^{(\infty)}}(a),a\geq 0$, 
 (\ref{eqn:limVarThetaAa}) 
is equivalent to    
\begin{align}
&\lim_{n \to \infty}\vartheta_{1,n}=0
\mbox{ or } \lim_{n \to \infty}\overline{\vartheta}_{1,n}=0.
\label{eqn:limVarThetaA}
\end{align}
If we have the first statement of (\ref{eqn:limVarThetaA}), we let 
$n\to \infty $ in both sides of 
(\ref{eqn:varThetaOneLb})
to obtain the following:  
\begin{align*}
& \lim_{n\to \infty}  
\Pr\Biggl\{
\frac{1}{n}
 \log\frac{1}{p_{{M}_1^{(n)}}({M}_1^{(n)})} 
\leq \underline{H}_{1}+{\ts \frac{1}{2}}\sigma 
\Biggr\}=0, 
\end{align*}
which contradicts the first equality in (\ref{eqn:LemSevenCondTwo}).
If we have the second statement of (\ref{eqn:limVarThetaA}), we let 
$n\to \infty $ in both sides of 
(\ref{eqn:AvarThetaOneLb})
to obtain the following:   
\begin{align}
& \lim_{n\to \infty}  
\Pr\Biggl\{
\frac{1}{n}
 \log\frac{1}{p_{{M}_1^{(n)}}({M}_1^{(n)})} 
> \underline{H}_{1}+\sigma \Biggr\}=0.
\label{eqn:varthetaLB}
\end{align}
Since we have the second equality in 
(\ref{eqn:LemSevenCondTwo}), the bound (\ref{eqn:varthetaLB})
implies $\underline{H}_1+{\sigma}
\geq\overline{H}_1$, which contradicts 
$\underline{H}_1+{\sigma}=\underline{H}_{1,\sigma}<\overline{H}_1$. 
Hence we do not have (\ref{eqn:limVarThetaAa}), implying that 
$\exists a \in (0,\frac{1}{4}]$ such that 
$\vartheta_{1,n}\overline{\vartheta}_{1,n}\geq a$ 
or equivalent to 
\begin{align}
&{\ts \frac{1}{2}}-\sqrt{{\ts \frac{1}{4}}-a} \leq \vartheta_{1,n}   
\leq {\ts \frac{1}{2}}+\sqrt{{\ts \frac{1}{4}}-a}
\label{eqn:BoundOnVarTheta}
\end{align}
for infinitely many $n$. 
By choosing $\varsigma$ 
$\in (0,\frac{1}{2}]$ so that 
it is equal to the first quantity in (\ref{eqn:BoundOnVarTheta}),
we have $\vartheta_{1,n}
\in [\varsigma,\overline{\varsigma}]$ 
for infinitely many $n$.   
\end{IEEEproof} 

 
\begin{IEEEproof}[Proof of Lemma \ref{lm:LimTheta}] 
We first consider the case of $\sigma=0$. In this 
case we have 
$$
\theta_n(\gamma)\leq \sum_{i=1}^2
\theta_{i,n}(\gamma).
$$
Then the bound (\ref{eqn:ThetaZero}) in 
Property \ref{pr:InfSpecLm} part d),
we have 
$$
\lim_{n\to\infty} \theta_n(\gamma)\leq \lim_{n\to\infty}
\sum_{i=1}^2\theta_{i,n}(\gamma)=0.
$$
We next consider the case of $\sigma>0$. In this 
case we assume Condition \ref{cond:CondThrA} 
and $F_{M_1^{(\infty)}}(a)$ exists for $a\geq 0$. 
We further assume that $p_{X_1X_2}$ is indecomposable.
We set 
\begin{align*}
\wt{\vartheta}_n(\sigma,\gamma)  
\defeq & \sum_{i=1}^2
\Pr\left\{\rvcxi         \in {\cal F}_i, 
          \rvcx_{3-i} \notin {\cal F}_{3-i}\right\}.
\end{align*}
On upper bounds of $\vartheta_n(\sigma,\gamma)$,
we have the following chain of inequalities:
\begin{align}
&\vartheta_n(\sigma,\gamma)=   
\Pr\left\{\rvcxi \in {\cal F}_i, i=1\mbox{ or }2   
\right\}
\notag\\
&= \wt{\vartheta}_{n}(\sigma,\gamma) +\Pr\left\{\rvcxi 
 \in {\cal F}_i, i=1,2\right\}
\notag\\
& \MLeq{a} \wt{\vartheta}_{n}(\sigma,\gamma)
           +\vartheta_{12,n}(\gamma).  
\label{eqn:UpVarThetaOne}
\end{align}
Step (a) follows from that 
\begin{align*}
&\rvcx_i \in {\cal F}_i,i=1,2
\\
&\Leftrightarrow \frac{1}{n}\log\frac{1}
{p_{{M}_i^{(n)}} 
   ({M}_i^{(n)} )}\leq H_{i,\sigma}-{\ts \frac{1}{4}}
   \gamma\:\mbox{ for }i=1,2,
\\
&\Rightarrow 
\sum_{i=1}^2
\frac{1}{n}\log\frac{1}{p_{{M}_i^{(n)}} ({M}_i^{(n)} )}
\leq \underline{H_1+H_2}-{\ts \frac{1}{2}}\gamma.
\end{align*}
From (\ref{eqn:LimVarTheta}),
     (\ref{eqn:LimtrThetaUpdTwo}), and
     (\ref{eqn:UpVarThetaOne}), we have
\begin{align}
\limsup_{n\to\infty}\theta_n(\sigma,\gamma)
&\leq \limsup_{n\to\infty}\wt{\vartheta}_n(\sigma,\gamma).
\label{eqn:LimVarThetaB}
\end{align} 
From (\ref{eqn:LimVarThetaB}), we can see that
it suffices to evaluate upper bounds of 
$\wt{\vartheta}_n(\sigma,\gamma)$
to derive the bound (\ref{eqn:LimThetaUpBd})
in Lemma \ref{lm:LimTheta}.  
According to Ahlswede and G\'acs \cite{ahlswede:76b},
under the condition that $p_{X_1X_2}$ 
is indecomposable we have for $i=1,2$,    
\begin{align}
&\Pr\left\{\rvcxi         \in {\cal F}_i, 
           \rvcx_{3-i} \notin {\cal F}_{3-i}\right\}
\notag\\
&\leq [\Pr\left\{\rvcxi
       \in {\cal F}_i\right\}]^{\lambda_i}
     [\Pr\left\{\rvcx_{3-i}
        \notin {\cal F}_{3-i}\right\}]^{\lambda_{3-i}},   
\label{eqn:PbdOne}
\end{align}
where $\lambda_1$ and $\lambda_2$ satisfy
$0<\lambda_1<1$, 
$0<\lambda_2<1$, and 
$\lambda_1+$ $\lambda_2>1$. Those are constants 
which depend only on $p_{X_1X_2}$. On upper bounds of 
$\wt{\vartheta}_n(\sigma,\gamma)$, 
we have the following chain of inequalities:  
\begin{align}
&\wt{\vartheta}_n(\sigma,\gamma)
\MLeq{a}
\sum_{i=1}^2(\vartheta_{i,n}^{\lambda_1+\lambda_2}
)^{\frac{\lambda_i}{\lambda_1+\lambda_2}} 
(\overline{\vartheta}_{3-i,n}
^{\lambda_1+\lambda_2})
^{\frac{\lambda_{3-i}}
  {\lambda_1+\lambda_2}}
\notag\\
 & 
 \MLeq{b}
 \prod_{i=1}^2 \left[
             \vartheta_{i,n}^{\lambda_1+\lambda_2}
+ \overline{\vartheta}_{i,n}^{\lambda_1+\lambda_2}
\right]^{\frac{\lambda_i}{\lambda_1+\lambda_2}}
\notag\\
&\MLeq{c}
  \left[    \vartheta_{1,n}^{\lambda_1+\lambda_2}
+ \overline{\vartheta}_{1,n}^{\lambda_1+\lambda_2}
\right]^{\frac{\lambda_1}{\lambda_1+\lambda_2}}.
\label{eqn:VarThetaUpdTwo}
\end{align}
Step (a) follows from (\ref{eqn:PbdOne}). 
Step (b) follows from the H\"older inequality.
Step (c) follows from 
$
             \vartheta_{2,n}^{\lambda_1+\lambda_2}
+ \overline{\vartheta}_{2,n}^{\lambda_1+\lambda_2}
\leq 1.
$
On the other hand, by Lemma \ref{lm:LimThetaIOneTwo} part b), 
we have that 
$\exists \{k_n\}_{n\geq 1}$ such that 
$\forall \gamma\in (0,2\sigma]$,
\begin{align}
 \lim_{n\to\infty} \vartheta_{1,k_n}(\sigma,\gamma)
 =\vartheta_{\ast}\in (0,1). 
\label{eqn:LimtrThetaUpdOneB} 
\end{align}
We choose $\omega \in (0,1)$ 
so that
$$
\theTa=
\left[  \vartheta_{\ast}^{\lambda_1+\lambda_2}
+ \overline{\vartheta}_{\ast}^{\lambda_1+\lambda_2}
\right]^{\frac{\lambda_1}{\lambda_1+\lambda_2}}.
$$
Then, from 
(\ref{eqn:VarThetaUpdTwo}) and    (\ref{eqn:LimtrThetaUpdOneB}), we have 
\begin{align*}
\limsup_{n\to\infty}\wt{\vartheta}_{k_n}(\sigma,\gamma) 
\leq {\theTa}\in (0,1),
\end{align*}
which together with 
(\ref{eqn:LimVarThetaB}), yields  
the bound (\ref{eqn:LimThetaUpBd}).
\end{IEEEproof}

}

\subsection{Error Probability Estimation}

According to Property \ref{pr:InfSpecLm} part e), we have 
the following two cases on $\underline{I}$:
\begin{itemize}
\item [1)] $\underline{I}>0$: i.e., 
$\underline{H}_{1,\sigma}+\underline{H}_{2,\sigma}> H(X_1X_2)$.
\item [2)] $\underline{I}=0$: i.e., 
$\underline{H}_{1,\sigma}+\underline{H}_{2,\sigma}=H(X_1X_2)$.
\end{itemize}
In the first case of $\underline{I}>0$, we have $\gamma_0>0$ 
such that 
\begin{align}
\underline{I}=  \underline{H}_{1,\sigma}
              + \underline{H}_{2,\sigma}
               -H(X_1X_2) \geq 4\gamma_0.
\label{eqn:GammaZero}
\end{align}
In the following arguments we fix such 
$\gamma_0$. Set 
\begin{align*}
{\cal V}_{\nu,(\underline{H}_{1,\sigma},
               \underline{H}_{2,\sigma})}
\defeq &\{(r_1,r_2):r_i\leq \underline{H}_{i,\sigma},i=1,2, 
\\
&\quad r_1+r_2\geq H(X_1X_2)+\nu\}.    
\end{align*}
Specifically, when $\nu=0$, we omit ``0," in 
the subscript of 
${\cal V}_{0,(\underline{H}_{1,\sigma},
               \underline{H}_{2,\sigma})}$ to simply write 
$ {\cal V}_{(\underline{H}_{1,\sigma},
             \underline{H}_{2,\sigma})}$.              
It is obvious that $\forall\nu\in [0,4\gamma_0]$, 
we have 
$$
{\cal V}_{4\gamma_0,(\underline{H}_{1,\sigma},
               \underline{H}_{2,\sigma})}
\subseteq {\cal V}_{\nu,(\underline{H}_{1,\sigma},
                         \underline{H}_{2,\sigma})}              
\subseteq {\cal V}_{(\underline{H}_{1,\sigma},
               \underline{H}_{2,\sigma})}.              
$$
In the second case of $\underline{I}=0$. We 
consider the following set:
\begin{align*}
\wt{\cal V}_{\nu, (\underline{H}_{1,\sigma},
                   \underline{H}_{2,\sigma})} 
=\{(r_1,r_2): r_i \geq \underline{H}_{i,\sigma}+\nu,i=1,2\}.
\end{align*}
The following lemma provides an upper bound 
of 
$
\Theta_n\bigl(\gamma,$ $r_1^{(n)},$ 
$r_2^{(n)}\bigr).
$
\begin{lemma}\label{lm:PrUbLm}
We have the following:  
\begin{itemize}
 \item[a)] We consider the case of 
 $
 \underline{I}=\underline{H}_{1,\sigma}
 +\underline{H}_{2,\sigma}-H(X_1X_2)>0.
 $ 
 In this case we choose $\gamma_0$ so 
 that $4\gamma_0\in (0,\underline{I}]$ 
 as stated in (\ref{eqn:GammaZero}). 
 Then $\forall \gamma \in (0,\gamma_0]$ and  
      $\forall(r_1^{(n)},r_2^{(n)})
       \in {\cal V}_{3\gamma,(\underline{H}_{1,\sigma},
                              \underline{H}_{2,\sigma})}$, we have   
\begin{align}
&\Theta_n\left(\gamma,r_1^{(n)},r_2^{(n)}\right)\leq \eta_n(\gamma)+\theta_{n}(\sigma,\gamma).
\label{eqn:ThetaUpbdOne}  
\end{align}
 \item[b)] 
  We consider the case of 
  $ \underline{H}_{1,\sigma}
   +\underline{H}_{2,\sigma}=H(X_1X_2)$. 
  In this case we choose 
  $\gamma>0$ sufficiently small. 
  Then for any $(r_1^{(n)},r_2^{(n)})$
  satisfying $(r_1^{(n)},r_2^{(n)})
  \in \wt{\cal V}_{3\gamma,(\overline{H}_{1,\sigma},\overline{H}_{2,\sigma})}$,
   we have 
\begin{align}
&\Theta_n\left(\gamma,r_1^{(n)},r_2^{(n)}\right)\leq \eta_n(\gamma)+\theta_{n}(\sigma,\gamma).
\label{eqn:ThetaUpbdTwo}
\end{align}
\end{itemize}
\end{lemma}

\begin{IEEEproof}
By definition, for $i=1,2$, we have the following:
\begin{align}
&(\wt{M}_1^{(n)},\wt{M}_2^{(n)})\in
 {\cal U}^{(n)}_{i,\gamma}
\notag\\
&\Leftrightarrow 
\frac{1}{n}
 \log\frac{1}
  {p_{\wt{M}_{i}^{(n)}|\wt{M}_{3-i}^{(n)}}(\wt{M}_i^{(n)}|\wt{M}_{3-i}^{(n)})}
\geq {r}_{i}^{(n)}-\gamma.
\label{eqn:PrBdThr}
\end{align}
For $i=3$, we have the following.  
\begin{align}
&(\wt{M}_1^{(n)},\wt{M}_2^{(n)})\in
 {\cal U}^{(n)}_{3,\gamma}
\notag\\
&\Leftrightarrow 
\frac{1}{n}
 \log\frac{1}
  {p_{\wt{M}_1^{(n)}\wt{M}_2^{(n)}}(\wt{M}_1^{(n)},\wt{M}_2^{(n)})}
\geq {r}_1^{(n)}+{r}_2^{(n)}-\gamma.
 \label{eqn:PrBdFou}
\end{align}
We first consider the general case of 
$\underline{H}_{1,\sigma}+\underline{H}_{2,\sigma}>H(X_1X_2)$.
For $i=1,2$, we have the following:
\begin{align}
&(\wt{M}_1^{(n)},\wt{M}_2^{(n)})\in
 {\cal U}^{(n)}_{i,\gamma}
\notag\\
&
\MLRarrow{a}
\frac{1}{n}
   \log\frac{1}{p_{\wt{\lrvcx}_1\wt{\lrvcx}_2}(
     \wt{\rvcx}_1 ,\wt{\rvcx}_2)}
-[ H(X_1X_2) +{\ts\frac{3}{2}}\gamma] 
\notag\\
&\qquad+\underline{H}_{3-i,\sigma}
  -{\ts\frac{1}{2}}\gamma-\frac{1}{n}
   \log\frac{1}
  {p_{\wt{M}_{3-i}^{(n)}} 
     (\wt{M}_{3-i}^{(n)})}
\notag\\
&\quad \qquad \geq r_{i}^{(n)}
+\underline{H}_{3-i,\sigma}-[H(X_1X_2)+3\gamma]\MGeq{b}0
\notag\\
&\Rightarrow
\frac{1}{n}
   \log\frac{1}{p_{\wt{\lrvcx}_1\wt{\lrvcx}_2}(
     \wt{\rvcx}_1 ,\wt{\rvcx}_2)}
 \geq[ H(X_1X_2) +{\ts\frac{3}{2}}\gamma]
\notag\\
&\quad\mbox{ or }\underline{H}_{3-i,\sigma}
  -{\ts\frac{1}{2}}\gamma\geq \frac{1}{n}
   \log\frac{1}
  {p_{\wt{M}_{3-i}^{(n)}} 
     (\wt{M}_{3-i}^{(n)})}.
\label{eqn:ThetaBdOne}     
\end{align}
Step (a) follows from the equality (\ref{eqn:PrBdThr})
and the equality (\ref{eqn:PrCKtoPrXb}) 
in Property \ref{pr:InfSpecLm} part a).  
Step (b) follows from 
$(r_1^{(n)},r_2^{(n)})
\in {\cal V}_{3\gamma,
    (\underline{H}_{1,\sigma},\underline{H}_{2,\sigma})}.
$
For $i=3$, we have the following:  
\begin{align}
& (\wt{M}_1^{(n)},\wt{M}_2^{(n)})\in
 {\cal U}^{(n)}_{3,\gamma}
\notag\\
&\MLRarrow{a}
\frac{1}{n}
   \log\frac{1}{p_{\wt{\lrvcx}_1\wt{\lrvcx}_2}(
     \wt{\rvcx}_1 ,\wt{\rvcx}_2)}
-[H(X_1X_2) +{\ts\frac{3}{2}}\gamma] 
\notag\\
&\quad \qquad \geq {r}_1^{(n)}+{r}_2^{(n)}
-[H(X_1X_2)+{\ts\frac{5}{2}}\gamma]
\MGeq{b}{\ts\frac{1}{2}}\gamma\geq 0.
\notag\\
& \Rightarrow
  \frac{1}{n}
   \log\frac{1}{p_{\wt{\lrvcx}_1\wt{\lrvcx}_2}(
     \wt{\rvcx}_1 ,\wt{\rvcx}_2)}
\geq H(X_1X_2) +{\ts\frac{3}{2}}\gamma.
\label{eqn:ThetaBdTwo}
\end{align}
Step (a) follows from the equality (\ref{eqn:PrBdFou}).
Step (b) follows from 
$
(r_1^{(n)},r_2^{(n)})
\in {\cal V}_{3\gamma,
    (\underline{H}_{1,\sigma},\underline{H}_{2,\sigma})}.
$
From (\ref{eqn:ThetaBdOne}) and (\ref{eqn:ThetaBdTwo}), we have 
\begin{align}
&(\wt{M}_1^{(n)},\wt{M}_2^{(n)})\in
 \bigcup_{i=1}^3{\cal U}^{(n)}_{i,\gamma}
\notag\\
&\Rightarrow
\frac{1}{n}
   \log\frac{1}{p_{\wt{\lrvcx}_1\wt{\lrvcx}_2}(
     \wt{\rvcx}_1 ,\wt{\rvcx}_2)}
 \geq[ H(X_1X_2) +{\ts\frac{3}{2}}\gamma]
\notag\\
&\quad\mbox{ or for }i=1\mbox{ or }2,\:
\notag\\
& \qquad \underline{H}_{i,\sigma}
  -{\ts\frac{1}{2}}\gamma\geq \frac{1}{n}
   \log\frac{1}
  {p_{\wt{M}_{i}^{(n)}} 
     (\wt{M}_{i}^{(n)})}.
\label{eqn:ThetaBdThr} 
\end{align}
From (\ref{eqn:ThetaBdThr}), we have the bound 
(\ref{eqn:ThetaUpbdOne}) in Lemma \ref{lm:PrUbLm}.  
We next consider the case of 
$\underline{H}_{1,\sigma}
+\underline{H}_{2,\sigma}=H(X_1X_2)$. 
For $i=1,2$, we have the following:
\begin{align}
& (\wt{M}_1^{(n)},\wt{M}_2^{(n)})\in
 {\cal U}^{(n)}_{i,\gamma}
\notag\\
&\MLRarrow{a}
\frac{1}{n}
   \log\frac{1}{p_{\wt{\lrvcx}_1\wt{\lrvcx}_2}(
     \wt{\rvcx}_1 ,\wt{\rvcx}_2)}
-[H(X_1X_2) +{\ts\frac{3}{2}}\gamma] 
\notag\\
&\qquad \quad+\underline{H}_{3-i,\sigma}
-{\ts\frac{1}{2}}\gamma-\frac{1}{n}
   \log\frac{1}
  {p_{\wt{M}_{3-i}^{(n)}} 
     (\wt{M}_{3-i}^{(n)})}
\notag\\
&\quad \qquad \geq {r}_{i}^{(n)}
        -\underline{H}_{i,\sigma}-3\gamma
\MGeq{b}0
\notag\\
&\Rightarrow \frac{1}{n}
   \log\frac{1}{p_{\wt{\lrvcx}_1\wt{\lrvcx}_2}(
     \wt{\rvcx}_1 ,\wt{\rvcx}_2)}
\geq [H(X_1X_2) +{\ts\frac{3}{2}}\gamma]  
\notag\\
&\quad\mbox{ or }
\underline{H}_{3-i,\sigma}
-{\ts\frac{1}{2}}\gamma
\geq \frac{1}{n}
    \log\frac{1}
     {p_{\wt{M}_{3-i}^{(n)}}(\wt{M}_{3-i}^{(n)})}.
\label{eqn:ThetaBdFou}     
\end{align}
Step (a) follows from the equality (\ref{eqn:PrBdThr}).
Step (b) follows from 
$(r_1^{(n)},r_2^{(n)})
 \in \wt{\cal V}_{3\gamma,
     (\underline{H}_{1,\sigma},\underline{H}_{2,\sigma})}.$
For $i=3$, we have the following:  
\begin{align}
& (\wt{M}_1^{(n)},\wt{M}_2^{(n)})\in
 {\cal U}^{(n)}_{3,\gamma}
\notag\\
&\MLRarrow{a}
\frac{1}{n}
   \log\frac{1}{p_{\wt{\lrvcx}_1\wt{\lrvcx}_2}(
     \wt{\rvcx}_1 ,\wt{\rvcx}_2)}
-[ H(X_1X_2)+{\ts\frac{3}{2}}\gamma] 
\notag\\
&\quad \qquad \geq {r}_1^{(n)}+{r}_2^{(n)}
-[\underline{H}_{1,\sigma}+
  \underline{H}_{2,\sigma}+{\ts\frac{5}{2}}\gamma]
\MGeq{b}{\ts\frac{7}{2}\gamma}\geq0
\notag\\
&\Rightarrow
\frac{1}{n}
   \log\frac{1}{p_{\wt{\lrvcx}_1\wt{\lrvcx}_2}(
     \wt{\rvcx}_1 ,\wt{\rvcx}_2)}
\geq H(X_1X_2)+{\ts\frac{3}{2}}\gamma  
\label{eqn:ThetaBdFiv}  
\end{align}
Step (a) follows from the equality (\ref{eqn:PrBdFou}).
Step (b) follows from 
$(r_1^{(n)},r_2^{(n)})
 \in \wt{\cal V}_{3\gamma,
     (\underline{H}_{1,\sigma},\underline{H}_{2,\sigma})}.$
From (\ref{eqn:ThetaBdFou}) and (\ref{eqn:ThetaBdFiv}), 
we have (\ref{eqn:ThetaBdThr}) also in the case of $\underline{I}=0$.
Hence we have the bound (\ref{eqn:ThetaUpbdTwo}) 
in Lemma \ref{lm:PrUbLm}. 
\end{IEEEproof}

\begin{figure}[t]
\centering
\includegraphics[width=0.47\textwidth]{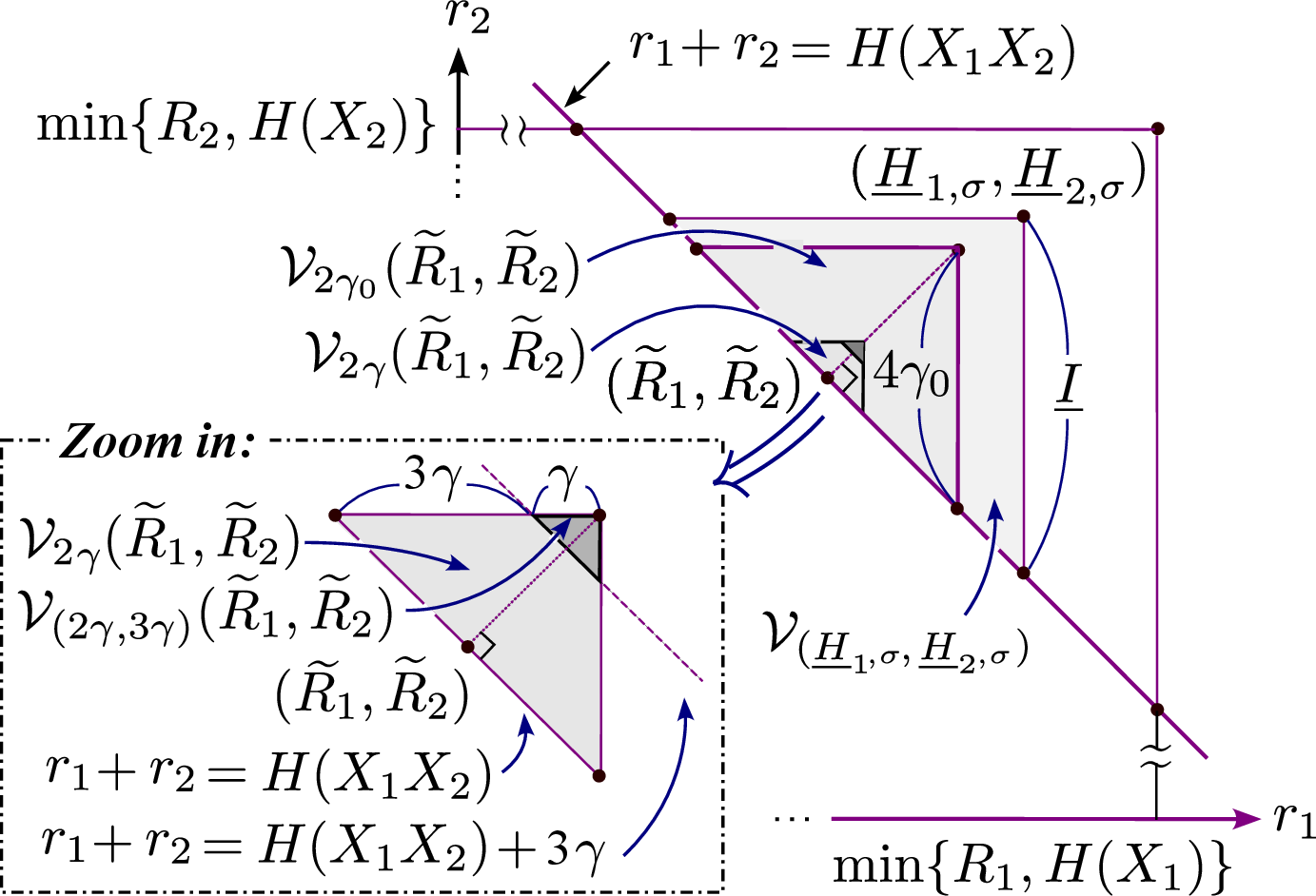}
\caption{
Shapes of the four sets ${\cal V}_{(\underline{H}_{1,\sigma},\underline{H}_{2,\sigma})},
{\cal V}_{2\gamma_0}(\wt{R}_1,\wt{R}_2),
{\cal V}_{2\gamma}(\wt{R}_1,\wt{R}_2),
$ and 
$ {\cal V}_{(2\gamma,3\gamma)}
(\wt{R}_1,\wt{R}_2)$ related to the case of $\underline{I}>0$.
}
\label{fig:VSets}
\end{figure}

\begin{figure}[t]
\centering
\includegraphics[width=0.35\textwidth]{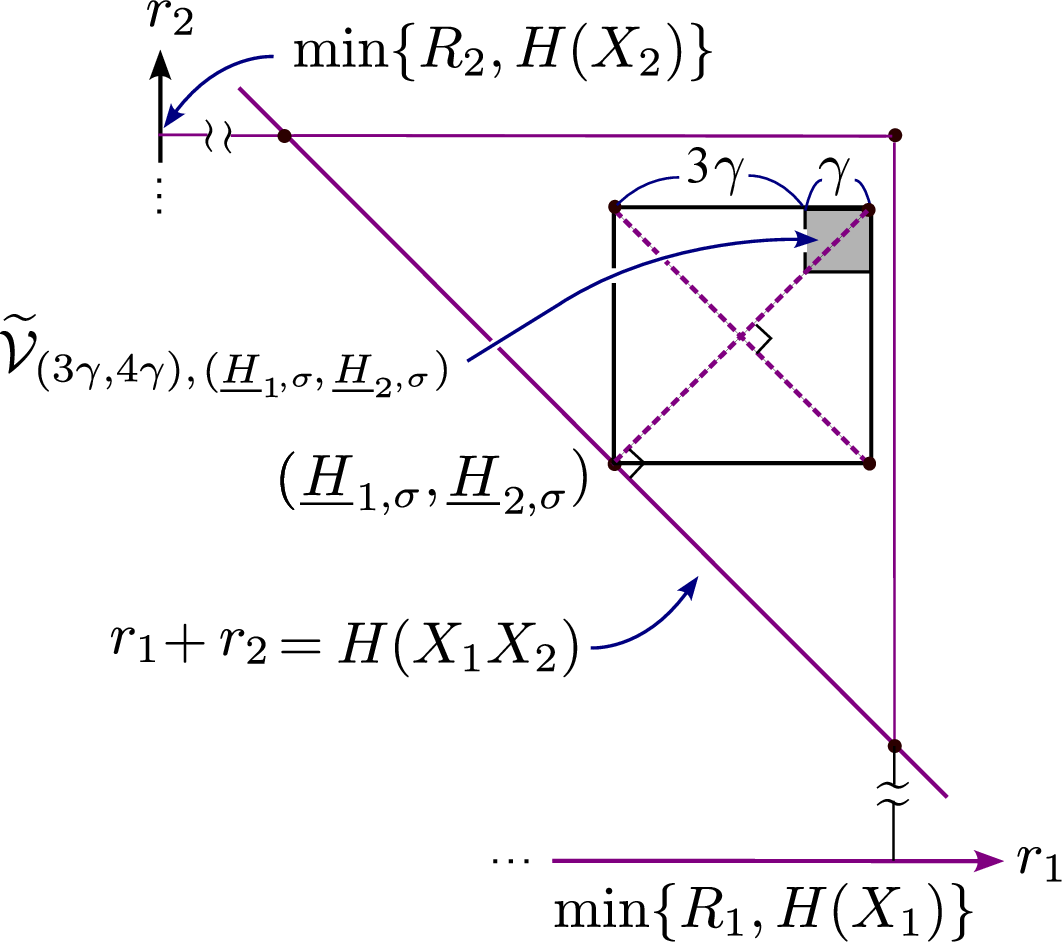}
\caption{The set 
$\wt{\cal V}_{(3\gamma,4\gamma),
(\underline{H}_{1,\sigma},\underline{H}_{2,\sigma})}
$ related to the case of $\underline{I}=0$.
}
\label{fig:tiVSet}
\end{figure}
Let $(\wt{R}_1,\wt{R}_2)\in {\cal S}_{\rm sw}(p_{X_1X_2})$.
For $0\leq \nu_2< 2\nu_1$, we set 
\begin{align*}
{\cal V}_{(\nu_1,\nu_2)}(\wt{R}_1,\wt{R}_2)
\defeq  &\{(r_1,r_2):  |r_i-\wt{R}_i|\leq\nu_1,\\
        & \quad r_1+r_2 \geq H(X_1X_2)+\nu_2\}.       
\end{align*} 
Specifically, when $\nu_2=0$, we write 
${\cal V}_{(\nu_1,0)}(\wt{R}_1,\wt{R}_2)$
as ${\cal V}_{\nu_1}($ $\wt{R}_1,\wt{R}_2)$.
When $\underline{I}>0$, we choose $\gamma_0$ 
so that $4\gamma_0\in(0,\underline{I}]$ as stated 
in (\ref{eqn:GammaZero}). 
Then $\exists (\wt{R}_1,\wt{R}_2)
\in {\cal S}_{\rm sw}(p_{X_1X_2})$ 
such that 
\begin{align}
{\cal V}_{2\gamma_0}(\wt{R}_1,\wt{R}_2)\subseteq 
{\cal V}_{(\underline{H}_{1,\sigma},\underline{H}_{2,\sigma})}.  
\label{eqn:VsetRelZero}
\end{align}
Furthermore, $\forall\gamma\in (0,\gamma_0]$, 
we have the following: 
\begin{align}
\{(\wt{R}_1,\wt{R}_2)\}
\subseteq & {\cal V}_{2\gamma}(\wt{R}_1,\wt{R}_2)
\notag\\
\subseteq & {\cal V}_{2\gamma_0}(\wt{R}_1,\wt{R}_2)
\subseteq {\cal V}_{(\underline{H}_{1,\sigma},\underline{H}_{2,\sigma})}.
\end{align}
Note that $\forall\gamma\in (0,\gamma_0]$, 
\begin{align}
{\cal V}_{(2\gamma,3\gamma)}(\wt{R}_1,\wt{R}_2)
=
{\cal V}_{2\gamma}(\wt{R}_1,\wt{R}_2)
\cap 
{\cal V}_{3\gamma,(\underline{H}_{1,\sigma},\underline{H}_{2,\sigma})}.
\label{eqn:VsetRel}
\end{align}
We show the four sets
${\cal V}_{(\underline{H}_{1,\sigma},\underline{H}_{2,\sigma})},
 {\cal V}_{2\gamma_0}(\wt{R}_1,\wt{R}_2),
 {\cal V}_{2\gamma}(\wt{R}_1,\wt{R}_2$ $)$,
and ${\cal V}_{(2\gamma,3\gamma)}
($$\wt{R}_1,\wt{R}_2)$ in Fig. \ref{fig:VSets}.
When $\underline{I}=0$, we set 
\begin{align*}
& \wt{\cal V}_{(\nu_1,\nu_2), (\underline{H}_{1,\sigma},
                            \underline{H}_{2,\sigma})} 
\\
& \defeq\{(r_1,r_2): 
      r_i- \underline{H}_{i,\sigma}\in [\nu_1,\nu_2], i=1,2 \}.
\end{align*}
By definition it is obvious that
\begin{align}
\wt{\cal V}_{(3\gamma,4\gamma),
(\underline{H}_{1,\sigma},\underline{H}_{2,\sigma})}
\subseteq
\wt{\cal V}_{3\gamma,
(\underline{H}_{1,\sigma},\underline{H}_{2,\sigma})}.
\label{eqn:tiVsetRel}
\end{align}
We show the set 
$\wt{\cal V}_{(3\gamma,4\gamma),
(\underline{H}_{1,\sigma},\underline{H}_{2,\sigma})}$ 
in Fig. \ref{fig:tiVSet}.

\subsection{
Proof of Proposition \ref{pro:BaseProb}
}

In this subsection we prove  Proposition \ref{pro:BaseProb}.
Considering (\ref{eqn:VsetRel}) and (\ref{eqn:tiVsetRel}), 
we obtain the following corollary from Lemma \ref{lm:PrUbLm}. 
\begin{corollary}\label{cor:PrUbLmB}
We have the following:  
\begin{itemize}
 \item[a)] We consider the case of 
 $
 \underline{I}=\underline{H}_{1,\sigma}+\underline{H}_{2,\sigma}-H(X_1X_2)>0.$ 
 In this case we choose $\gamma_0>0$ so that  
 $4\gamma_0\in (0, \underline{I}]$ as 
 stated in (\ref{eqn:GammaZero}).
 We further choose 
 $(\wt{R}_1,\wt{R}_2) 
  \in {\cal S}_{\rm sw}(p_{X_1X_2})$ 
  so that 
$
{\cal V}_{2\gamma_0}(\wt{R}_1,\wt{R}_2)\subseteq 
{\cal V}_{(\underline{H}_{1,\sigma},\underline{H}_{2,\sigma})}.  
$
As shown in (\ref{eqn:VsetRelZero}), this choice of 
$(\wt{R}_1,\wt{R}_2) \in {\cal S}_{\rm sw}(p_{X_1X_2})$   
is possible. 
Then $\forall \gamma \in (0,\gamma_0]$ and  
      $\forall(r_1^{(n)},r_2^{(n)})
       \in {\cal V}_{(2\gamma,3\gamma)} 
       (\wt{R}_1,\wt{R}_2)$, 
       we have
\begin{align}
&\Theta_n\left(\gamma,r_1^{(n)},r_2^{(n)}\right)
\leq \eta_n(\gamma)+\theta_{n}(\sigma,\gamma).
\notag  
\end{align}
 \item[b)] 
  We consider the case of 
  $\underline{H}_{1,\sigma}+\underline{H}_{2,\sigma}=H(X_1X_2)$. 
  In this case we choose 
  $\gamma>0$ sufficiently small. 
  Then for any $(r_1^{(n)},r_2^{(n)})$
  satisfying $(r_1^{(n)},r_2^{(n)})
  \in \wt{\cal V}_{(3\gamma,4\gamma),
   (\overline{H}_{1,\sigma},\overline{H}_{2,\sigma})}$,
   we have 
\begin{align}
&\Theta_n\left(\gamma,r_1^{(n)},r_2^{(n)}\right)
\leq \eta_n(\gamma)+\theta_{n}(\sigma,\gamma).
\notag  
\end{align}
\end{itemize}
\end{corollary}

\begin{IEEEproof}[Proof of Proposition \ref{pro:BaseProb}]
Fix a pair $({\errP},{\secP}) 
        \in  (0,1)\times [0,{\secP}_0]$, arbitrary.  
We start from the assumption that 
$(R_1,R_2)\in {\cal R}_{\rm II}^{\ast}({\secP}$ $|
 p_{X_1X_2},p_{K_1K_2})$. 
Under this assumption we have a sequence 
  $\{(\Phi_1^{(n)},\Phi_2^{(n)},$ $\Psi^{(n)})\}_{n \geq 1}$
  such that $\forall \gamma >0$ and $\forall \errP >0$,
  $\exists n_0=n_0(\gamma,\errP) \in \mathbb{N}$, 
  $\forall n\geq n_0$, we have 
\begin{align*}
   &\frac{1}{n} \log |{\cal M}_i^{(n)}| 
          \leq  R_i+ \gamma,\: i=1,2,
\\  				
  & p_{{\rm e}}^{(n)}(
  \phi_1^{(n)},
  \phi_2^{(n)},
  \psi^{(n)}|{p}_{X_1X_2}^n) \leq \errP,
\\
   & I(C_1^{(n)}C_2^{(n)};\rvcxone\rvcxtwo)\leq \secP.
\end{align*}
We define a {\it new data transmission scheme} 
based on the above sequence 
$\{(\Phi_1^{(n)},\Phi_2^{(n)}$, 
$\Psi^{(n)})\}_{n \geq 1}$ which attains  
the rate pair $(R_1,R_2)$ belonging to the 
$\secP$-secure rate region. 
By Lemma \ref{lm:LemRcBdB}, 
we have that there exists at least one deterministic 
code $(\wt{\phi}_1^{(n)},
       \wt{\phi}_2^{(n)},$
     $\wt{\psi}^{(n)})$ such that
$\forall\gamma>0,\forall \errP>0,$
and $\forall n\geq n_0(\gamma,\errP)$, 
\begin{align}
& \wt{p}_{\rm e}^{(n)}=
  \wt{p}_{\rm e}^{(n)}\Bigl(
                \wt{\phi}_1^{(n)}\circ\phi_1^{(n)},
                \wt{\phi}_2^{(n)}\circ\phi_2^{(n)}, 
\notag\\
& \qquad\qquad \psi^{(n)}\circ\wt{\psi}^{(n)}
   \Big|p_{X_1X_2}\Bigr)
\notag\\   
&\leq 3\cdot{\rm 2}^{-n\gamma}+
 \nu_n(\gamma,\errP)
 +\Theta_n\left(\gamma,r_1^{(n)},r_2^{(n)}\right).
\label{eqn:ErUbzOne}
\end{align}
We consider the following two cases: 
\begin{itemize}
\item [$\:$] Case 1: $\underline{I}>0$, i.e., 
$\underline{H}_{1,\sigma}+\underline{H}_{2,\sigma}> H(X_1X_2)$.
\item [$\:$] Case 2: $\underline{I}=0$, i.e., 
$\underline{H}_{1,\sigma}+\underline{H}_{2,\sigma}=H(X_1X_2)$.
\end{itemize}

\noindent 
\underline{\it Case 1:} \ 
We choose $\gamma_0$ so that $4\gamma_0\in (0,\underline{I}]$. We further choose $(\wt{R}_1,\wt{R}_2) \in {\cal S}_{\rm sw}(p_{X_1X_2})$ so that 
${\cal V}_{2\gamma_0}(\wt{R}_1,\wt{R}_2)\subseteq 
{\cal V}_{(\underline{H}_{1,\sigma},
\underline{H}_{2,\sigma})}.$ 
We choose $\wt{\gamma}$ so that $\wt{\gamma}=2\gamma$. 
Then we have $\gamma=\frac{1}{2}\wt{\gamma}$.
We choose 
$\left\{(r_1^{(n)},r_2^{(n)})\right\}_{n\geq 1}$ 
so that 
\begin{align}
  (r_1^{(n)},r_2^{(n)}) 
 \in {\cal V}_{(2\gamma,3\gamma)} 
        (\wt{R}_1,\wt{R}_2)
   = {\cal V}_{(\wt{\gamma},\frac{3}{2}\wt{\gamma})} 
       (\wt{R}_1,\wt{R}_2).  
\label{eqn:CaseOneRateSet}
\end{align}
From (\ref{eqn:CaseOneRateSet}), we have 
\begin{align}
\frac{1}{n}\log|{\cal L}^{(n)}_i|
=r_i^{(n)}\leq \wt{R}_i
+\wt{\gamma},\:i=1,2.
\label{eqn:RateFin}
\end{align}
By Corollary \ref{cor:PrUbLmB} part a), we have 
that $\forall \wt{\gamma} \in (0,2\gamma_0]$ 
and $\forall (r_1^{(n)},r_2^{(n)})$ satisfying 
(\ref{eqn:CaseOneRateSet}), 
\begin{align*}
\Theta_n\left({\ts\frac{1}{2}}\wt{\gamma},
        r_1^{(n)},r_2^{(n)}\right)&
 \leq \eta_n\left({\ts\frac{1}{2}}\wt{\gamma}\right)
 +\theta_{n}\left(\sigma,{\ts\frac{1}{2}}\wt{\gamma}\right),
\end{align*}
which together with the bound (\ref{eqn:ErUbzOne}) 
with the choice $\gamma=\frac{1}{2}\wt{\gamma}$ 
yields that for $n\geq n_0$, 
\begin{align}
   \wt{p}_{\rm e}^{(n)}
&\leq  3\cdot{\rm 2}^{-\frac{n}{2} \wt{\gamma}}+
 \nu_n\left({\ts\frac{1}{2}}\wt{\gamma},\errP 
       \right)
     +\eta_n\left({\ts\frac{1}{2}}\wt{\gamma}
              \right)
\notag\\
& \quad +\theta_{n}\left(\sigma,
                    {\ts\frac{1}{2}}\wt{\gamma}
                    \right).
\label{eqn:ErBdTheta}
\end{align}
According to Property \ref{pr:InfSpecLm} part b), we have the following upper bound of $\eta_n\left({\ts\frac{1}{2}}\wt{\gamma}\right)$: 
\begin{align}
\eta_n\left({\ts\frac{1}{2}}\wt{\gamma}\right)
\leq  \frac{8}{n\wt{\gamma}}\log\left(\frac{{\rm e}^{2{\rm e}^{-1}}}
    {1-\nu_n\left(\frac{1}{2}\wt{\gamma},
    \varepsilon\right)}\right).
\label{eqn:ErBdEta}    
\end{align}
From (\ref{eqn:ErBdTheta}) and (\ref{eqn:ErBdEta}), 
we have for $n\geq n_0$, 
\begin{align}
   \wt{p}_{\rm e}^{(n)}
&\leq  3\cdot{\rm 2}^{-\frac{n}{2} \wt{\gamma}}+
 \nu_n\left({\ts\frac{1}{2}}\wt{\gamma},\errP 
       \right)
     +\frac{8}{n\wt{\gamma}}\log\left(\frac{{\rm e}^{2{\rm e}^{-1}}}
     {1-\nu_n\left(\frac{1}{2}\wt{\gamma},\varepsilon\right)}\right)
\notag\\
& \quad
 +\theta_{n}
 \left(\sigma,{\ts\frac{1}{2}}\wt{\gamma}\right)
=\varepsilon +\xi_n(\sigma, \wt{\gamma},{\errP}).
\label{eqn:ErBdAll}
\end{align}
Here we set 
\begin{align}
&\xi_n(\sigma,\wt{\gamma},{\errP})
 \defeq 3\cdot{\rm 2}^{-\frac{n}{2} \wt{\gamma}}+
 \nu_n\left({\ts\frac{1}{2}}\wt{\gamma}
       \right)  
\notag\\  
&\quad +\frac{8}{n\wt{\gamma}}\log\left(\frac{{\rm e}^{2{\rm e}^{-1}}}
     {1-{\errP}-\nu_n\left(\frac{1}{2}\wt{\gamma}\right)}\right)
+\theta_{n}
 \left(\sigma,{\ts\frac{1}{2}}\wt{\gamma}\right).
\notag
\end{align}
The first three terms of $\xi_n(\sigma,\wt{\gamma},{\errP})$
vanish as $n\to\infty$. 
Then, by Lemma \ref{lm:LimTheta}, 
under the condition that $p_{X_1X_2}$ 
is indecomposable, $\exists \{k_n\}_{n\geq 1}$ and $\omega\in (0,\frac{1}{4}]$ such that  
for each fixed $\wt{\gamma}\in (0, 4\sigma)$ 
and $\errP>0$, we have   
\begin{align}
& \lim_{n\to\infty} \xi_{k_n}(\sigma,\wt{\gamma},{\errP})
 =\lim_{n\to\infty} 
\theta_{k_n}(\sigma,{\ts \frac{1}{2}}\wt{\gamma})
\leq {\theTa}\in (0,1).    
\end{align}
Fix $\kappa\in(0,1)$ arbitrary and choose 
$\tau>0$ such that 
$\tau\in (0,\kappa(1-{\theTa})]$.
We choose $\errP={\ts \frac{1}{2}}\tau$. 
We further choose $n_1=n_1(\wt{\gamma},\tau)\in \mathbb{N}$ 
such that $\forall n\geq n_1$, 
\begin{align}
 \xi_{k_n}(\sigma,\wt{\gamma},{\errP})
 = \xi_{k_n}(\sigma,\wt{\gamma},{\ts \frac{1}{2}}\tau)
\leq {\ts \frac{1}{2}}\tau + {\theTa}. 
\label{eqn:limxi}
\end{align}
We set $n_2=n_2(\wt{\gamma},\tau)$ such that
$$
n_2(\wt{\gamma},\tau)=\max\{
n_0\left({\ts\frac{1}{2}}\wt{\gamma},{\ts \frac{1}{2}\tau}\right),n_1(\wt{\gamma},\tau)\}.$$ 
Hence, by (\ref{eqn:ErBdAll}) and (\ref{eqn:limxi}), 
for each fixed $\kappa\in(0,1)$, 
$\forall\tau\in (0,\kappa(1-{\theTa})]$ and    
$\forall n\geq n_2$, we have that $k_n\geq n\geq n_2$ and 
\begin{align}
&\wt{p}_{\rm e}^{(k_n)}=\Pr\Bigl\{
\psi^{(k_n)}\circ\wt{\psi}^{(k_n)}
(\wt{\phi}_1^{(k_n)}(M_1^{(k_n)}),
 \wt{\phi}_2^{(k_n)}(M_2^{(k_n)}))
\notag\\
&\qquad \quad 
\neq (\rvcxone,\rvcxtwo)\Bigr\}
\leq \errP+{\ts \frac{1}{2}}\tau+{\theTa}
= \tau +{\theTa}.
\label{eqn:ErBdAllFin}
\end{align}
From (\ref{eqn:SecBdFinO}), (\ref{eqn:RateFin}), 
and  (\ref{eqn:ErBdAllFin}), we have that 
$\forall\tau\in (0,\kappa(1-{\theTa})]$, 
$$
(\wt{R}_1,\wt{R}_2)\in
\overline{\cal S}^{\ast}(\tau+{\theTa},{\secP}|p_{X_1X_2},p_{K_1K_2}).
$$

\noindent
\underline{\it Case 2:} \ We choose 
$\wt{R}_i=\underline{H}_{i,\sigma},i=1,2$.
Since 
\begin{align*}
&\wt{R}_i=\underline{H}_{i,\sigma}
\leq
\overline{H}_{i}\leq \min\{R_i,H(X_i)\},i=1,2,  
 \\
&\wt{R}_1+\wt{R}_2=H(X_1X_2),
\end{align*}
$(\wt{R}_1,\wt{R}_2)\in {\cal S}_{\rm sw}(p_{X_1X_2})$. 
We choose $\wt{\gamma}$ so that $\wt{\gamma}=4\gamma$. 
Then we have $\gamma=\frac{1}{4}\wt{\gamma}$.  
We choose $\left\{(r_1^{(n)},r_2^{(n)})\right\}_{n\geq 1} $ 
so that 
\begin{align}
   (r_1^{(n)},r_2^{(n)}) 
   &\in {\cal V}_{(3\gamma,4\gamma),
      (\underline{H}_{1,\sigma},\underline{H}_{2,\sigma})}   
\notag\\      
  &\quad = {\cal V}_{(\frac{3}{4}\wt{\gamma},\wt{\gamma}),
      (\underline{H}_{1,\sigma},\underline{H}_{2,\sigma})}.     
\label{eqn:CaseOneRateSetb}
\end{align}
From (\ref{eqn:CaseOneRateSetb}), we have 
\begin{align}
\frac{1}{n}\log|{\cal L}^{(n)}_i|
=r_i^{(n)}\leq \wt{R}_i
+\wt{\gamma},\:i=1,2.
\label{eqn:RateFinb}
\end{align} 
By Corollary \ref{cor:PrUbLmB} part b), we have 
that $\forall \wt{\gamma} >0$ and 
$\forall (r_1^{(n)},r_2^{(n)})$ satisfying 
(\ref{eqn:CaseOneRateSetb}), 
\begin{align*}
\Theta_n\left({\ts\frac{1}{4}}\wt{\gamma},r_1^{(n)},r_2^{(n)}\right)&
 \leq \eta_n\left({\ts\frac{1}{4}}\wt{\gamma}\right)
+\theta_{n}\left(\sigma,{\ts\frac{1}{4}}\wt{\gamma}\right),
\end{align*}
which together with the bound (\ref{eqn:ErUbzOne}) 
with the choice $\gamma=\frac{1}{4}\wt{\gamma}$ 
yields that for $n\geq n_0$,   
\begin{align}
   \wt{p}_{\rm e}^{(n)}
&\leq  3\cdot{\rm 2}^{-\frac{n}{4} \wt{\gamma}}+
  \nu_n\left({\ts\frac{1}{4}}\wt{\gamma},\errP 
       \right)
     +\eta_n\left({\ts\frac{1}{4}}\wt{\gamma}
              \right)
\notag\\
& \quad +\theta_{n}\left(\sigma,
                    {\ts\frac{1}{4}}\wt{\gamma}
                    \right).
\label{eqn:ErBdThetabc}
\end{align}
According to Property \ref{pr:InfSpecLm} part b), we have the following upper bound of $\eta_n\left({\ts\frac{1}{4}}\wt{\gamma}\right)$: 
\begin{align}
\eta_n\left({\ts\frac{1}{4}}\wt{\gamma}\right)
\leq  \frac{16}{n\wt{\gamma}}\log\left(\frac{{\rm e}^{2{\rm e}^{-1}}}
    {1-\nu_n\left(\frac{1}{4}\wt{\gamma},\varepsilon\right)}\right).
   \label{eqn:ErBdEtabc}    
\end{align}
From (\ref{eqn:ErBdThetabc}) and (\ref{eqn:ErBdEtabc}), 
we have for $n\geq n_0$,
\begin{align}
   \wt{p}_{\rm e}^{(n)}
&\leq  3\cdot{\rm 2}^{-\frac{n}{4} \wt{\gamma}}+
 \nu_n\left({\ts\frac{1}{4}}\wt{\gamma},\errP 
       \right)
     +\frac{16}{n\wt{\gamma}}\log
      \left(\frac{{\rm e}^{2{\rm e}^{-1}}}
     {1-\nu_n\left(\frac{1}{4}\wt{\gamma},\varepsilon\right)}\right)
\notag\\
& \quad
 +\theta_{n}
 \left(\sigma,{\ts\frac{1}{4}}\wt{\gamma}\right)
=\varepsilon +\wt{\xi}_n(\sigma, \wt{\gamma},{\errP}).
\label{eqn:ErBdAllbc}
\end{align}
Here we set 
\begin{align}
&\wt{\xi}_n(\sigma,\wt{\gamma},{\errP})
 \defeq 3\cdot{\rm 2}^{-\frac{n}{4} \wt{\gamma}}+
 \nu_n\left({\ts\frac{1}{4}}\wt{\gamma}
       \right)
  \notag\\  
&\quad +\frac{16}{n\wt{\gamma}}\log\left(\frac{{\rm e}^{2{\rm e}^{-1}}}
     {1-{\errP}-\nu_n\left(\frac{1}{4}\wt{\gamma}\right)}\right)
+\theta_{n}
 \left(\sigma,{\ts\frac{1}{4}}\wt{\gamma}\right).
\notag
\end{align}
Similarly to (\ref{eqn:limxi}), by Lemma \ref{lm:LimTheta}, 
$\exists \{k_n\}_{n\geq 1}$  and $\omega\in (0,1)$ such that for each fixed $\wt{\gamma}\in (0,8\sigma)$ 
and $\errP>0$, 
\begin{align}
 \lim_{n\to\infty} \wt{\xi}_{k_n}(\sigma,\wt{\gamma},{\errP})
=\lim_{n\to\infty} 
\theta_{k_n}(\sigma,{\ts \frac{1}{4}}\wt{\gamma})
\leq {\theTa}\in (0,1).
 \label{eqn:limxib}  
\end{align}
In a manner quite similar to the Case 1, we can 
show that for each fixed $\kappa\in(0,1)$ and  
$\forall\tau\in (0,\kappa(1-{\theTa})]$, 
$\exists n_3=n_3(\wt{\gamma},\tau)\in \mathbb{N}$ 
such that $\forall n\geq n_3$, 
\begin{align}
&\wt{p}_{\rm e}^{(k_n)}=\Pr\Bigl\{
\psi^{(k_n)}\circ\wt{\psi}^{(k_n)}
(\wt{\phi}_1^{(k_n)}(M_1^{(k_n)}),
\wt{\phi}_2^{(k_n)}({M}_2^{(k_n)}))
\notag\\
&\qquad \quad \quad \neq (\rvcxone,\rvcxtwo)\Bigr\}
\leq \tau+{\theTa}.
\label{eqn:ErBdAllFinb}
\end{align}
From (\ref{eqn:SecBdFinO}), 
     (\ref{eqn:RateFinb}), 
and  (\ref{eqn:ErBdAllFinb}), we have that 
$\forall\tau
\in (0,\kappa(1-{\theTa})]$, 
$$
(\wt{R}_1,\wt{R}_2)\in
\overline{\cal S}^{\ast}
(\tau+{\theTa},{\secP}|p_{X_1X_2},p_{K_1K_2}).
$$
Thus Proposition \ref{pro:BaseProb} is proved. 
\end{IEEEproof}


\subsection{
Proof of Proposition \ref{pro:BasePro}}
\begin{figure}[t]
\centering
\includegraphics[width=0.47\textwidth]{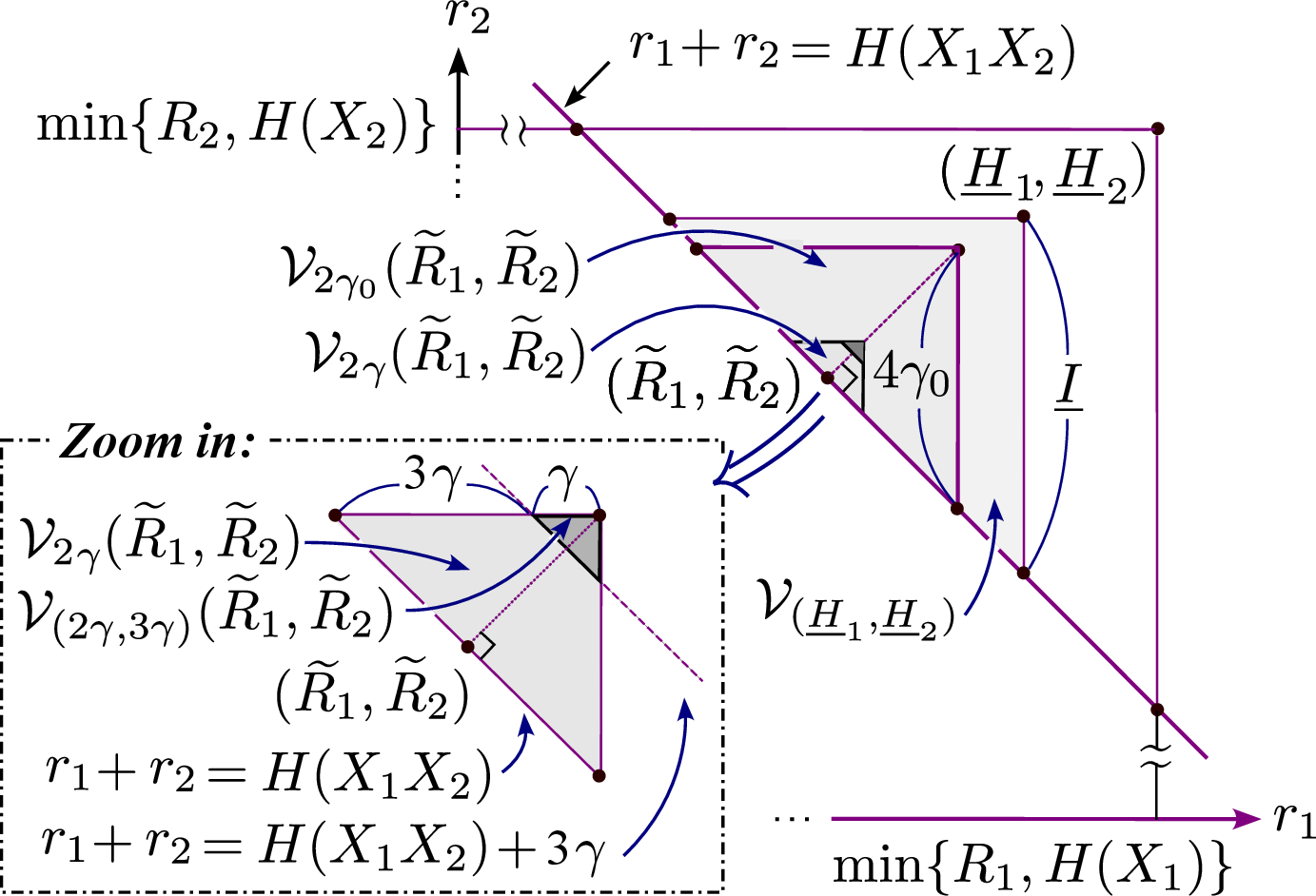}
\caption{
Shapes of the four sets ${\cal V}_{(\underline{H}_{1},\underline{H}_{2})},
{\cal V}_{2\gamma_0}(\wt{R}_1,\wt{R}_2),
{\cal V}_{2\gamma}(\wt{R}_1,\wt{R}_2),
$ and 
$ {\cal V}_{(2\gamma,3\gamma)}
(\wt{R}_1,\wt{R}_2)$ related to the case of $\underline{I}>0$.
}
\label{fig:VSetsA}
\end{figure}

\begin{figure}[t]
\centering
\includegraphics[width=0.35\textwidth]{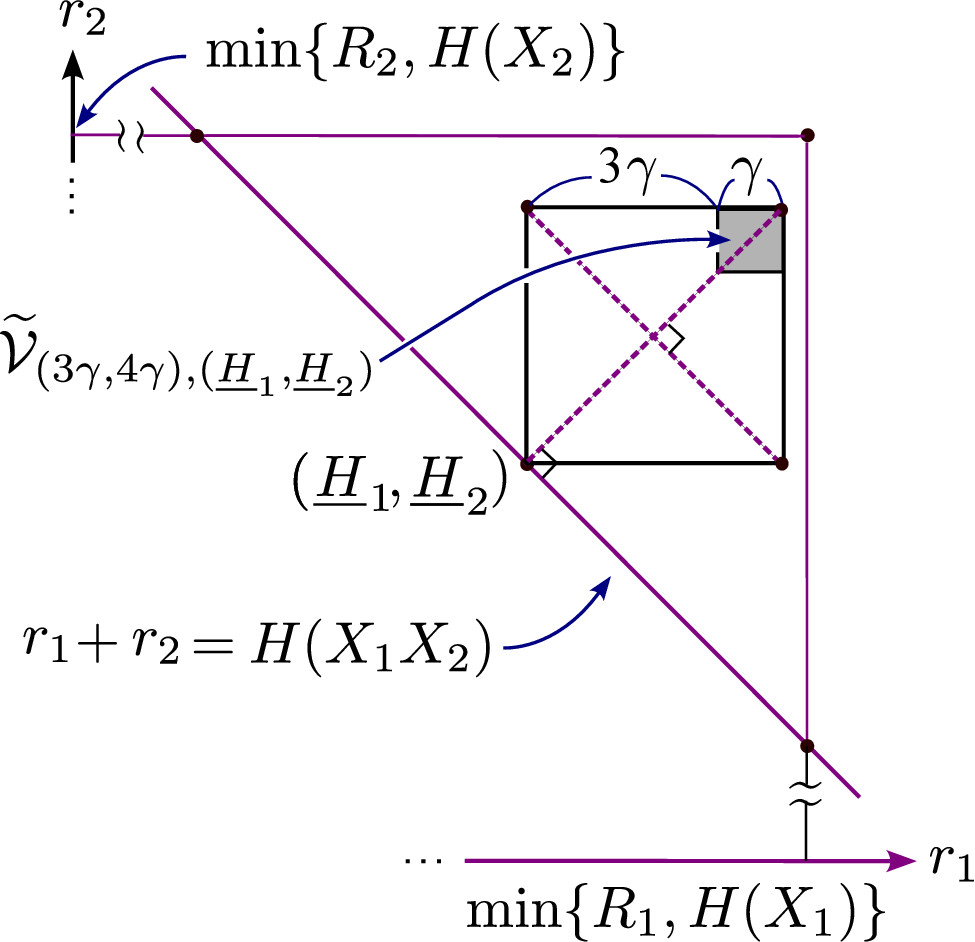}
\caption{The set 
$\wt{\cal V}_{(3\gamma,4\gamma),
(\underline{H}_{1},\underline{H}_{2})}
$ related to the case of $\underline{I}=0$.
}
\label{fig:tiVSetA}
\end{figure}

In this subsection we prove Proposition \ref{pro:BasePro}.
It suffices to consider the case of $\sigma=0$, 
to prove this proposition. 

When $\underline{I}>0$, 
we choose $\gamma_0>0$ so that
\begin{align}
\underline{I}=  \underline{H}_{1}
              + \underline{H}_{2}
               -H(X_1X_2) \geq 4\gamma_0.
\label{eqn:GammaZeroB}
\end{align}
Then $\exists (\wt{R}_1,\wt{R}_2)
\in {\cal S}_{\rm sw}(p_{X_1X_2})$ 
such that 
\begin{align}
{\cal V}_{2\gamma_0}(\wt{R}_1,\wt{R}_2)\subseteq 
{\cal V}_{(\underline{H}_{1},\underline{H}_{2})}.  
\label{eqn:VsetRelZerob}
\end{align}
Furthermore, $\forall\gamma\in (0,\gamma_0]$, 
we have the following: 
\begin{align}
\{(\wt{R}_1,\wt{R}_2)\}
\subseteq & {\cal V}_{2\gamma}(\wt{R}_1,\wt{R}_2)
\notag\\
\subseteq & {\cal V}_{2\gamma_0}(\wt{R}_1,\wt{R}_2)
\subseteq 
{\cal V}_{(\underline{H}_{1},\underline{H}_{2})}.
\end{align}
Note that $\forall\gamma\in (0,\gamma_0]$, 
\begin{align}
{\cal V}_{(2\gamma,3\gamma)}(\wt{R}_1,\wt{R}_2)
=
{\cal V}_{2\gamma}(\wt{R}_1,\wt{R}_2)
\cap 
{\cal V}_{3\gamma,(\underline{H}_{1},\underline{H}_{2})}.
\label{eqn:VsetRelb}
\end{align}
\newcommand{\FigofHOne}{}{
We show the four sets
${\cal V}_{(\underline{H}_{1},\underline{H}_{2})},
 {\cal V}_{2\gamma_0}(\wt{R}_1,\wt{R}_2),
 {\cal V}_{2\gamma}(\wt{R}_1,\wt{R}_2)$,
and ${\cal V}_{(2\gamma,3\gamma)}
($$\wt{R}_1,\wt{R}_2)$ in Fig. \ref{fig:VSetsA}.
}
When $\underline{I}=0$, we consider the set 
\begin{align*}
& \wt{\cal V}_{(\nu_1,\nu_2), (\underline{H}_{1},
                            \underline{H}_{2})} 
=\{(r_1,r_2): 
      r_i- \underline{H}_{i}\in [\nu_1,\nu_2], i=1,2 \}.
\end{align*}
By definition it is obvious that
\begin{align}
\wt{\cal V}_{(3\gamma,4\gamma),
(\underline{H}_{1},\underline{H}_{2})}
\subseteq
\wt{\cal V}_{3\gamma,
(\underline{H}_{1},\underline{H}_{2})}.
\label{eqn:tiVsetRelZ}
\end{align}
\newcommand{\FigofHtwo}{}{
We show the set 
$\wt{\cal V}_{(3\gamma,4\gamma),
(\underline{H}_{1},\underline{H}_{2})}$ 
in Fig. \ref{fig:tiVSetA}.
}

Considering (\ref{eqn:VsetRelb}) and (\ref{eqn:tiVsetRelZ}), 
we obtain the following corollary from Lemma \ref{lm:PrUbLm}. 
\begin{corollary}\label{cor:PrUbLmC}
We have the following:  
\begin{itemize}
 \item[a)] We consider the case of 
 $\underline{I}=\underline{H}_1
 +\underline{H}_2-H(X_1X_2)>0.$ 
 In this case we choose $\gamma_0>0$ so that  
 $4\gamma_0\in (0, \underline{I}]$ as stated 
 in (\ref{eqn:GammaZeroB}). We further choose 
 $(\wt{R}_1,\wt{R}_2) 
  \in {\cal S}_{\rm sw}(p_{X_1X_2})$ 
  so that
$ {\cal V}_{2\gamma_0}(\wt{R}_1,\wt{R}_2)\subseteq 
{\cal V}_{(\underline{H}_1,\underline{H}_2)}.$
As shown in (\ref{eqn:VsetRelZerob}), this 
choice of $(\wt{R}_1,\wt{R}_2) \in 
{\cal S}_{\rm sw}(p_{X_1X_2})$ is possible. 
Then $\forall \gamma \in (0,\gamma_0]$ 
and  $\forall(r_1^{(n)},r_2^{(n)})
     \in {\cal V}_{(2\gamma,3\gamma)} 
        (\wt{R}_1,\wt{R}_2)$, we have 
\begin{align}
&\Theta_n\left(\gamma,r_1^{(n)},r_2^{(n)}\right)
 \leq \eta_n(\gamma)+\theta_{n}(\gamma). \notag  
\end{align}
 \item[b)] 
  We consider the case of 
  $\underline{H}_1+\underline{H}_2=H(X_1X_2)$. 
  In this case we choose 
  $\gamma>0$ sufficiently small. 
  Then for any $(r_1^{(n)},r_2^{(n)})$
  satisfying $(r_1^{(n)},r_2^{(n)})
  \in \wt{\cal V}_{(3\gamma,4\gamma),
   (\overline{H}_1,\overline{H}_2)}$,
   we have 
\begin{align}
&\Theta_n\left(\gamma,r_1^{(n)},r_2^{(n)}\right)
\leq \eta_n(\gamma)+\theta_{n}(\gamma).
\notag  
\end{align}
\end{itemize}
\end{corollary}

\begin{IEEEproof}[Proof of Proposition \ref{pro:BasePro}]
Fix a pair $({\errP},{\secP}) 
        \in  (0,1)\times [0,{\secP}_0]$, arbitrary.  
We start from the assumption that 
$(R_1,R_2)\in {\cal R}_{\rm III}^{\ast}
($ ${\errP}, {\secP}|
 p_{X_1X_2},p_{K_1K_2})$. 
Under this assumption we have a sequence 
  $\{(\Phi_1^{(n)},\Phi_2^{(n)},$ $\Psi^{(n)})\}_{n \geq 1}$
  such that $\forall \gamma >0$,
  $\exists n_0=n_0(\gamma) \in \mathbb{N}$, 
  $\forall n\geq n_0$, we have 
\begin{align*}
   &\frac{1}{n} \log |{\cal M}_i^{(n)}| 
          \leq  R_i+ \gamma,\: i=1,2,
\\  				
  & p_{{\rm e}}^{(n)}(
  \phi_1^{(n)},
  \phi_2^{(n)},
  \psi^{(n)}|{p}_{X_1X_2}^n) \leq \errP,
\\
   & I(C_1^{(n)}C_2^{(n)};\rvcxone\rvcxtwo)\leq \secP.
\end{align*}
We define a {\it new data transmission scheme} 
based on the above sequence 
$\{(\Phi_1^{(n)},\Phi_2^{(n)}$, 
$\Psi^{(n)})\}_{n \geq 1}$ which attains the rate 
pair $(R_1,R_2)$ belonging to the $(\errP,\secP)$-reliable and secure rate region. 
By Lemma \ref{lm:LemRcBdB}, 
we have that there exists at least one deterministic 
code $(\wt{\phi}_1^{(n)},
       \wt{\phi}_2^{(n)},$
     $\wt{\psi}^{(n)})$ such that
$\forall\gamma>0$
and $\forall n\geq n_0(\gamma)$,      
\begin{align}
& \wt{p}^{(n)}_{\rm e}=
  \wt{p}^{(n)}_{\rm e}\Bigl(
                \wt{\phi}_1^{(n)}\circ\phi_1^{(n)},
                \wt{\phi}_2^{(n)}\circ\phi_2^{(n)}, 
\notag\\
& \qquad\qquad \psi^{(n)}\circ\wt{\psi}^{(n)}
   \Big|p_{X_1X_2}\Bigr)
\notag\\   
&\leq 3\cdot{\rm 2}^{-n\gamma}+
 \nu_n(\gamma,\errP)
 +\Theta_n\left(\gamma,r_1^{(n)},r_2^{(n)}\right).
\label{eqn:ErUbzOneb}
\end{align}
We consider the following two cases: 
\begin{itemize}
\item [$\:$] Case 1: $\underline{I}>0$, i.e., 
$\underline{H}_{1}+\underline{H}_{2}> H(X_1X_2)$.
\item [$\:$] Case 2: $\underline{I}=0$, i.e., 
$\underline{H}_{1}+\underline{H}_{2}=H(X_1X_2)$.
\end{itemize}

\noindent 
\underline{\it Case 1:} \ 
We choose $\gamma_0$ so that $4\gamma_0\in (0,\underline{I}]$. We further choose $(\wt{R}_1,\wt{R}_2) \in {\cal S}_{\rm sw}(p_{X_1X_2})$ so that 
$
{\cal V}_{2\gamma_0}(\wt{R}_1,\wt{R}_2)\subseteq 
{\cal V}_{(\underline{H}_1,\underline{H}_2)}.  
$
We choose $\wt{\gamma}$ so that $\wt{\gamma}=2\gamma$. 
Then we have $\gamma=\frac{1}{2}\wt{\gamma}$.
We choose 
$\left\{(r_1^{(n)},r_2^{(n)})\right\}_{n\geq 1}$ 
so that 
\begin{align}
  (r_1^{(n)},r_2^{(n)}) 
 \in {\cal V}_{(2\gamma,3\gamma)} 
        (\wt{R}_1,\wt{R}_2)
   = {\cal V}_{(\wt{\gamma},\frac{3}{2}\wt{\gamma})} 
       (\wt{R}_1,\wt{R}_2).  
\label{eqn:CaseOneRateSetaZ}
\end{align}
From (\ref{eqn:CaseOneRateSetaZ}), we have 
\begin{align}
\frac{1}{n}\log|{\cal L}^{(n)}_i|
=r_i^{(n)}\leq \wt{R}_i
+\wt{\gamma},\:i=1,2.
\label{eqn:RateFinbb}
\end{align}
By Corollary \ref{cor:PrUbLmC} part a), we have 
that $\forall \wt{\gamma} \in (0,2\gamma_0]$ 
and $\forall (r_1^{(n)},r_2^{(n)})$ satisfying 
(\ref{eqn:CaseOneRateSetaZ}), 
\begin{align*}
\Theta_n\left({\ts\frac{1}{2}}\wt{\gamma},
        r_1^{(n)},r_2^{(n)}\right)&
 \leq \eta_n\left({\ts\frac{1}{2}}\wt{\gamma}\right)
 +\theta_{n}\left({\ts\frac{1}{2}}\wt{\gamma}\right),
\end{align*}
which together with the bound (\ref{eqn:ErUbzOneb}) 
with the choice $\gamma=\frac{1}{2}\wt{\gamma}$ 
yields that for $n\geq n_0$,  
\begin{align}
   \wt{p}_{\rm e}^{(n)}
&\leq  3\cdot{\rm 2}^{-\frac{n}{2} \wt{\gamma}}+
 \nu_n\left({\ts\frac{1}{2}}\wt{\gamma},\errP 
       \right)
     +\eta_n\left({\ts\frac{1}{2}}\wt{\gamma}
              \right)
+\theta_{n}\left(
                    {\ts\frac{1}{2}}\wt{\gamma}
                    \right).
\label{eqn:ErBdThetab}
\end{align}
According to  Property \ref{pr:InfSpecLm} part b),
we have the following upper bound of $\eta_n\left({\ts\frac{1}{2}}\wt{\gamma}\right)$: 
\begin{align}
\eta_n\left({\ts\frac{1}{2}}\wt{\gamma}\right)
\leq  \frac{8}{n\wt{\gamma}}\log\left(\frac{{\rm e}^{2{\rm e}^{-1}}}
    {1-\nu_n\left(\frac{1}{2}\wt{\gamma},
    \varepsilon\right)}\right).
\label{eqn:ErBdEtab}    
\end{align}
From (\ref{eqn:ErBdThetab}) and (\ref{eqn:ErBdEtab}), 
we have for $n\geq n_0$,
\begin{align}
   \wt{p}_{\rm e}^{(n)}
&\leq  
\varepsilon +\xi_n(\wt{\gamma},{\errP}).
\label{eqn:ErBdAllb}
\end{align}
Here we set 
\begin{align}
&\xi_n(\wt{\gamma},{\errP})
 \defeq 3\cdot{\rm 2}^{-\frac{n}{2} \wt{\gamma}}+
 \nu_n\left({\ts\frac{1}{2}}\wt{\gamma}
       \right)  
\notag\\  
&\quad +\frac{8}{n\wt{\gamma}}\log\left(\frac{{\rm e}^{2{\rm e}^{-1}}}
     {1-{\errP}-\nu_n\left(\frac{1}{2}\wt{\gamma}\right)}\right)
+\theta_{n}
 \left({\ts\frac{1}{2}}\wt{\gamma}\right).
\notag
\end{align}
For each fixed $\wt{\gamma}>0$, we have that  
$$
\lim_{n\to\infty} \xi_n(\wt{\gamma},{\errP})=0.
$$
Hence for some fixed $\kappa\in(0,1)$, we have that
$\forall\tau\in (0,\kappa(1-{\errP})]$, 
$\exists n_1(\wt{\gamma})\in \mathbb{N}$ 
such that $\forall n\geq n_1$, 
\begin{align}
&\wt{p}_{\rm e}^{(n)}=\Pr\Bigl\{
\psi^{(n)}\circ\wt{\psi}^{(n)}
(\wt{\phi}_1^{(n)}(M_1^{(n)}),\wt{\phi}_2^{(n)}(M_2^{(n)}))
\notag\\
&\qquad \quad \quad 
\neq (\rvcxone,\rvcxtwo)\Bigr\}
\leq \errP+\tau.
\label{eqn:ErBdAllFinbb}
\end{align}
From (\ref{eqn:SecBdFinO}), (\ref{eqn:RateFinbb}), and (\ref{eqn:ErBdAllFinbb}), we have that 
$\forall\tau\in (0,\kappa(1-{\errP})]$, 
$$
(\wt{R}_1,\wt{R}_2)\in
{\cal S}^{\ast}({\errP}+\tau,{\secP}|p_{X_1X_2},p_{K_1K_2}).
$$

\noindent
\underline{\it Case 2:} \ We choose 
$\wt{R}_i=\underline{H}_{i},i=1,2$.
Since 
\begin{align*}
&\wt{R}_i=\underline{H}_{i}\leq \min\{R_i,H(X_i)\},i=1,2,  
 \\
&\wt{R}_1+\wt{R}_2=H(X_1X_2),
\end{align*}
$(\wt{R}_1,\wt{R}_2)\in {\cal S}_{\rm sw}(p_{X_1X_2})$. 
We choose $\wt{\gamma}$ so that $\wt{\gamma}=4\gamma$. 
Then we have $\gamma=\frac{1}{4}\wt{\gamma}$.  
We choose $\left\{(r_1^{(n)},r_2^{(n)})\right\}_{n\geq 1} $ 
so that 
\begin{align}
  (r_1^{(n)},r_2^{(n)}) 
 \in {\cal V}_{(3\gamma,4\gamma),
      (\underline{H}_{1},\underline{H}_{2})}    
   = {\cal V}_{(\frac{3}{4}\wt{\gamma},\wt{\gamma}),
      (\underline{H}_{1},\underline{H}_{2})}     
\label{eqn:CaseOneRateSetZ}
\end{align}
From (\ref{eqn:CaseOneRateSetZ}), we have 
\begin{align}
\frac{1}{n}\log|{\cal L}^{(n)}_i|
=r_i^{(n)}\leq \wt{R}_i
+\wt{\gamma},\:i=1,2.
\label{eqn:RateFinZ}
\end{align} 
By Corollary \ref{cor:PrUbLmC} part b), we have 
that $\forall \wt{\gamma} >0$ and 
$\forall (r_1^{(n)},r_2^{(n)})$ satisfying 
(\ref{eqn:CaseOneRateSetZ}), 
\begin{align*}
\Theta_n\left({\ts\frac{1}{4}}\wt{\gamma},r_1^{(n)},r_2^{(n)}\right)&
 \leq \eta_n\left({\ts\frac{1}{4}}\wt{\gamma}\right)
+\theta_{n}\left({\ts\frac{1}{4}}\wt{\gamma}\right),
\end{align*}
which together with the bound (\ref{eqn:ErUbzOneb}) 
with the choice $\gamma=\frac{1}{4}\wt{\gamma}$ 
yields that for $n\geq n_0$,   
\begin{align}
   \wt{p}_{\rm e}^{(n)}
&\leq  3\cdot{\rm 2}^{-\frac{n}{4} \wt{\gamma}}+
  \nu_n\left({\ts\frac{1}{4}}\wt{\gamma},\errP 
       \right)
     +\eta_n\left({\ts\frac{1}{4}}\wt{\gamma}
              \right)
+\theta_{n}\left(
                    {\ts\frac{1}{4}}\wt{\gamma}
                    \right).
\label{eqn:ErBdThetaZ}
\end{align}
According to  Property \ref{pr:InfSpecLm} part b), 
we have the following upper bound 
of $\eta_n\left({\ts\frac{1}{4}}\wt{\gamma}\right)$: 
\begin{align}
\eta_n\left({\ts\frac{1}{4}}\wt{\gamma}\right)
\leq  \frac{16}{n\wt{\gamma}}\log\left(\frac{{\rm e}^{2{\rm e}^{-1}}}
    {1-\nu_n\left(\frac{1}{4}\wt{\gamma},\varepsilon\right)}\right).
   \label{eqn:ErBdEtaZ}    
\end{align}
From (\ref{eqn:ErBdThetaZ}) and (\ref{eqn:ErBdEtaZ}), 
we have for $n\geq n_0$,  
\begin{align}
   \wt{p}_{\rm e}^{(n)}
&\leq  
%
\varepsilon +\wt{\xi}_n(\wt{\gamma},{\errP}).
\label{eqn:ErBdAllZ}
\end{align}
Here we set 
\begin{align}
&\wt{\xi}_n(\wt{\gamma},{\errP})
 \defeq 3\cdot{\rm 2}^{-\frac{n}{4} \wt{\gamma}}+
 \nu_n\left({\ts\frac{1}{4}}\wt{\gamma}
       \right)
  \notag\\  
&\quad +\frac{16}{n\wt{\gamma}}
     \log\left(\frac{{\rm e}^{2{\rm e}^{-1}}}
     {1-{\errP}-\nu_n\left(\frac{1}{4}
      \wt{\gamma}\right)}\right)
+\theta_{n}
 \left({\ts\frac{1}{4}}\wt{\gamma}\right).
\notag
\end{align}
For each fixed $\wt{\gamma}>0$, we have that  
$$
\lim_{n\to\infty} \wt{\xi}_n(\wt{\gamma},{\errP})=0.
$$
Hence for some fixed $\kappa\in(0,1)$, we have that
$\forall\tau\in (0,\kappa(1-{\errP})]$, 
$\exists n_1(\wt{\gamma})\in \mathbb{N}$ 
such that $\forall n\geq n_1$, 
\begin{align}
&\wt{p}_{\rm e}^{(n)}=\Pr\Bigl\{
\psi^{(n)}\circ\wt{\psi}^{(n)}
(\wt{\phi}_1^{(n)}(M_1^{(n)}),\wt{\phi}_2^{(n)}({M}_2^{(n)}))
\notag\\
&\qquad \quad \quad \neq (\rvcxone,\rvcxtwo)\Bigr\}
\leq \errP+\tau.
\label{eqn:ErBdAllFinZ}
\end{align}
From (\ref{eqn:SecBdFinO}), (\ref{eqn:RateFinZ}), and (\ref{eqn:ErBdAllFinZ}), we have that 
$\forall\tau\in (0,\kappa(1-{\errP})]$, 
$$
(\wt{R}_1,\wt{R}_2)\in
{\cal S}^{\ast}({\errP}+\tau,{\secP}|p_{X_1X_2},p_{K_1K_2}).
$$
Thus Proposition \ref{pro:BasePro} is proved. 
\end{IEEEproof}

\appendix

\ProofPrOne
\ProofLemOne
\ProofBaseLm
\ProofPrtiRvAndRv
\PrfLemRcBd
\ProofPrInfSpec
\ProofPrInfSpecB
\PrfLimTheta

\bibliographystyle{IEEEtran}
\bibliography{Isit2026}
\end{document}